\documentclass[10pt,aps,prd,amsmath,amssymb,superscriptaddress,twocolumn,nofootinbib,showpacs,preprintnumbers,amsfonts, notitlepage]{revtex4-1}
\usepackage{XYZ}

\begin{document}
\preprint{JLAB-THY-20-3231}

\title{\boldmath \XYZ spectroscopy at electron-hadron facilities:\\
Exclusive processes}

\author{M.~Albaladejo}
\email{albalade@jlab.org}
\affiliation{\jlab}

\author{A.~N.~Hiller Blin}
\email{ahblin@jlab.org}
\affiliation{\jlab}

\author{A.~Pilloni}
\email{pillaus@jlab.org}
\affiliation{\ect}
\affiliation{\genova}

\author{D.~Winney}
\email{dwinney@iu.edu}
\affiliation{\ceem}
\affiliation{\indiana}

\author{C.~Fern\'andez-Ram\'irez}
\affiliation{\icn}

\author{V.~Mathieu}
\affiliation{\ucm}

\author{A.~Szczepaniak}
\affiliation{\jlab}
\affiliation{\ceem}
\affiliation{\indiana}

\collaboration{Joint Physics Analysis Center}


\newcommand{\jlab}{Theory Center,
Thomas  Jefferson  National  Accelerator  Facility,
Newport  News,  VA  23606,  USA}
\newcommand{\ceem}{Center for  Exploration  of  Energy  and  Matter,
Indiana  University,
Bloomington,  IN  47403,  USA}
\newcommand{\indiana}{Physics  Department,
Indiana  University,
Bloomington,  IN  47405,  USA}
\newcommand{\icn}{Instituto de Ciencias Nucleares, 
Universidad Nacional Aut\'onoma de M\'exico, Ciudad de M\'exico 04510, Mexico}
\newcommand{\ect}{European Centre for Theoretical Studies in Nuclear Physics and related Areas (ECT$^*$) and Fondazione Bruno Kessler, Villazzano (Trento), I-38123, Italy}
\newcommand{\genova}{INFN Sezione di Genova, Genova, I-16146, Italy}
\newcommand{\ucm}{Departamento de F\'isica Te\'orica, Universidad Complutense de Madrid and IPARCOS, 28040 Madrid, Spain}
\begin{abstract}
The next generation of electron-hadron facilities
has the potential for significantly improving our understanding of exotic hadrons. 
The \XYZ states have not been seen in photon-induced reactions so far. Their observation in such processes would provide an independent confirmation of their existence  and  offer new  insights into their internal structure.
Based on the known experimental data and the well-established quarkonium and Regge phenomenology, we give estimates for the exclusive cross sections of
several \XYZ states.
For energies near threshold we expect cross sections of few nanobarns for the \Z  and upwards of tens of nanobarn for the \X, which are well within reach of new facilities. 
\end{abstract}

\maketitle

\section{Introduction}

Since 2003, a plethora of new resonance candidates, commonly  referred to as the 
  \XYZ, appeared in the heavy quarkonium spectrum. Their properties do not fit the expectations for heavy $Q \bar Q$ bound states as predicted by the conventional phenomenology. 
  An exotic composition is most likely required~\cite{Esposito:2016noz,*Guo:2017jvc,*Olsen:2017bmm,*Brambilla:2019esw}. 
  Having a comprehensive description of these states  
   will improve our
understanding of the nonperturbative features of Quantum Chromodynamics. 
The majority of these has been observed in specific  
  production channels, most notably in heavy hadron decays and  direct production in $e^+e^-$ collisions. Exploring alternative production mechanisms would provide complementary information, that can further shed light on their nature. In particular, photoproduction at high energies is not affected by 3-body dynamics which complicates the determination of the resonant nature of several \XYZ~\cite{Szczepaniak:2015eza,*Albaladejo:2015lob,*Guo:2016bkl,*Pilloni:2016obd,*Nakamura:2019btl,*Guo:2019twa}. 

Photons are efficient probes of  the internal structure of hadrons, and their collisions with hadron targets result in a copious production of meson and baryon resonances. 
Searches for \XYZ in existing experiments, \ie COMPASS~\cite{Adolph:2014hba,*Aghasyan:2017utv} or the Jefferson Lab~\cite{Meziani:2016lhg,Ali:2019lzf,Brodsky:2020vco}, have produced limited results so far. However the situation can change 
   significantly if higher 
 luminosity is reached in the appropriate energy range.

\begin{table*}
    \begin{ruledtabular}
    \begin{tabular}{c c c c c}
         $V$ & $m_V$  (\nsmev) & $\Gamma_V$ (\nskev) & $\mB(V \to e^+e^-)$ (\%)& $f_V$  (\nsmev) 
         \\
         \hline
         $J/\psi$ & $3096.900 \pm 0.006$ & $92.9 \pm 2.8$ & $5.971 \pm 0.032$ & $ 277.5 \pm 4.2$
         \\ 
         $\Upsilon(1S)$ & $9460.30 \pm 0.26$ & $54.02 \pm 1.25$ & $2.38 \pm 0.11$ & $233.45 \pm 6.03$ 
         \\
         $\Upsilon(2S)$ & $10023.26\pm 0.31$ & $31.98\pm2.63$ & $1.91 \pm 0.16$ & $165.63 \pm 9.72$
         \\
         $\Upsilon(3S)$ & $10355.2\pm 0.5$ & $20.32 \pm 1.85$ & $2.18 \pm 0.20$ & $143.1 \pm 9.7$
    \end{tabular}
    \end{ruledtabular}
    \caption{Input parameters for VMD ($\gamma V$) couplings in \cref{eq:amplitude}}
    \label{tab:VMD}
\end{table*}
 The next generation of lepton-hadron facilities includes, for example,  the 
 Electron-Ion Collider (EIC)~\cite{Accardi:2012qut,*eic}
 that is projected to have 
  the center-of-mass energy per  electron-nucleon collision in the range from 20 to 140\gev, and  a peak luminosity of $1.2\E{34}~\text{cm}^{-2}~\text{s}^{-1}$  in the middle of this range.
 The ion beam can cover a large number of species, from proton to uranium. Both the electron and ion beam can be polarized. An Electron-Ion Collider in China (EicC) has also been proposed~\cite{Chen:2018wyz}. 

In this paper, we aim at providing estimates for exclusive photoproduction cross sections of \XYZ states in a wide kinematic range, from near threshold to that expected to be covered by the EIC. While cross sections of exclusive reactions are expected to be smaller than the inclusive ones, the constrained kinematics makes the identification of the signal less ambiguous and can 
determine precisely
  the  production mechanism. The analysis of semi-inclusive processes will be the subject  of a forthcoming work~\cite{jpacinpreparation}. Since the many \XYZ  states have been seen with a varying  degree of significance, we present numerical estimates for the few that
  are considered more robust, \ie seen in more than one channel with high significance. The possible extensions to other states are commented in the text. 
To make our predictions as agnostic as possible to the nature of the \XYZ, we rely on their measured branching fractions and infer other 
  properties from well-established quarkonium phenomenology.
  A brief description of each state, together with the motivation for why a specific decay channel is chosen, is given at the beginning of each section. 
 The details of the formalism 
are discussed in section~\ref{sec:formalism}.
In section~\ref{sec:zc3900} we present the production of the charged $Z$ states. Section~\ref{sec:x3872} is devoted to the \X and compared to the production of the ordinary $\chi_{c1}(1P)$. Speculations about the newly seen di-\jpsi resonance are in Section~\ref{sec:x6900}. Predictions for the vector $Y$ states, specifically of the \Y, and the comparison with the $\psi(2S)$ are given in section~\ref{sec:y4220}. 
Possible detection of exclusive processes with hidden charm pentaquarks is discussed 
in Section~\ref{sec:pc}. 
In section~\ref{sec:conclusions}  we present our conclusions, and comment on the significance of the cross sections by estimating the yields expected at a hypothetical fixed-target photoproduction experiment.

\section{Formalism}
\label{sec:formalism}
\begin{figure}[b]
    \centering
    \includegraphics[width=.5\columnwidth]{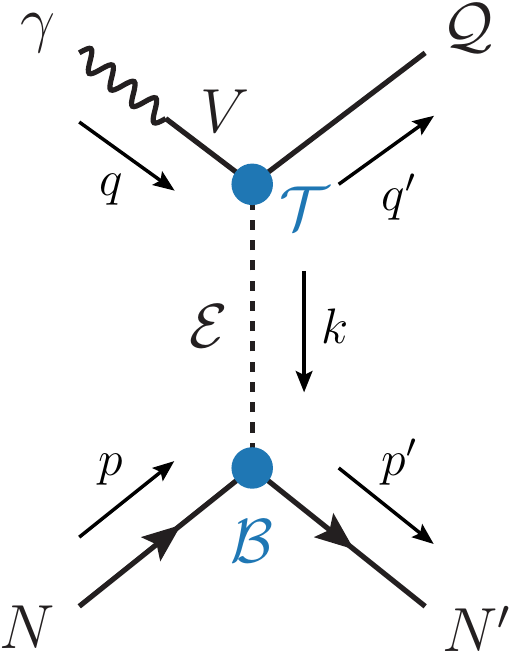}
    \caption{Photoproduction of a quarkonium-like meson, \mQ via an exchange \mE in the $t$-channel. }
    \label{fig:photo_reaction_s}
\end{figure}

We consider the process $\gamma N \to \mQ N'$, with $\mQ$ a heavy quarkonium or quarkoniumlike meson. 
At the energies of interest, the process is dominated by photon fragmentation,  as represented in Fig.~\ref{fig:photo_reaction_s}. The amplitude $T_{\lambda_i/\mu_i}(s,t)$ depends on the standard Mandelstam variables, $s$ being the total center-of-mass energy squared and $t$ the momentum transferred squared, with $\lambda_i$ and $\mu_i$ denoting the helicities of particle $i$ in the $s$- or $t$- channel frame, respectively.

    \begin{table*}[]
    \begin{ruledtabular}
        \begin{tabular}{c c c c c c c}
           $Z$ & $m_Z$ (\nsmev) & $\Gamma_Z$ (\nsmev)  & $V$ & $\mB(Z\to V \pi)$ (\%)  & $g_{V Z \pi}$ & $g_{\gamma Z \pi}$ ($\times10^{-2}$)
           \\
           \hline
            $Z_c(3900)^+$  & $3888.4 \pm 2.5$ & $28.3 \pm 2.5$ & $J/\psi$ & $10.5 \pm 3.5$ & 1.91 & $5.17$
            \\
            \hline
            \multirow{3}{*}{$Z_b(10610)^+$} & \multirow{3}{*}{$10607.2\pm2.0$} & \multirow{3}{*}{$18.4 \pm 2.4$} &  $\Upsilon(1S)$ & $0.54^{+0.19}_{-0.15} $ & $0.49$ & \multirow{3}{*}{$5.8$}
            \\ 
            & & & $\Upsilon(2S)$ & $3.6^{+1.1}_{-0.8} $ & $3.30$ & 
            \\
            & & & $\Upsilon(3S)$ & $2.1^{+0.8}_{-0.6} $ & $9.22$ & 
            \\
            \hline       
            \multirow{3}{*}{$Z^\prime_b(10650)^+$} & \multirow{3}{*}{$10652.2\pm1.5$} & \multirow{3}{*}{$11.5 \pm 2.2$} &  $\Upsilon(1S)$ & $0.17^{+0.08}_{-0.06} $ & 0.21 & \multirow{3}{*}{$2.9$}
            \\ 
            & & & $\Upsilon(2S)$ &  $1.4^{+0.6}_{-0.4} $ & $1.47$ &
            \\
            & & & $\Upsilon(3S)$ &  $1.6^{+0.7}_{-0.5} $ & $4.8$ & 
            \\
        \end{tabular}
        \caption{Parameters used for $Z$ production. Couplings are calculated with central values of branching fractions. The coupling radiative coupling is calculated via $g_{\gamma Z \pi} = \sum_V e f_V g_{V Z \pi}/m_V$.}
        \label{tab:Z_couplings}
    \end{ruledtabular}
    \end{table*}
Crossing symmetry relates the $s$-channel amplitude $\gamma N \to \mQ N'$ to that of the $t$-channel  $\bar N N \to \mQ \gamma$:
\begin{widetext}
\begin{align}
    \mel{\mu_\mQ \mu_\gamma}{T}{\mu'_{\bar N} \mu_N} =
    -\!\!\!\!\sum_{\lambda_\mQ\lamNp\lamgam\lamN}\!\!\!\! \delta_{\lamgam,-\mu_\gamma} \,
    d^{1/2}_{\lamN\mu_N}(-\chi_N) \; 
    d^{J_\mQ}_{\lamQ\mu_\mQ}(-\chi_\mQ)\;
    d^{1/2}_{\lamNp\mu'_N}(-\chi'_N) \; 
    \mel{\lambda_\mQ \lamNp}{T}{\lamgam \lamN}~,
    \label{eq:crossed}
\end{align}
\end{widetext}
 with $\chi_i$ the crossing angles whose explicit expressions are given in~\cite{Martin:1970}. Because of the orthogonality of the Wigner-$d$ matrices, 
 \begin{align}
    &
    \sum_\mu
    \left|\mel{\mu_\mQ \mu_\gamma}{T}{\mu'_{\bar N} \mu_N}\right|^2 =
    \sum_\lambda
    \left|\mel{\lambda_\mQ, \lamNp}{T}{\lamgam, \lamN}\right|^2~,
\end{align}
we can use either one to compute the cross sections.

Specifically, the $s$-channel amplitude can be written as 
\begin{align}
    &\mel{\lambda_\mQ \lamNp}{T}{\lamgam \lamN} =
    \nn \\
&\qquad \qquad\sum_{V, \,\mE} \frac{e f_V}{m_V} \,  \mT^{\alpha_1\cdots\alpha_j}_{\lambda_V=\lamgam,\lambda_\mQ}
	\, \mP_{\alpha_1\cdots\alpha_j;\beta_1\cdots\beta_j} \,
	\mB^{\beta_1\cdots\beta_j}_{\lambda_N \lambda'_N},\label{eq:amplitude}
\end{align}
where $j$ is the spin of the exchanged particle \mE, and 	$\mP$ is its propagator. More complicated exchanges are discussed later.
We assume vector-meson dominance (VMD) to estimate the coupling between the incoming photon and the intermediate vector quarkonia $V = \jpsi$ or $\Upsilon(nS)$ which \mQ couples to. The decay constant $f_V$ is related to the $V$ electronic width by $\Gamma\!\left(V \to e^+e^-\right) = 4\pi \alpha^2 f^2_V / 3 \, m_V$. Masses, widths and decay constants of the vectors of interest are reported in \cref{tab:VMD}. 

The top vertex \mT is related to the partial decay width
\begin{align}
    \Gamma\!\left(\mQ \to V\, \mE\right) &= \frac{1}{2J_\mQ + 1} \frac{\lambda^{1/2}\!\left(m_\mQ^2,m_V^2,m_\mE^2\right)}{16\pi  \, m_\mQ^3}\nn\\ 
    &\quad\times\sum_{\lambda_\mQ\lambda_V \lambda_\mE}
    \left|\mT^{\alpha_1\cdots\alpha_j}_{\lambda_V\lambda_\mQ} \, \varepsilon^*_{\alpha_1\cdots\alpha_j}(k, \lambda_\mE)\right|^2,
    \label{eq:partialwidth}
\end{align}
with $\lambda(a,b,c)=a^2+b^2+c^2 -2ab-2ac-2bc$ the usual \Kallen function, $J_\mQ$ the spin of the produced quarkonium, $m_i$ the mass of particle $i$, and $\varepsilon(k,\lambda_\mE)$ the polarization tensor of particle $\mE$. The bottom vertex \mB describes the interaction $N \mE \to N'$, and is discussed in the following sections.

We expect a model with fixed-spin exchange to be valid from threshold to moderate values of $s$. However, it can be shown that the $t$-channel amplitude in~\eqref{eq:crossed} behaves as
\begin{align}
\label{eq:dtchannel}
	\mel{\mu_\mQ \mu_\gamma}{T}{\mu'_{\bar N} \mu_N} &\propto \frac{d^{j}_{\mu'_{\bar N} - \mu_N,\mu_\mQ - \mu_\gamma}(\theta_t)}{t-m_\mE^2}
\end{align}
where $\cos \theta_t$ is the $t$-channel scattering angle, and depends linearly on $s$.  At high energies, this expression grows as $s^{j}$, which exceeds the unitarity bound. The reason for this is that the amplitude in~\eqref{eq:amplitude} with fixed-spin exchange is not analytic in angular momentum. Assuming that the large-$s$ behavior is dominated by a Regge pole rather than a fixed pole, 
we obtain the amplitude with the standard form of the Regge propagator. This can be interpreted as originating from the resummation of the leading powers  of $s^{j}$ in the $t$-channel amplitude, which originate  from the exchange of a tower of particles with increasing spin, 
\begin{widetext}
        \begin{equation}
        \left(\frac{4 \, p(t) \, q(t)}{s_0}
        \right)^{j-M}
       \mathcal{N}^j_{\mu\mu'} \, \frac{d_{\mu\mup}^j(\theta_t)}{\xi^{(t)}_{\mu\mup}(s,t)}  \, 
        \frac{1}{t-m_\mE^2} 
        \longrightarrow - \alpha^\prime \,\Gamma\!\big(j- \alpha(t)\big)
         \bigg[\frac{1+ \tau \, e^{-i\pi\alpha(t)}}{2}\bigg]  \,\left(\frac{s}{s_0}\right)^{\alpha(t)-M} 
        ~.        \label{eq:regge_replacement}
    \end{equation}
\end{widetext}
Here,  $\mathcal{N}^j_{\mu\mup}=(-)^{\tfrac{1}{2}\left(|\mu-\mup|+\mu-\mup\right)}\frac{\sqrt{(j-M)!(j+M)!(j-N)!(j+N)!}}{(2j)!}$, $\xi^{(t)}_{\mu\mup}(s,t) = \left(\tfrac{1-\cos\theta_t}{2}\right)^{|\mu-\mup|/2}\left(\tfrac{1-\cos\theta_t}{2}\right)^{|\mu+\mup|/2}$, $p(t)$ and $q(t)$ the incoming and outgoing 3-momenta in the $t$-channel frame, $M=\max\{|\mu|,|\mup|\}$, $N=\min\{|\mu|,|\mup|\}$, and $\tau = (-)^j$ the signature factor~\cite{Collins:1977jy,Nys:2018vck}. The hadronic scale $s_0$ is set to $1\gevsq$. The Regge trajectory satisfies $\alpha(t = m_\mE^2) = j$, and $\alpha' = \tfrac{d}{dt}\alpha(t = m_\mE^2)$, and the normalization is such that at the pole $t=m_\mE^2$ the right-hand side becomes $(s/s_0)^{j-M}\big /(t - m_\mE^2)$, which coincides with the leading $s$ power of the left-hand side.

From this discussion, it follows that at low energies the fixed-spin exchange amplitude contains the full behavior in $s$, and is more reliable than the Regge one, which is practical only for the leading power. Conversely, at high energies where the leading $s$ power dominates, the fixed-spin amplitude becomes unphysical, while the Regge one has the correct behavior. For this reason, we will show results based on the fixed-spin amplitudes in the region close to threshold, and the predictions from the Regge amplitudes at  asymptotic energies.

Since the systematic uncertainties related to our predictions are much larger than the uncertainties of the couplings the models depend upon, we do not perform the usual error propagation, and just consider the qualitative behavior and the order of magnitude of these simple estimates. For this reason, we will not add error bands to our curves.

\section{\boldmath $Z_c(3900)^+$, $Z_b(10610)^+$, and $Z^\prime_b(10650)^+$}
\label{sec:zc3900}
%
    \begin{figure*}
        \centering
        \includegraphics[width=0.48\textwidth]{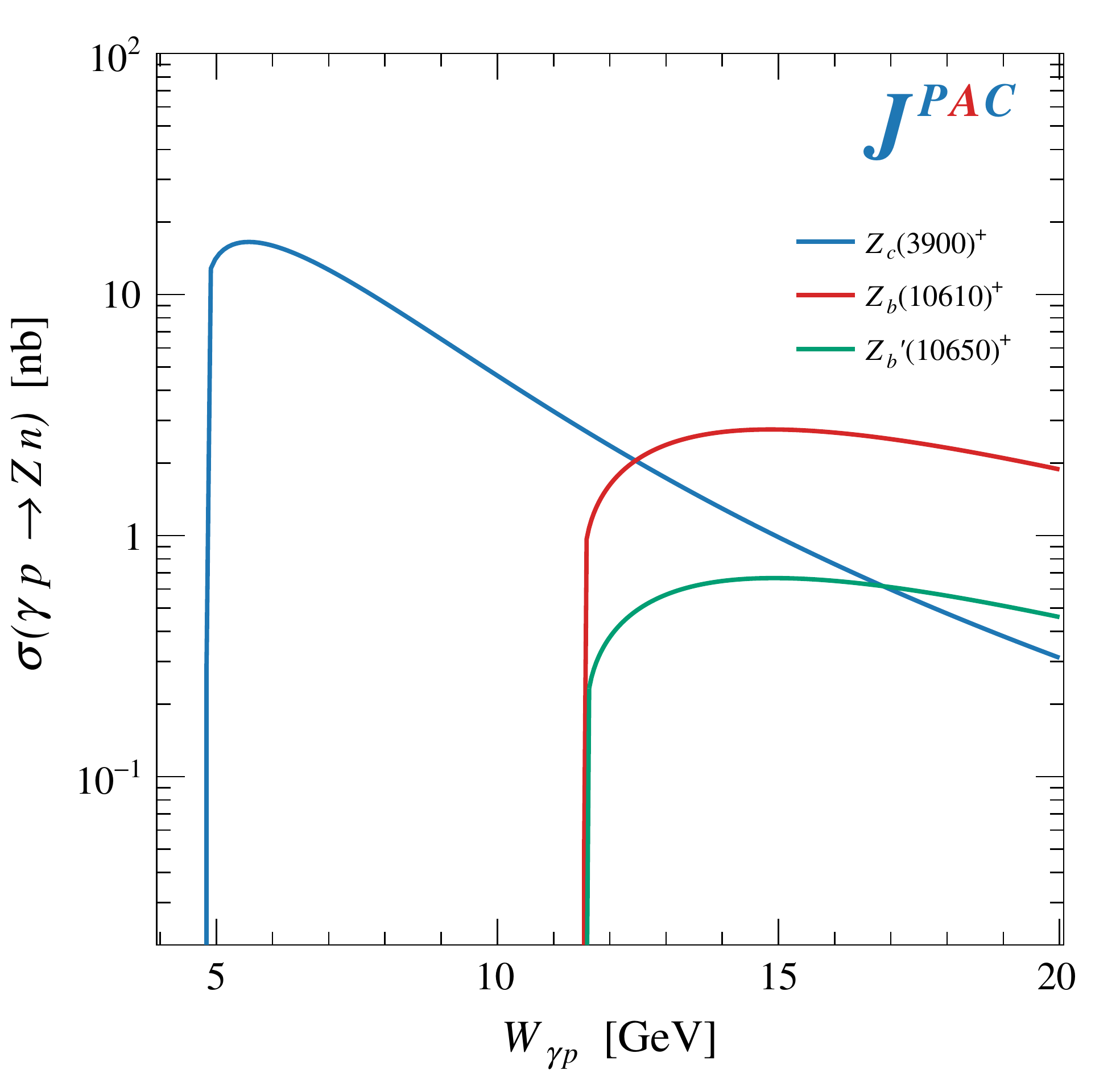}
        \includegraphics[width=0.48\textwidth]{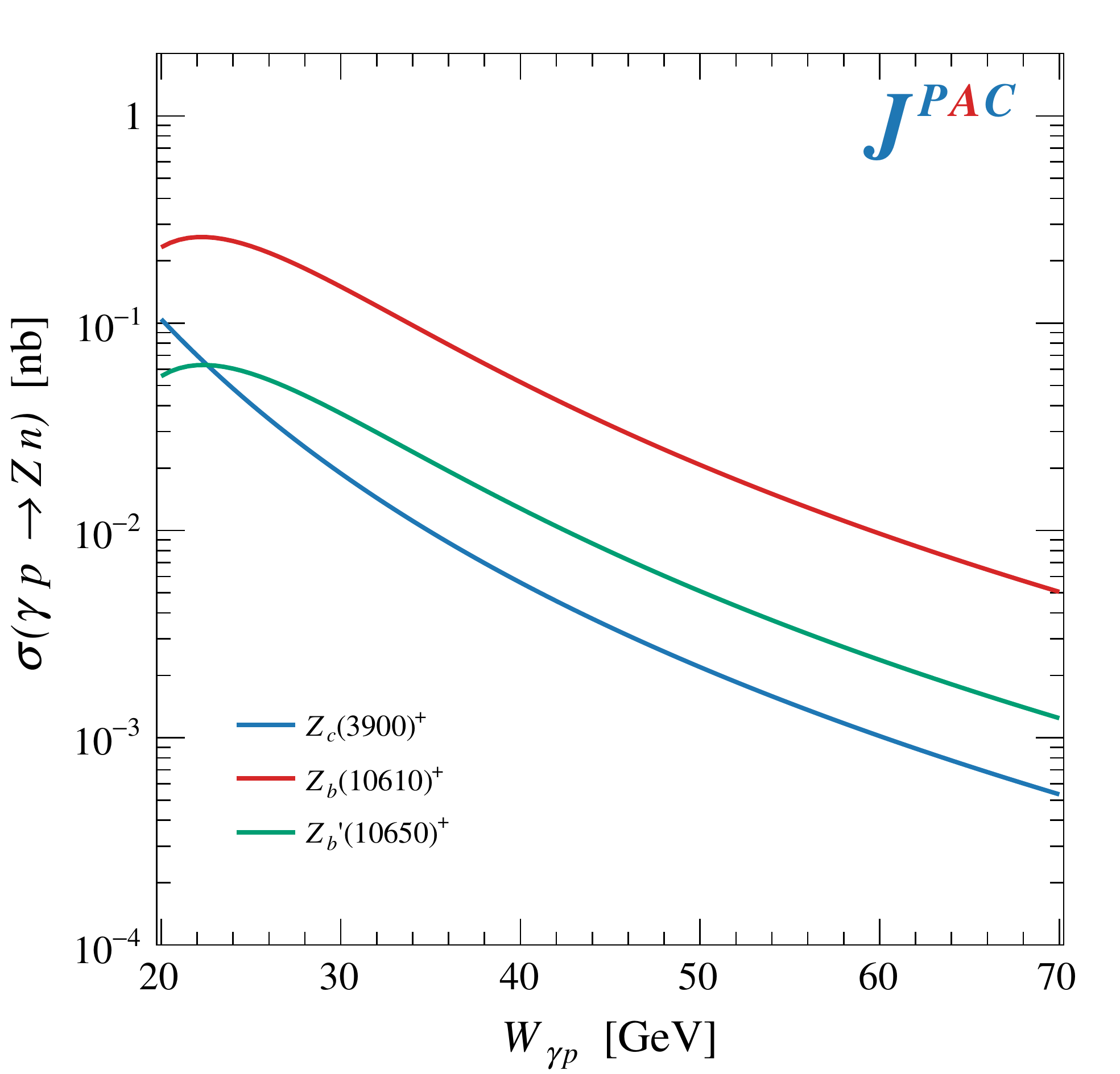}
        \caption{Integrated cross sections for the three $Z$ states considered. Left panel: predictions for fixed-spin exchange, which we expect to be valid up to approximately 10\gev above each threshold. Right panel: predictions for Regge exchange, valid at high energies.}
        \label{fig:Zplots}
    \end{figure*}
We start from the production of charged $Z$ states. We focus on the narrow ones seen in $e^+e^-$ collisions that lie close to the open flavor thresholds: the hidden charm \Z  and hidden bottom $Z_b(10610)^+$ and $Z_b'(10650)^+$.  They all have sizeable branching fractions to $V \pi^+$, with $V=\jpsi,\Upsilon(nS)$~\cite{pdg}, which makes them relatively easy to detect. We do not consider the  narrow $Z_c'(4020)$, which decays mostly into $h_c(1P) \pi^+$ and $\bar D^{*0} D^{*+}$ and is therefore more difficult to reconstruct. These four states have the same quantum numbers $J^{PC} = 1^{+-}$~\cite{Belle:2011aa,Collaboration:2017njt},\footnote{As customary, by $C$ we mean the charge conjugation quantum number of the neutral isospin partner.} 
 and the absolute branching fractions can be calculated by assuming that the several observed decay modes saturate the total width. 
 Obviously, reaching the \Zbs  requires higher energy and an optimal setup for the \Z and \Zbs may not be the same.
 The same amplitudes can in principle be extended to the broad $Z$ states seen in $B$ decays. However, their branching ratio to $V\pi^+$ is unknown, and their broad width would make the separation from the background more challenging. 
 Predictions for some of them have already been given in~\cite{Wang:2015lwa,*Galata:2011bi}, while the \Z was studied previously in~\cite{Lin:2013mka} on the basis of outdated estimates for the branching ratios. The \Zbs have recently been studied in~\cite{Wang:2020stx}.

    \begin{table*}[]
    \begin{ruledtabular}
        \begin{tabular}{ccccccc}
             $X$ & $m_X$ (\nsmev) & $\Gamma_X$ (\nsmev) & $\mE$ & \multicolumn{2}{c}{$\mB(X\to\gamma \, \mE)$ (\%)}   & $g_{\gamma X \mE}$ ($\E{-3}$) 
           \\
           \hline
            \multirow{4}{*}{$\chi_{c1}(1P)$} & \multirow{4}{*}{$3510.67 \pm 0.05$} & \multirow{4}{*}{$0.84 \pm 0.04$} 
            & $\rho$ & \multicolumn{2}{c}{$(2.16 \pm 0.17)\E{-4}$} & $0.92$
            \\
            &&& $\omega$ & \multicolumn{2}{c}{$(6.8 \pm 0.8) \E{-5}$} & $0.52$ 
            \\
            &&& $\phi$ & \multicolumn{2}{c}{$(2.4 \pm 0.5) \E{-5}$} & $0.42$
            \\
            &&& $\jpsi$ & \multicolumn{2}{c}{$34.3 \pm 1.0$} & $1.0\E{3}$
            \\
           \hline
                         &  &  &  &  $\mB(X\to \jpsi\, \mE)$ (\%) & $g_{\psi X \mE}$ & $g_{\gamma X \mE}$ ($\E{-3}$) \\ \hline
            \multirow{2}{*}{$X(3872)$} & \multirow{2}{*}{$3871.69 \pm 0.17$} & \multirow{2}{*}{$1.19 \pm 0.19$} & $\rho$ & $4.1^{+1.9}_{-1.1}$ & $0.13$ & $3.6$ \\ 
            &&& $\omega$ & $4.4^{+2.3}_{-1.3}$ & $0.30$ & $8.2$
             \\
        \end{tabular}
        \end{ruledtabular}
        \caption{Parameters used for \X and $\chi_{c1}(1P)$ production. Couplings are calculated with central values of branching fractions.}
        \label{tab:X_couplings}
    \end{table*}
   \begin{table}[b]
        \begin{ruledtabular}
        \begin{tabular}{c c c c}
            $\mE$ & $m_\mE$ 
            (\nsmev) & $g_{\mE NN}$ & $g^\prime_{\mE NN}$ 
            \\
            \hline
            $\rho$ & $775.26 \pm 0.25$ & $2.4$ & $14.6$ \\
            $\omega$ & $782.65 \pm 0.12$ & $16$ & $0$ \\
            $\phi$ & $1019.461 \pm 0.016$ & $-6.2$ & $2.1$ \\
            $\jpsi$ & $3096.900 \pm 0.006$ & $1.6\E{-3}$ & 0 
        \end{tabular}
        \end{ruledtabular}
        \caption{Couplings of the bottom vertex for the vector exchanges considered in Eq.~\eqref{eq:X_bot_vert}. Light vectors couplings are taken from~\cite{Chiang:2001as,Blin:2017hez}, while the $\jpsi$ coupling is calculated from the $\jpsi\to p \bar p$ braching ratio. }
        \label{tab:X_nucleon_couplings}
    \end{table}

The production of these $Z$ states proceeds primarily through a charged pion exchange. A 
minimal 
parameterization of the top vertex in Eq.~\eqref{eq:amplitude}, consistent with gauge invariance is given by
    \begin{align}
        \label{eq:Z_top}
        \mT_{\lamV\lamZ} =  \frac{g_{V Z \pi}}{m_Z}&  
        \varepsilon_\mu(q, \lamV) \, \varepsilon^*_\nu(\qp, \lamZ) \nn \\
         &\qquad\times  \left[
        (q \cdot k) \; g^{\mu\nu} - k^\mu \, q^{\nu} 
        \right] 
        ~.
    \end{align}
The coupling $g_{V Z \pi}$ is calculated from the partial decay width $\Gamma(Z\to V \pi)$ using Eq.~\eqref{eq:partialwidth}. 
For the \Z we assume that the width is saturated by the three decay modes $\jpsi\,\pi^+$, $(\bar D D^*)^+$, and $\eta_c \rho^+$. A similar assumption was made in~\cite{Belle:2011aa} for the \Zbs, the width being saturated by the $\Upsilon(nS)$ ($n=1,2,3$), $h_b(mP)$ ($m=1,2$) and $(\bar B^{(*)} B^*)^+$ modes. The couplings are summarized in \cref{tab:Z_couplings}. For the bottom $\pi N N$ vertex we take\footnote{ An explicit factor of  $\sqrt{2}$ is considered for the charged pion exchange.}
:
    \begin{equation}
        \mB_{\lambda_N \lambda_\Np} = \sqrt{2} \, g_{\pi NN} \, \beta(t)\, \bar{u}(\pp, \lamNp) \, \gamma_5 \, u(p, \lamN)
    ~,\label{eq:piNN_vert}
    \end{equation} 
with $g_{\pi NN}^2/(4\pi)\simeq 13.81\pm 0.12$~\cite{Matsinos:2019kqi}. 
Away from the pole, the residue $\beta(t)$ is unconstrained in Regge theory and accounts for  the suppression at large $t$  visible in data.
We use $\beta(t) = \exp(t'/\Lambda_\pi^2)$, with $t' = t - t(\cos\theta_s = 1)$, and $\Lambda_\pi = 0.9\gev$~\cite{Gasparyan:2003fp} (monopole form factors were used in~\cite{Lin:2013mka}). 
For the Reggeized amplitude of Eq.~\eqref{eq:regge_replacement}, we use the pion trajectory~\cite{Irving:1977ea}:
    \begin{align}
        \alpha_\pi(t) = \alpha_\pi'(t- m_\pi^2)\text{ with }\alpha'_\pi = 0.7\gev^{-2}\,.
    \end{align}
The results for the fixed-spin and Regge amplitudes are shown in \cref{fig:Zplots}. Note that the fixed-spin results are expected to be valid up to approximately 10\gev above threshold.
In particular, the range of validity of \Z and \Zbs are different.

\section{\boldmath \X and $\chi_{c1}(1P)$} 
\label{sec:x3872}
    \begin{figure*}
        \centering
        \includegraphics[width=0.48\textwidth]{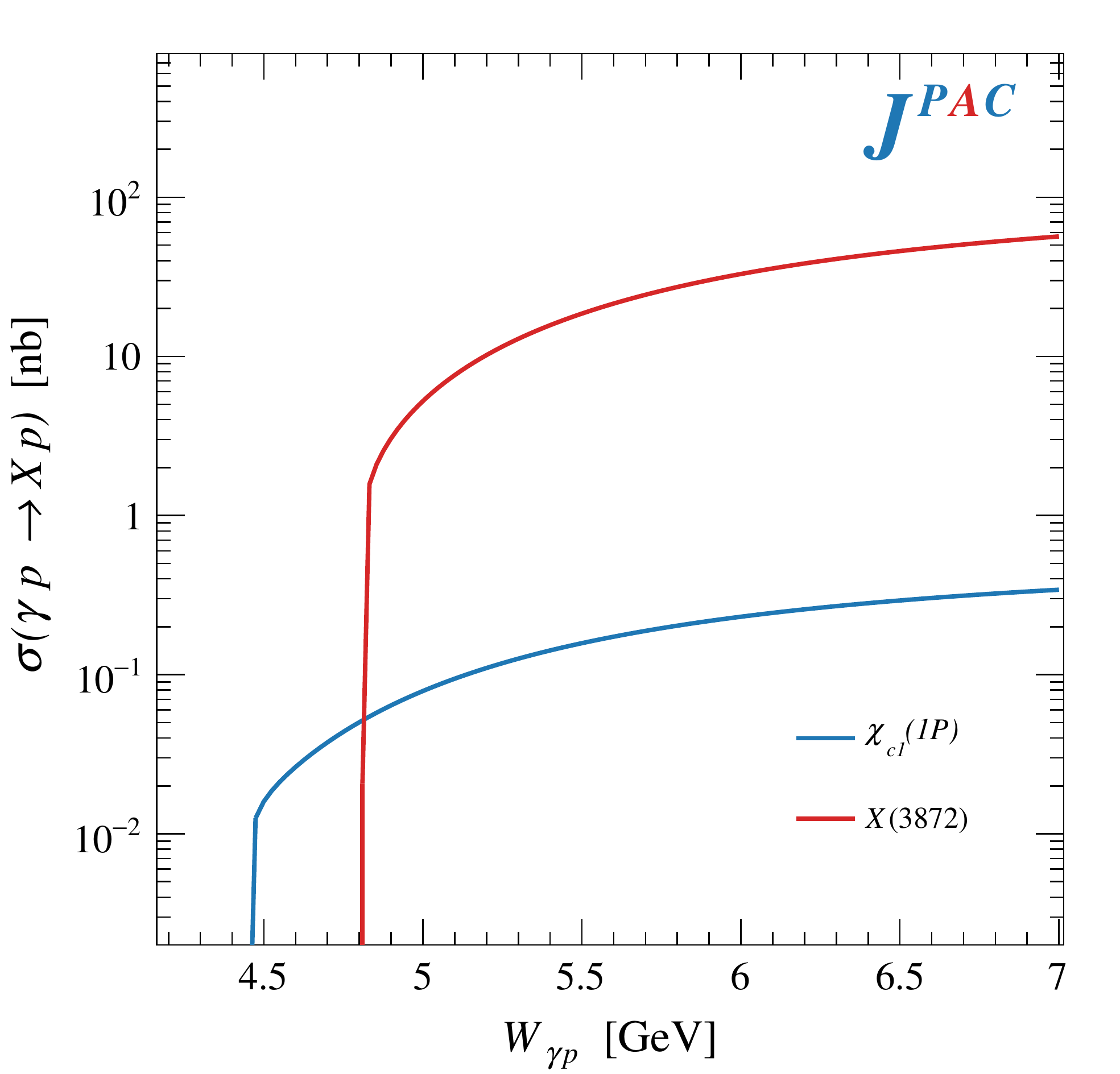}
        \includegraphics[width=0.48\textwidth]{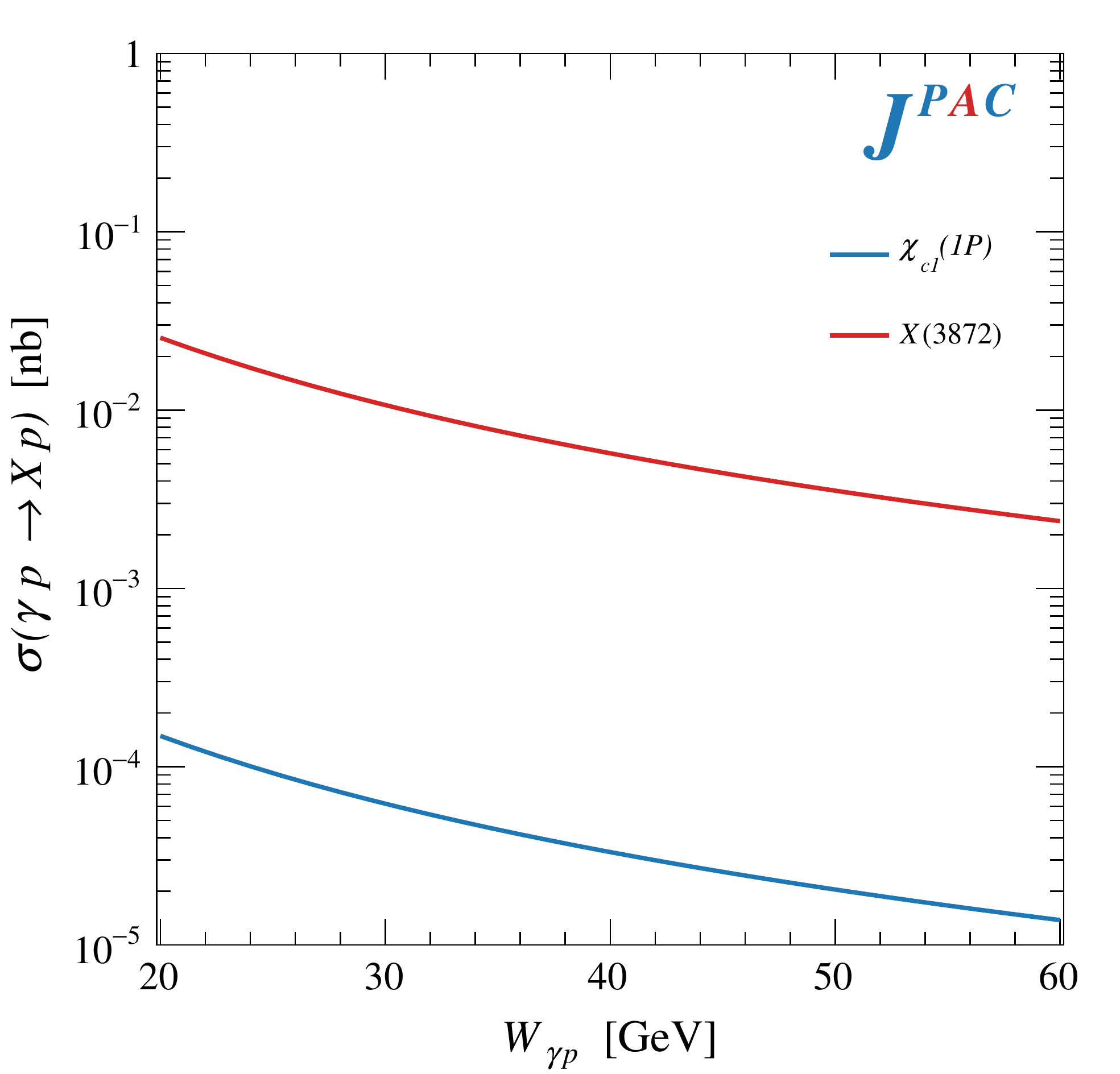}
        \caption{Integrated cross sections for the axial $\chi_{c1}(1P)$ and $X(3872)$. Left panel: predictions for fixed-spin exchange, valid at low energies. Right panel: predictions for Regge exchange, valid at high energies.}
        \label{fig:Xplots}
    \end{figure*}
The \X is by far the best known exotic meson candidate. It has been observed in several different decay modes and production mechanisms~\cite{pdg}. The Breit-Wigner mass and width have been recently been measured to be $M_X = 3872.62 \pm 0.08 \mev$ and $\Gamma_X = 1.19 \pm 0.19 \mev$, although significant deviations of the lineshape are expected because of the proximity to the $\bar D^0 D^{*0}$ threshold~\cite{Aaij:2020qga,*Aaij:2020xjx}. Quantum numbers have been measured to be $1^{++}$~\cite{Aaij:2013zoa,*Aaij:2015eva}. The most exotic feature of the $X$ is the strength of isospin violation, which is manifested in the decays $\mB(X\to \jpsi\,\omega)/\mB(X\to \jpsi\,\pi^+\pi^-) = 1.1\pm 0.4$~\cite{pdg}. The inclusive measurement of $B^+ \to K^+ \X (\to \text{anything})$~\cite{Lees:2019xea} allows for the estimation of the absolute branching fractions, and thus of the couplings of \X to its decay products~\cite{Li:2019kpj}.

Since \X has sizeable branching fractions to $\jpsi \,\rho$ and $\jpsi \,\omega$, light vector exchanges will provide the main production mechanism. The state can be detected in the $\jpsi \,\rho^0 (\to \pi^+\pi^-)$ final state, which is relatively easy to reconstruct. Similarly to Eq.~\eqref{eq:Z_top}, the top vertex is parameterized by:
    \begin{equation}
        \label{eq:X_top}
        \mT^\mu_{\lambda_\mE\lamX} = 
        g_{\psi X \mE} \,
        \epsilon_{\sigma\nu\alpha\beta}
         \left[ 
            g^{\sigma\mu}  \,
            {\varepsilon^*}^\nu(\qp, \lamX) \,
            q^\alpha \, \varepsilon^\beta(q, \lambda_\mE) 
        \right]
        ~,
    \end{equation}
with the coupling $g_{\psi X \mE}$ obtained from the partial decay width $\Gamma(X\to \jpsi \,\mE)$ with $\mE=\rho,\omega$ using 
 Eq.~\eqref{eq:partialwidth}. 
Since the mass of the $X$ is below the nominal $\jpsi \,\omega$ threshold, and almost exactly at the $\jpsi \rho$ one, we need to take into account the \mE widths to extract meaningful couplings. Details are left to the Appendix~\ref{appendix:x3872couplings}. 
The resulting values for the couplings are summarized in \cref{tab:X_couplings} (\cf the extractions in~\cite{Hanhart:2011tn,*Brazzi:2011fq,*Faccini:2012zv}). 
 The bottom vertex is described by the standard vector meson-nucleon interaction:
    \begin{align}
        \label{eq:X_bot_vert}
        \mB^\mu_{\lamN \lamNp} &= \beta(t)\, \bar{u}(\pp, \lamNp) \nn\\
        &\quad\times\left(
        g_{\mE NN} \, \gamma^\mu + g^\prime_{\mE NN} \, \frac{\sigma^{\mu \nu} k_\nu}{2 m_N} 
        \right)
         u(p, \lamN)
        ~.
    \end{align}
The  numerical values 
 of the  vector and tensor couplings $g^{(\prime)}_{\mE NN}$ are tabulated in \cref{tab:X_nucleon_couplings}. For the $\omega$ and the $\rho$ they are extracted from nucleon-nucleon potential models~\cite{Chiang:2001as}. The $\phi$ coupling is then estimated with the help of $SU(3)$ considerations as done in~\cite{Blin:2017hez}. These values are compatible with the ones used in Regge fits~\cite{Irving:1977ea,Nys:2018vck}. The  $\jpsi$ coupling is obtained from the $\jpsi \to p\bar{p}$ decay width using:
    \begin{equation}
    \mel{\lambda_N\lamNp}{T}{\lambda_\psi} =  
    g_{\psi NN} \, 
    \bar{u}(p, \lambda_{N}) \;
    \gamma^\mu \;
    v(\pp, \lambda_{\bar N})
    ~,
    \end{equation}
assuming vanishing tensor coupling. The resulting coupling is so small that the contribution of the $\jpsi$ is hardly relevant, despite the large top coupling. 

We use $\beta(t) = \exp(t'/\Lambda_{\mE}^2)$ with $\Lambda_\rho = 1.4\gev$ and $\Lambda_\omega = 1.2\gev$~\cite{Gasparyan:2003fp}. For $\phi$ and $\jpsi$, we set the form factor $\beta(t) = 1$.
For the Reggeized amplitude, $\rho$ and $\omega$ have degenerate trajectories,
    \begin{equation}
        \alpha_V(t) = 1 + \alpha_V' (t - m_\rho^2)\text{ with }\alpha'_V = 0.9\gev^{-2}\,.
    \end{equation}

For comparison with the exotic \X, we also consider the photoproduction of the ordinary axial charmonium, $\chi_{c1}(1P)$. The radiative decay branching fractions for $\chi_{c1} \to \gamma \mE$ are available for $\mE =\rho, \omega, \phi, \psi$, so that the coupling in the top vertex can be readily calculated without assuming VMD, \ie by setting $e f_V/m_V \to 1 $ in Eq.~\eqref{eq:amplitude}, and replacing $g_{\psi X \mE} \to g_{\gamma \chi \mE}$ in Eq.~\eqref{eq:X_top}. 

In the Reggeized amplitude, the $\phi$ and $\jpsi$ trajectories are subleading to the $\rho$ and $\omega$ ones, the intercept $\alpha(0)$ being roughly $\sqrt{\alpha'}$ times twice the heavy quark mass, and can be safely neglected at high energies. The results for the fixed-spin and Regge amplitudes are shown in \cref{fig:Xplots}. It is worth noting the mismatching strengths of the amplitudes in the two regimes. The fixed-spin one describes correctly the size of the cross section at threshold. However, the saturation observed is unphysical and entirely due to the fixed-spin approximation. The physical amplitude is expected to start decreasing faster and match the Regge prediction at $W_{\gamma p}\sim 20\gev$. 

\section{\boldmath Production of \X via Primakoff effect}
\begin{figure}[b]
    \centering
    \includegraphics[width=.5\columnwidth]{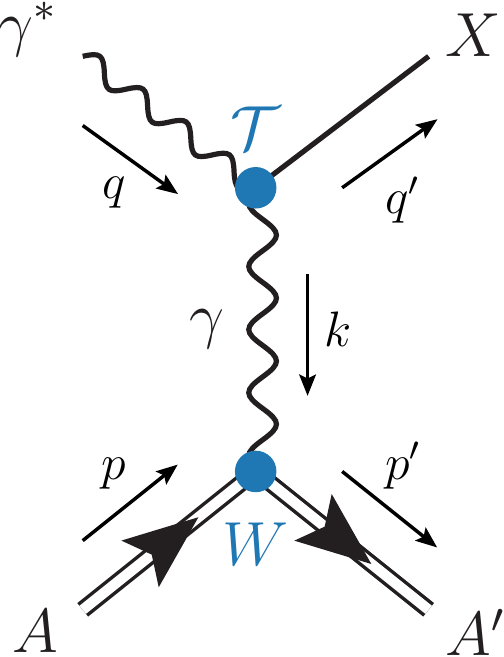}
    \caption{Electroproduction of \X via Primakoff effect. The exchanged photon is quasi-real.}
    \label{fig:Primakoff_diagram}
\end{figure}
    \begin{figure*}
        \includegraphics[width=.48\textwidth]{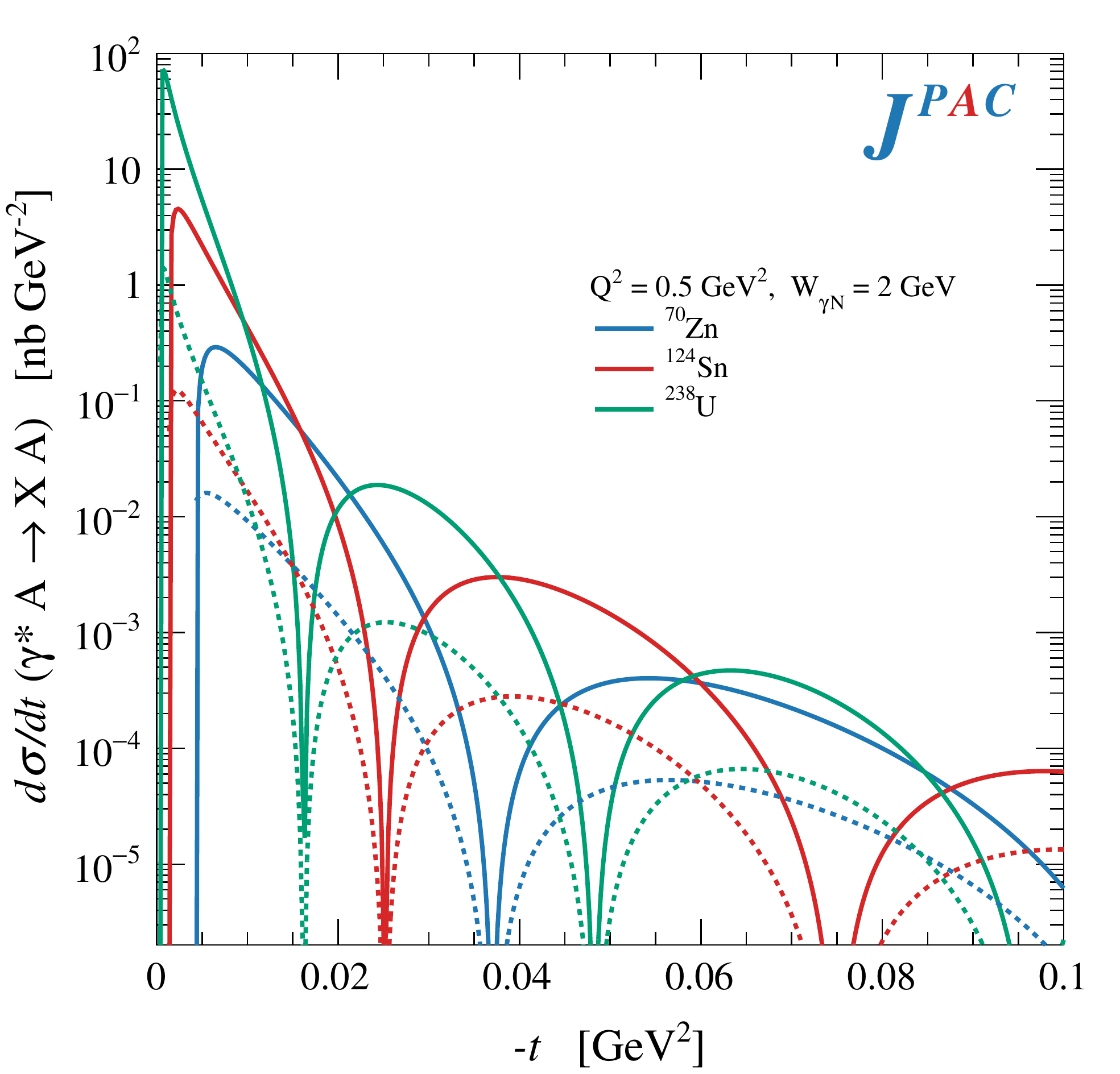}
         \includegraphics[width=.48\textwidth]{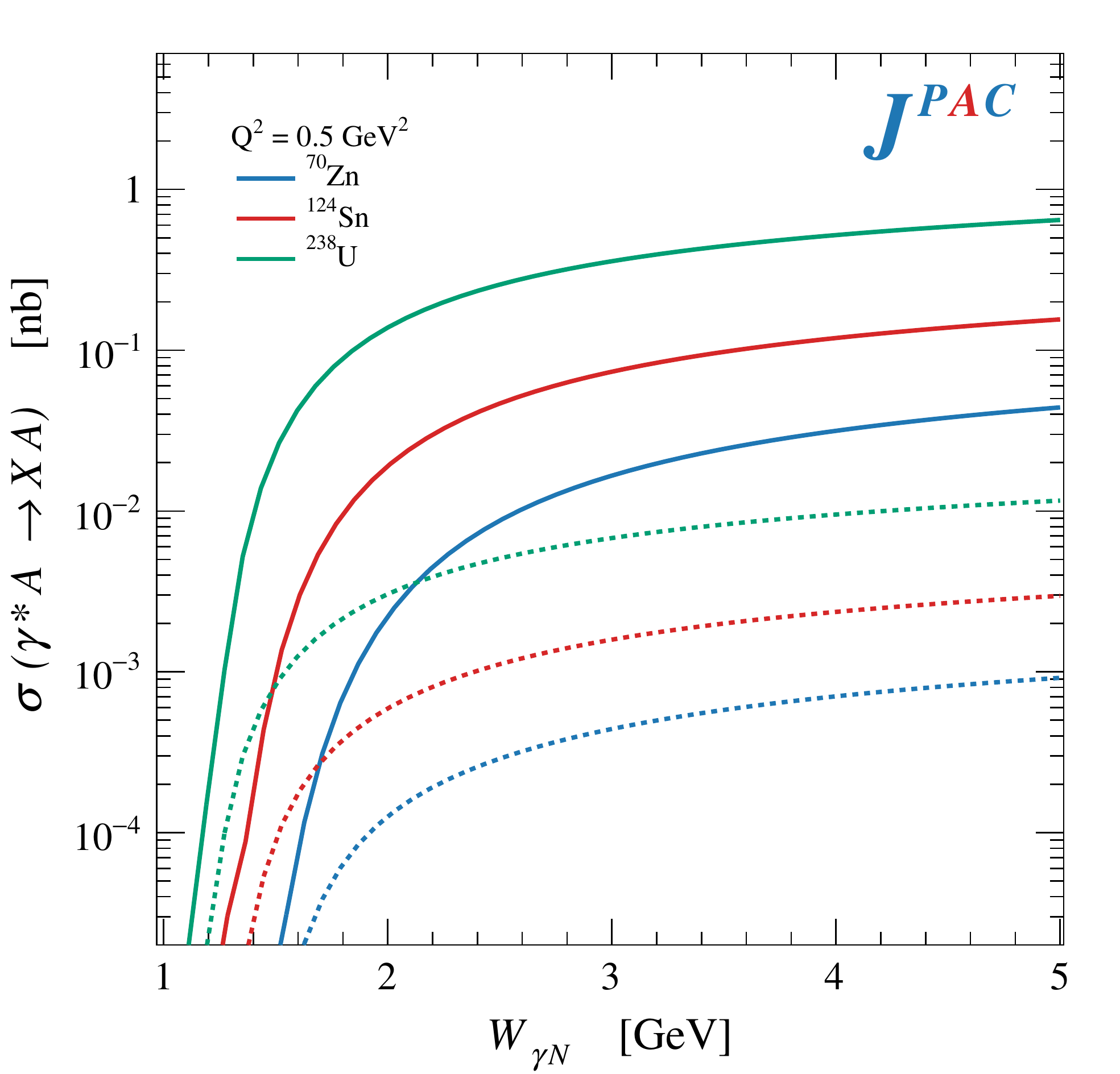}
        \caption{Cross sections for Primakoff production of \X off various nuclei. Solid and dashed curves correspond to longitudinal and transverse photons respectively. Left panel: differential cross sections for  $W_{\gamma N} = 2\gev$. Right panel: integrated cross sections}
        \label{fig:primakoff}
    \end{figure*}

Another possible mechanism to produce the \X at the EIC is in two-photon collisions through the Primakoff effect.  Because of the Landau-Yang theorem~\cite{Landau:1948kw,*Yang:1950rg}, the \X cannot couple to two real photons. Nevertheless one can define
\begin{equation}
    \tilde\Gamma^X_{\gamma \gamma} = \lim_{Q^2\to 0} \frac{m_X^2}{Q^2} \,  \Gamma^X_{\gamma\gamma^*}(Q^2)~,
\end{equation}
that defines the coupling of $X$ to a real and a virtual photons. A recent measurement by Belle gives indeed
$\tilde\Gamma^X_{\gamma \gamma} \times \mB(\X\to \jpsi\,\pi^+\pi^-) = 5.5^{+4.1}_{-3.8}\pm 0.7\eV$~\cite{Teramoto:2020ezr}. 
At the EIC, one can consider the photon emitted by the electron beam, scattering onto the photon emitted by the nuclear beam,
\begin{align}
        \label{eq:Primakoff_top}
        \mT_{\lamgam\lamX}^\mu &=  \frac{g_{X \gamma \gamma^*}}{m_X^2}
        \varepsilon^{\alpha\nu\rho\sigma} \, \varepsilon^*_\sigma(\qp, \lambda_X)\nn\\
        &\quad\times \Big[ g_\alpha^\mu\,k_\nu q^\tau \left[\varepsilon_\rho(q, \lamgam)\,q_\tau - \varepsilon_\tau(q, \lamgam)\,q_\rho\right]\nn\\
        &\quad+ \varepsilon_\alpha(q, \lamgam)\,q_\nu k^\tau \left[g_\rho^\mu\,k_\tau - g_\tau^\mu\,k_\rho\right] \Big]~.
\end{align}
The virtuality of the exchanged photon is suppressed for $t \gg R^{-2} \sim \mathcal{O}(10^{-3})\gev^2$, so that the exchanged photon is quasi-real, and we need to consider finite virtualities of the incoming photon. We use the standard notation $Q^2 = - q^2$.

Using Belle's reduced width and the estimate for the absolute branching ratios in~\cite{Li:2019kpj}, we get $g_{X \gamma \gamma^*} \sim 3.2\E{-3}$.

From this top vertex, we can define the matrix element squared of the quasi-elastic process $\gamma^*A \to \X \,A$, where $A$ indicates a nucleus of atomic number $Z$~\cite{Donnelly:2017aaa,*Aloni:2019ruo}:
\begin{align}
        \label{eq:Primakoff_msq}
        \overline{\sum_{\!\!\!\!\lambda_X,\lambda_A^{(\prime)}\!\!\!\!}}\left|\mel{\lamX\lamAp}{T}{\lamgam, \lamA}\right|^2 &=  \frac{e^2}{t^2}\mT^{\mu\nu}_{\lambda_\gamma} W_{\mu\nu}\,.
\end{align}
where the factor $e^2$ is factored out from the bottom vertex, and $t^{-2}$ comes from the exchanged photon propagator. The nuclear tensor $W_{\mu\nu}$ is dominated by the electric field of the nucleus,
\begin{align}
  \label{eq:Primakoff_W}
        W_{\mu\nu} & \simeq 16\pi \left(p_\mu + \tfrac{1}{2}k_\mu\right)\left(p_\nu + \tfrac{1}{2}k_\nu\right)\,\frac{Z^2}{4\pi} \frac{16m_A^4 \,F^2_0(t)}{(4m_A^2-t)^2}\nn\\
        & \to 16\pi\, p_\mu p_\nu \,\frac{Z^2}{4\pi} \frac{16m_A^4 \,F^2_0(t)}{(4m_A^2-t)^2}
        \,,
\end{align}
where we drop the components proportional to $k$, since we keep $\mT^{\mu\nu}_{\lambda_\gamma}$ explicitly transverse, and we give the nuclear charge distribution in the Fermi model by
\begin{equation}
    F_0(t) = \frac{\rho_0}{Z} \int d^3 x \, \frac{\sin |\vec k| |\vec x|}{|\vec k| |\vec x|}
\bigg[1 + \exp\left(\frac{|\vec x| - R}{a}\right)\bigg]^{-1} 
~,
\end{equation}
with $|\vec k|=\left[\left(4m_A^2-t\right)(-t)\right]^{1/2}/(2m_A)$, and the normalization $\rho_0 = Z \Big/ \int d^3 x \left[1 + \exp\left(\frac{|\vec x| - R}{a}\right)\right]^{-1}$. The parameters $R$, $a$ and $Z$ for various nuclei can be found in~\cite{DeJager:1974liz}, and we quote in \cref{tab:primakoff_params} a few examples.
Details on the top tensor $\mathcal{T}^{\mu\nu}$ and the calculation of the cross sections are in Appendix~\ref{app:Primakoff_kinematics}.

Although the coupling in \cref{eq:Primakoff_top} is small, the cross section is enhanced by the atomic number of the nuclear beam, thus we expect this production mechanism to be viable for high $Z$ beams. 
The coupling $g_{X\gamma\gamma^*}$ in principle depends on $Q^2$. At large virtualities, one should match the expectations from perturbative QCD (in analogy to the pion transition form factor), and the coupling can no longer be approximated as a constant. For $Q^2 < 1\gevsq$, we believe that this approximation is safely under control.  Predictions for sample ion species are shown in \cref{fig:primakoff}, for $Q^2 = 0.5\gev^2$, for an average photon nucleon energy $W_{\gamma N} = W_{\gamma A} / A=2\gev$, $A$ being the mass number of the nucleus.

   \begin{table}[t]
        \begin{ruledtabular}
        \begin{tabular}{c c c c c}
            $A$ & $Z$ & $m_A$ 
            (\nsgev) & $R$ ($\nsgev^{-1}$)& $a$ ($\nsgev^{-1}$)
            \\
            \hline
            $^{70}$Zn & $30$ & $65.12$ & $22.34$ & $2.95$ \\
            $^{124}$Sn & $50$ & $115.39$ & $27.56$ & $2.73$ \\
            $^{238}$U & $92$ & $221.70$ & $34.48$ & $3.07$
        \end{tabular}
        \end{ruledtabular}
        \caption{Atomic numbers, masses, and Fermi model parameters for three example nuclei, from~\cite{DeJager:1974liz}.}
        \label{tab:primakoff_params}
    \end{table}

\section{\boldmath $X(6900)$}
\label{sec:x6900}
\begin{figure}[b]
    \centering
    \includegraphics[width=\columnwidth]{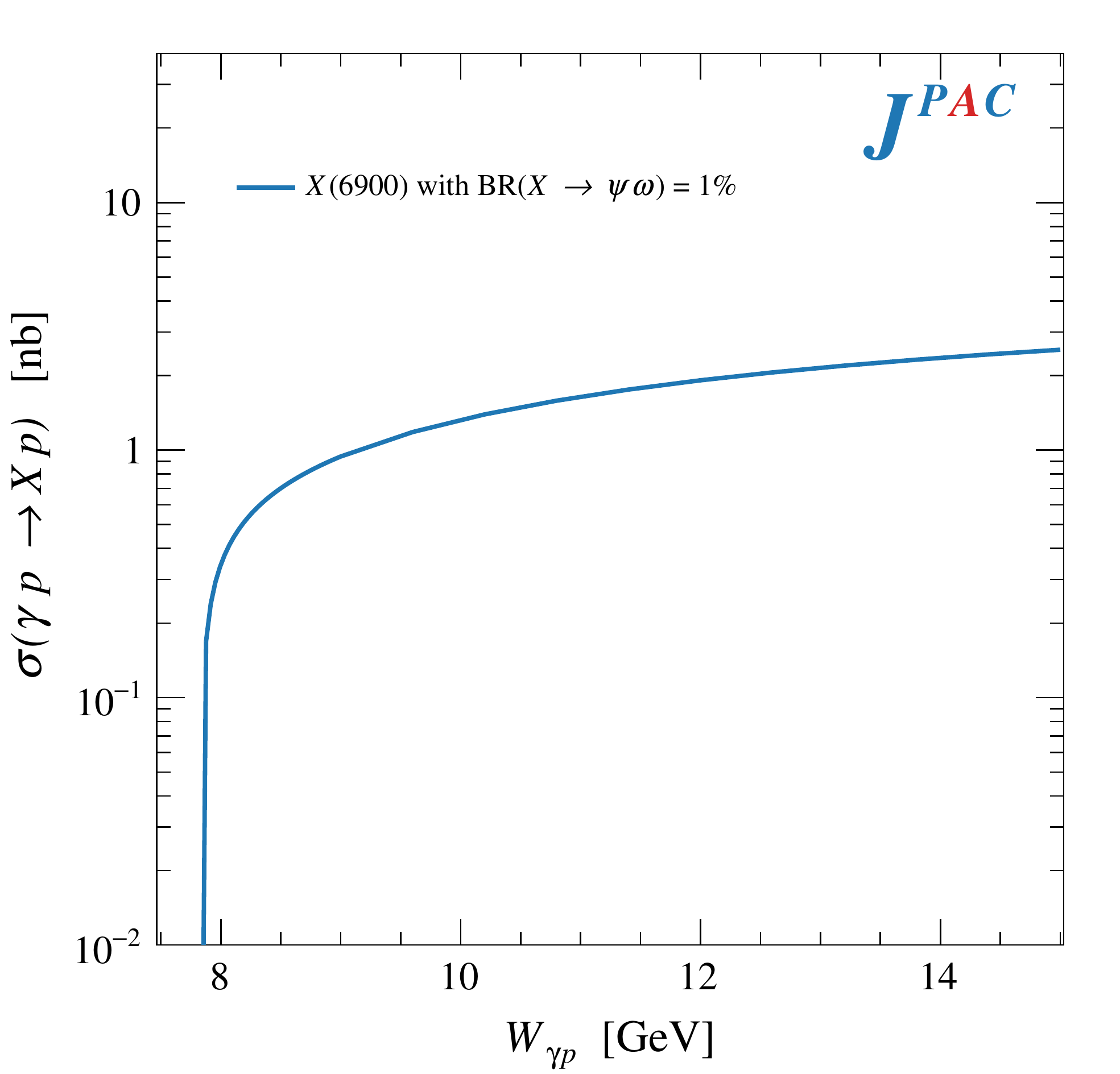}
    \caption{The production of the $X(6900)$ based on $\omega$ exchange, assuming $\mB[X(6900) \to \psi\omega]\sim 1\%$. The $\jpsi$ exchange is negligible even for large $\mB[X(6900) \to \psi\psi]$.}
    \label{fig:X6900}
\end{figure}
    \begin{table*}
        \begin{ruledtabular}
        \begin{tabular}{ccccccc}
             $Y$ &  $m_Y$ (\nsmev) & $\Gamma_Y$ (\nsmev) & $\mB(Y\to\gamma gg)$ (\%)& $\mB(Y\to\psi gg)$ (\%)& $\mB(Y\to\psi \pi \pi)$ (\%)& $R_Y$
            \\
            \hline
            $J/\psi$ & $3096.900\pm0.006$ & $0.0929\pm0.0028$ & $8.8\pm1.1$ & $-$ & $-$ & $1.0$
            \\
            $\psi(2S)$ & $3686.10 \pm 0.06$ & $0.294\pm 0.008$ & $1.03 \pm 0.29$ & $61.4 \pm0.6$ & $34.68 \pm 0.30$ & $0.55$
            \\
            $Y(4260)$ & $4220 \pm 15$ & $44 \pm 9$ & $-$ & $-$ & $3.2$ & $0.84$
        \end{tabular}
        \end{ruledtabular}
        \caption{Parameters for $Y$ production. The branching ratio of $\mB(\Y\to\psi \pi \pi)$ is obtained assuming $\Gamma_{ee}^\Y = \Gamma_{ee}^{\psi(3770)} = 262\eV$.}
        \label{tab:Y_couplings}
    \end{table*}
Recently the LHCb collaboration reported the observation of a narrow $X(6900)$ in the di-\jpsi mass spectrum~\cite{Aaij:2020fnh}. This structure is consistent with a $cc\bar c\bar c$ state with mass $m_X = 6886 \pm 22\mev$ and width $\Gamma_X = 168 \pm 102\mev$. We provide an estimate of the exclusive photoproduction cross section  near threshold, assuming a vector meson exchange, in analogy to the $\chi_{c1}(1P)$ in \cref{sec:x3872}.

The spin-parity assignment of the $X(6900)$ is still unknown, we will assume $J^{PC} = 0^{++}$ (\cf \cite{Karliner:2016zzc,*Debastiani:2017msn,*Bedolla:2019zwg,*Becchi:2020uvq}). This leads to the top vertex:
    \begin{equation}
        \mT_{\lamgam}^\mu = \frac{g_{X\psi\psi}}{m_X} \,
        \big[
        (k\cdot q) \, \epsilon^\mu(q,\lamgam) - (\epsilon(q,\lamgam) \cdot k) \, q^\mu
        \big]
    ~.
    \end{equation}
We use Eq.~\eqref{eq:partialwidth} to place an upper bound on the coupling, by assuming the total width to be saturated by the di-\jpsi final state. The central value $\Gamma_X = 168\mev$ leads to $g_{X\psi\psi} \sim 3.2$. However, the bottom vertex  remains the same as Eq.~\eqref{eq:X_bot_vert}, meaning the amplitude is limited by the tiny $\jpsi \to p\bar{p}$ decay width. 
Moreover, the heavy mass of the exchange further suppresses the cross section, yielding $\sigma = \mathcal{O}(10^{-6}~\text{nb})$ even for a $100\%$ branching ratio.

However, if the $X(6900)$ has a sizeable branching fraction, \ie $\gtrsim 1\%$, to a final state involving light mesons, such as the $\jpsi\, \omega$, observation in photoproduction could be possible. Even though these decays are OZI-suppressed, they can be estimated by comparing to the $\psi(3770) \to \jpsi\,\eta$ and $\phi\, \eta$ decay modes~\cite{pdg},
\begin{align}
    \frac{\mB(X \to \psi\, \omega)}{\mB(X \to \psi\, \psi)}  &=\frac{\text{PS}(X \to \psi\,\omega)}{\text{PS}(X \to \psi\,\psi)} \frac{g^2(\psi'' \to \phi\,\eta)}{g^2(\psi''\to \psi\,\eta)}\sim (1\text{--}4)\%
    ~,
\end{align}
with the same notation as \cref{sec:y4220}.
A prediction for the cross section assuming a nominal $\mB(X\to \jpsi \,  \omega) = 1\%$ is shown in \cref{fig:X6900}.

\section{\boldmath \Y and $\psi(2S)$} 
\label{sec:y4220}

\begin{table}[b]
    \begin{ruledtabular}
    \begin{tabular}{ccccc}
          & $A_\psi$ &  $\alpha_0$ & $\alpha^\prime$ ($\nsgev^{-2}$) & $b_0$ ($\nsgev^{-2}$) \\\hline
        HE~\cite{Blin:2016dlf} & $0.16$ &  1.15 & 0.11 & 1.01  \\
        LE~\cite{Winney:2019edt} & $0.38$ & 0.94 & 0.36 & 0.12  \\
        \end{tabular}
    \end{ruledtabular}
    \caption{Summary of numerical values of relevant Pomeron exchange parameters from fits to high-energy (``HE") and low-energy (``LE") \(J/\psi\) photoproduction data. 
   }
    \label{tab:pomeron_couplings}
\end{table}
    \begin{figure*}
        \includegraphics[width=0.48\textwidth]{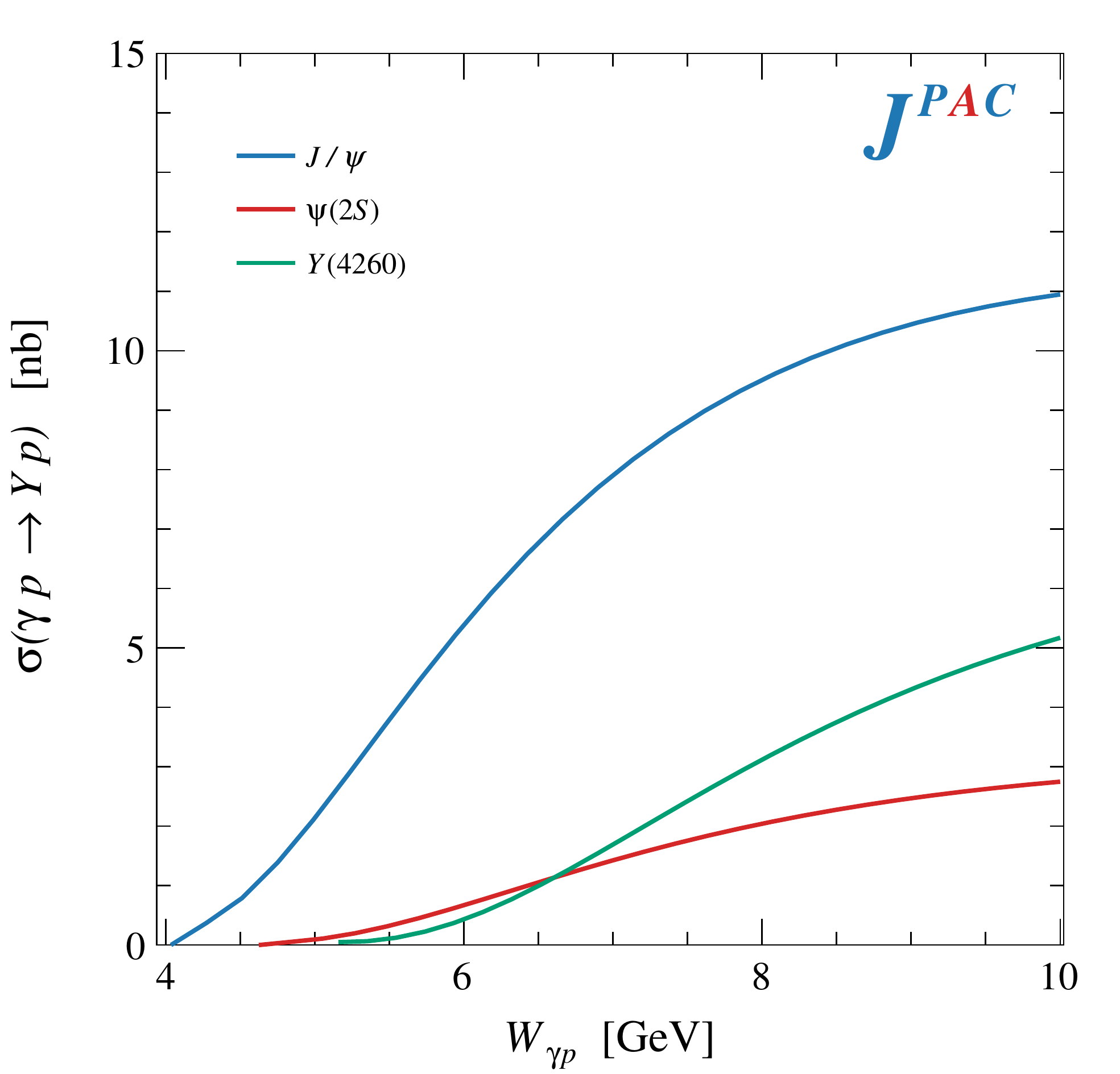}
        \includegraphics[width=0.48\textwidth]{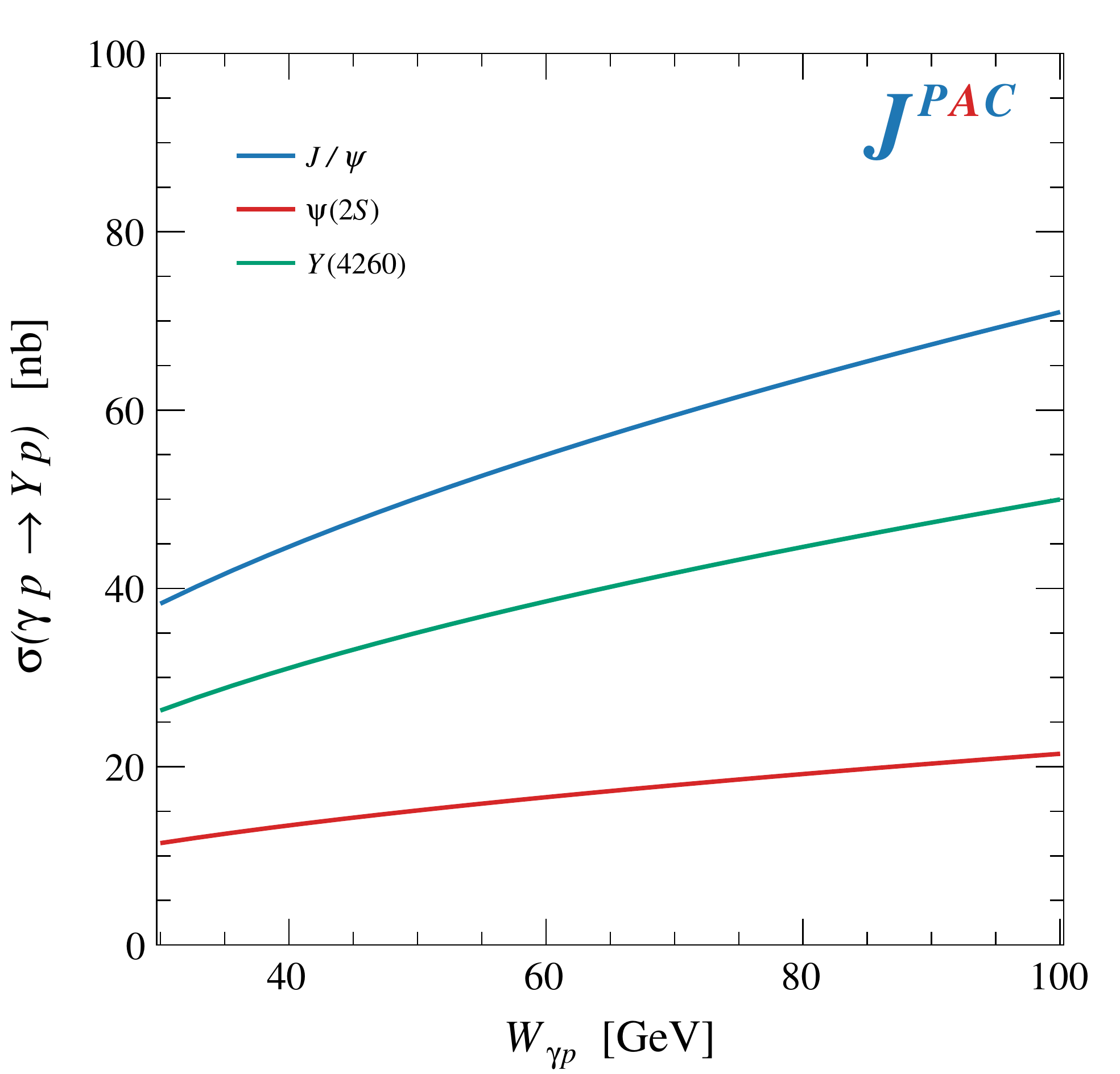}
        \caption{Cross sections for \Y photoproduction compared to the $J/\psi$ and $\psi(2S)$ at low (left) and high (right) energies.}
        \label{fig:Yplots}
    \end{figure*}

The \Y is one of the several $J^{PC}=1^{--}$ supernumerary states seen in direct $e^+e^-$ production. The detailed study of the $\jpsi\, \pi^+\pi^-$ lineshape by BESIII suggests a lighter and narrower state than the previous estimates~\cite{Ablikim:2016qzw}, which seems to be compatible with the signals seen in $\psi(2S)\, \pi^+\pi^-$, $h_c\, \pi^+\pi^-$, $\chi_{c0}\,\omega$, $\jpsi \,\eta$, and $\pi D^* \bar D$~\cite{BESIII:2016adj,*Ablikim:2017oaf,*Ablikim:2018vxx,*Ablikim:2019apl,*Ablikim:2020cyd}. The PDG average of mass and width is $M_Y= 4220\pm 15\mev$, $\Gamma_Y = 44\pm 9\mev$. The main motivation for an exotic assignment is that a fit to the inclusive $e^+e^-$ data provides three ordinary $\psi$ states that fulfill quark model predictions, and do not seem compatible with any of the $Y$ states~\cite{Ablikim:2007gd}.

Since the \Y has been seen in $e^+ e^-$ collisions only, it is only possible to measure the branching ratios times its electronic width $\Gamma^Y_{ee}$, which is presently unknown. An upper limit based on the inclusive data in~\cite{Ablikim:2007gd} gives  $\Gamma^Y_{ee} < 580 \eV$ at 90\% C.L.~\cite{Mo:2006ss}. A recent global analysis suggests $\Gamma^Y \sim \mathcal{O}(10^2)$--$\mathcal{O}(10^3)\eV$~\cite{Cao:2020vab}. 

At high energies, vector meson photoproduction is well described by Pomeron exchange, which is expected to be related to the spectrum of glueballs~\cite{Donnachie:2002en,*Laget:2019tou,*Mathieu:2018xyc}. At threshold, a model that realizes the Pomeron as an explicit 2- or 3-gluon exchange was given in~\cite{Brodsky:2000zc} (an alternative description via low-energy open charm exchanges is found in~\cite{Du:2020bqj}).
Given the uncertainties brought by this relation, we consider the two effective Pomeron models for $\jpsi$ photoproduction used in~\cite{Blin:2016dlf,Winney:2019edt} to interpolate the high and low energy regions.
The former model was fitted to the high energy data from HERA (hereafter ``HE",  $W_{\gamma p} \gtrsim 20\gev$~\cite{Chekanov:2002xi,*Aktas:2005xu}).
The latter  was fitted to the lower energy data from SLAC and the   newest from GlueX  (hereafter ``LE",  $W_{\gamma p} \lesssim 7\gev$~\cite{Camerini:1975cy,Ali:2019lzf}). This model has a $t$-dependence somewhat different from that of  HE when extrapolated to lower energies. 
The cross sections for the \Y and $\psi(2S)$ are obtained by replacing the {\jpsi} couplings, mass and width by those of the \Y and the $\psi(2S)$. This is further  detailed  below.

The HE model has a helicity-conserving amplitude~\cite{Blin:2016dlf},
    \begin{equation}
          \mel{\lamY\lamNp}{T^\text{(HE)}}{\lamgam\lamN} = F(s,t) \, \delta_{\lamgam\lamY} \delta_{\lamN\lamNp},
    \end{equation}
while the LE model is based on the vector Pomeron model~\cite{Close:1999bi,*Lesniak:2003gf,Winney:2019edt},
    \begin{align}
&\mel{\lamY\lamNp}{T^\text{(LE)}}{\lamgam\lamN} = \frac{F(s,t)}{s} \, 
        \left[\bar{u}(\pp,\lamNp) \, \gamma_\mu  \, u(p,\lamN)\right]\nn \\
        &\quad\times\, \varepsilon_\nu^*(\qp, \lamY) \,
        \big[
        \varepsilon^\mu(q, \lamgam) \, q^\nu - \varepsilon^\nu (q,\lamgam) \, q^\mu
        \big]
    ~.
    \end{align}
The function $F(s,t)$ is the same for both models, and contains the dynamical $s,t$ dependence of the Pomeron:
    \begin{equation}
        F(s,t) = i \, e \, A_\psi \, \bigg(\frac{s-s_\text{th}}{s_0}\bigg)^{\alpha(t)} \, e^{b_0 \, t'},
    \end{equation}
where $A_\psi$ is the product of the top and bottom couplings for $\jpsi$ photoproduction,  $s_\text{th}$ is an effective threshold fitted from data for HE, and fixed to the $\jpsi\,p$ threshold for LE. The slope $b_0$ further suppresses the amplitude at large values of $t$. The scale is $s_0 = 1\gevsq$ as customary.
The parameters $b_0$, $\alpha_0$, and $\alpha^\prime$ are assumed to be intrinsic to the Pomeron and do not depend on the vector particle produced. Values for all parameters are shown in \cref{tab:pomeron_couplings}.

For the \Y and $\psi(2S)$, we set $s_\text{th}$ to the physical $Y\,p$ threshold. If one 
considers  the Pomeron as an approximate 2-gluon exchange, the relative strength $R_{\psi'} = A_{\psi'} / A_\psi $ of the $\psi(2S)$ and \jpsi  couplings is given by the ratio of couplings to a photon and two gluons,
    \begin{equation}
        R_{\psi^\prime} = 
        \sqrt{
        \frac{g^2(\psi^\prime \to \gamma gg)}{g^2(\psi \to \gamma gg)}
        }
        ~.
    \end{equation}
The couplings $g^2$ can be computed form the known  
 partial widths $\mB\,\Gamma$ divided  by the corresponding 3-body phase space (PS), 
    \begin{equation}
g^2(Y\to \gamma g g) = \frac{6 m_Y \mB(Y\to \gamma g g) \Gamma_Y}{\text{PS}(Y\to \gamma g g)}.
\end{equation} 
The energy dependence of the underlying matrix element is neglected.
Using the branching ratios $\mB(\jpsi \to \gamma gg)$ and $\mB(\psi(2S) \to \gamma gg)$  extracted by CLEO~\cite{Besson:2008pr,*Libby:2009qb} we obtain  $R_{\psi^\prime}= 0.55$, which is comparable with the ratio of $\jpsi$ and $\psi(2S)$ quasi-elastic photoproduction cross sections in~\cite{Adloff:1997yv,*Grzelak:2018srl}, $\sqrt{\sigma_{\psi'}/\sigma_\psi} \sim 0.39$.

For the \Y, such radiative decays have not been seen. However, we resort to the arguments of \cite{Voloshin:1980zf,*Novikov:1980fa}, which assume that the matrix element of a vector $Y \to \jpsi \,\pi \pi$ factorizes into a hard $Y \to \jpsi gg$ process, calculable with QCD multipole expansion, and a hadronization process $gg \to \pi\pi$, which is universal and does not depend on the particular $Y$ state. Using VMD one can further relate the $Y \to \jpsi \,gg$ process to $Y \to \gamma gg$. If the energy dependence of the matrix elements is neglected, one gets:
    \begin{equation}
        R_Y = \frac{e \,f_\psi}{m_\psi} \sqrt{\frac{g^2(Y\to\psi\pi\pi)}{g^2(\psi \to \gamma g g)} \frac{g^2(\psi^\prime\to\psi g g)}{g^2(\psi^\prime \to \psi \pi\pi)}}~,
    \end{equation}
with $g(Y\to\psi\pi\pi) \simeq 120 \times \sqrt{\mB(Y\to\psi\pi\pi)}$.
The analysis of diffractive photoproduction at HERA shows the $\ell^+\ell^- \pi^+\pi^-$ invariant mass, where the strong signal of the $\psi(2S)$ appears on top of a small background that could be due to a \Y. In Appendix~\ref{app:hera} we put a 95\% C.L. upper limit on the ratio of the \Y and $\psi(2S)$ signals, $R_\text{HERA} = 6.5\%$. We can thus impose
\begin{equation}
    \left(\frac{R_Y}{R_{\psi'}}\right)^2 = R_\text{HERA} \frac{\mB(\psi'\to\psi \pi\pi)}{\mB(Y\to\psi \pi\pi)}~,
\end{equation}
and obtain $\mB(Y\to \psi \,\pi^+\pi^-)= 1\%$ and $R_Y = 0.84$. Incidentally, this branching ratio leads to a leptonic width $\Gamma_{ee}^Y\simeq 150$--$1350\eV$, depending on the specific solution extracted in~\cite{Ablikim:2016qzw},\footnote{The PDG reports $\Gamma^Y_{ee}\times \mB(Y\to \psi \pi^+ \pi^-) = 9.2 \pm 1.0\eV$ for the \Y~\cite{pdg}, averaging the BaBar and CLEO extractions based on a single resonance fit~\cite{He:2006kg,Lees:2012cn}. The most recent BESIII analysis, which resolves a composite structure of the peak, reports eight different couplings, depending on the relative phases of the $Y(4220)$ and $Y(4320)$~\cite{Ablikim:2016qzw}. Half of the solutions are roughly compatible with the old estimate, while the other half prefer smaller values $\sim 2\eV$.} 
We remind that a large $R_Y$ would suggest a larger affinity to gluons than ordinary charmonia, as expected for a heavy gluonic hybrid~\cite{Kou:2005gt,*Liu:2012ze,*Guo:2014zva,*Oncala:2017hop}.
We show the tabulated values for all couplings in \cref{tab:Y_couplings}.

The resulting cross sections for the \jpsi, $\psi(2S)$, and $Y(4260)$ are plotted for the low and high-energy regions  in \cref{fig:Yplots}. Both $\psi(2S)$ and $Y(4260)$ can
be measured in a clean $\jpsi\,\pi^+\pi^-$ final state, with branching ratios $34.68\pm 0.30\%$ and $1\%$ (in our estimate), respectively. The $\psi(2S)$ can also be reconstructed in a lepton pair, with branching ratio $\mB(\psi(2S)\to e^+ e^-) = (7.93 \pm 0.17)\E{-3}$.

\section{\boldmath $P_c$ Reggeons in backward $\jpsi$ photoproduction}
\label{sec:pc}
\begin{figure}[t]
    \centering
    \includegraphics[width=.5\columnwidth]{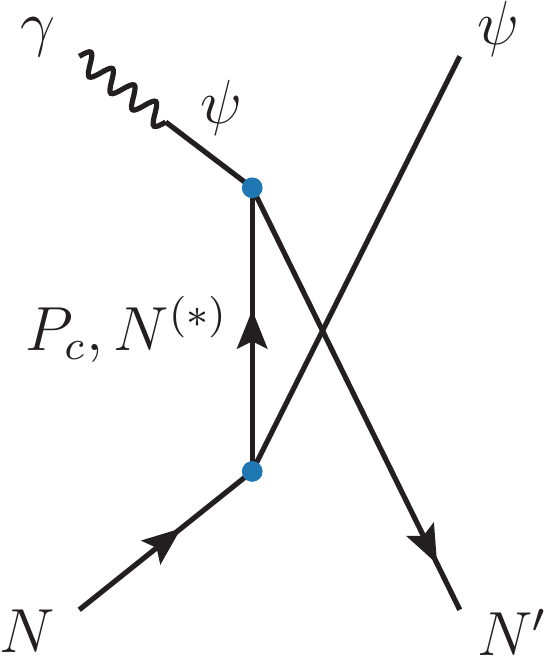}
    \caption{The photoproduction of $\jpsi$ at backward angles is populated by Reggeons having the $P_c$ quantum numbers, as well as ordinary $N^*$ trajectories. Unfortunately, the latter dominate, making the extraction of pentaquarks from this reaction impossible.}
    \label{fig:photo_pc}
\end{figure}
Photoproduction of hidden charm pentaquarks has extensively been discussed in~\cite{Blin:2016dlf,Winney:2019edt,Karliner:2015voa,*Kubarovsky:2015aaa,*Wang:2015jsa,*Cao:2019kst}. These studies consider the direct production of pentaquark resonances in the $s$-channel, which requires $W_{\gamma p}\sim m_{P_c} \sim 4.5\gev$. Such low energies will hardly be explored at the EIC. One could consider the associated production of pentaquarks with other pions. However, reliable predictions can be made for soft pions only (see \eg~\cite{Braaten:2019sxh}), which do not contribute significantly to the total energy. 

\begin{table*}
    \begin{ruledtabular}
    \begin{tabular}{cccccc}
        &  $W_{\gamma p}$ (\nsgev) & $\sigma$ (nb) & $\mB(\mQ \to \ell^+\ell^-\,n\pi)$ ($\times 10^{-3}$) & Counts & Comparison \\\hline
        \X &  \multirow{2}{*}{6} & 33.1 & $5.3$ & 877 & $\sim 90$~\cite{Ablikim:2020xpq}\\
        \Z &  & $15.9$ & $12.5$ & 994 & $\sim 1300$~\cite{Collaboration:2017njt} \\ \hline
         $Z_b(10610)^+$ & \multirow{2}{*}{15} & $2.8$ & $2.6$ &  36 & $\sim 750$~\cite{Garmash:2014dhx} \\
        $Z_b'(10650)^+$ & & $0.66$ & $2.1$ & 7 & $\sim 200$~\cite{Garmash:2014dhx} \\ \hline\hline
                &  &  & $\mB(\jpsi \to \ell^+\ell^-)^2$ ($\times 10^{-3}$) &  & \\\hline
                $X(6900)$ & 12 & 1.9  &  14 & 133 & $\sim 800$~\cite{Aaij:2020fnh} \\
        \end{tabular}
    \end{ruledtabular}
    \caption{Estimates of yields for one year of data taking under the conditions described in the text. The branching ratios $\mB(\mQ \to \ell^+\ell^-\,n\pi)$ are given by $\sum_V \mB(\mQ \to V \,n\pi)\times \mB(V \to \ell^+\ell^-)$.  Comparison with existing datasets are also given.  The efficiency is assumed to be $1\%$. Higher efficiencies are certainly possible,  \eg $50\%$ for the \Z at BESIII. The results for the $X(6900)$ must be rescaled by the yet unknown $\mB[X(6900) \to \psi \psi]$.}
    \label{tab:yields}
\end{table*}
Alternatively, one can search for the presence of $P_c$ trajectories in backward \jpsi  photoproduction, as shown in \cref{fig:photo_pc}.
In the backward region (small $u$ and large $t$), the contribution of Pomeron exchange in the $t$-channel, which represents the main background in~\cite{Blin:2016dlf,Winney:2019edt,Karliner:2015voa,*Kubarovsky:2015aaa,*Wang:2015jsa,*Cao:2019kst}, becomes negligible with respect to $u$-channel exchanges. 
These are populated by $P_c$ resonances, as well as ordinary $N^{(*)}$ trajectories. If the latter were to be negligible, a signal of \jpsi in the backward region will be unambiguously due to the existence of pentaquarks.
Here, we provide a rough estimate of the relative size between the two. 
Up to kinematic factors, the main differences are the couplings to the photon and \jpsi, and the different trajectories.
The couplings of $N^{(*)}$ can be simply taken as the ones of proton exchange. As shown in \cref{tab:X_couplings}, the coupling of $\jpsi$ to the proton
$\mathcal{O}(10^{-3})$ and the coupling to the photon is given by the  electric charge $e$. 
VMD relates the electromagnetic transition $P_c\to\gamma\,p$ to $P_c\to\jpsi\,p$.
The only input needed is the branching ratio $\mB(P_c\to \jpsi\,p)$. In~\cite{Winney:2019edt} we found upper limits for the branching fraction of roughly $1$--$5\%$, depending on the $P_c$ quantum numbers. Using a branching fraction of 1\%, and the typical width of the pentaquark signals found so far of the order of 10\mev~\cite{Aaij:2019vzc}, we obtain for the product of couplings values $\mathcal{O}(10^{-3}) \times e$.

This is the same order of magnitude as the product of couplings for the proton exchange. At high energies however, reggeization  will suppress the $P_c$ exchange  due to its larger mass and therefore smaller intercept for the trajectory. We conclude that searches of hidden-charm pentaquarks in this way are hindered by a large $N^{(*)}$ background. 
The photoproduction of hidden-bottom pentaquarks, were  they to exist, could still be possible, and has been discussed in~\cite{Cao:2019gqo,*Paryev:2020jkp}.

\section{Conclusions}
\label{sec:conclusions}
In this paper we provide estimates for photoproduction rates of various charmonia and exotic charmonium-like states in the kinematic regimes relevant to future electron-hadron colliders. We focus on a few states as benchmarks based on the availability of experimental information, \eg decay widths. However, the formalism presented here is readily applicable to other \XYZ states when more measurements will become available.

In the low-energy regime, with $W_{\gamma p}$ close to threshold, fixed-spin particle exchanges are expected to provide a realistic representation of the amplitude. As such, we give estimates for exclusive charged $Z_c(3900)^+$, $Z_b(10610)^+$, and $Z_b^\prime(10650)^+$ production via pion exchange, as well as $X(3872)$ and $\chi_{c1}(1P)$ production via vector meson exchange. For energies near threshold we expect cross-sections of the order of a few nanobarns for the \Z  and upwards of tens of nanobarn for the \X.
We remark on the possibility of exploring the recently observed $X(6900)$ in photoproduction. Production mechanisms involving possible OZI-suppressed couplings to light vector mesons yield 
to cross sections of a fraction of a nanobarn.

At high energies, the correct behavior is captured by (continuous spin) Regge exchange.
Based on standard Regge phenomenology, we extend our results for $Z_c$, $Z^{(\prime)}_b$ and \X production to center-of-mass energies where the EIC is expected to reach peak luminosity.
For the vector meson states, we build upon existing models to provide estimates of diffractive \Y and $\psi(2S)$ production. Unlike production of $XZ$, diffractive production increases as a function of energy, making high-energy colliders such as the EIC a preferable laboratory for the spectroscopy of the $Y$ states.
We further discuss the feasibility of indirect detection of $P_c$ states in backward $\jpsi$ photoproduction. However, we find that the contribution of $P_c$ states is hindered by the ordinary $N^{(*)}$ exchanges.

To further motivate the \XYZ spectroscopy program at high-energy electron-hadron facilities, it is important to translate the cross section predictions into expected yields. A detailed study, \eg for the EIC, would require details of the detector geometry. Nevertheless, one can have a rough idea of the number of events, by considering a hypothetical setup based on the existing GlueX detector~\cite{gluextcr} but higher energies. Specifically, assuming photon beam of the order of $E_\gamma^\text{lab} = 20\gev$, an intensity of $10^8 \,\gamma/\text{s}$, and a typical hydrogen target, one could reach a luminosity of $\sim 500\text{~pb}^{-1}$ for a year of data taking. For the yield estimates, one needs to multiply the cross section by the appropriate branching ratios $\mB(\XYZ \to \jpsi\,n\pi) \sim 5\%$ and by $\mB(\jpsi \to \ell^+\ell^-) = 12\%$. Even with a low $1\%$ detector efficiency, assuming $\sigma = 10\text{~nb}$, we estimate 300 events per year.
The expectations for the individual \XYZ are given in \cref{tab:yields}, together with the comparison with the existing datasets by BESIII.
While production of $Y$ states benefits from higher energies, lower $W_{\gamma p} \lesssim 7\gev$ are much more efficient in producing $X$ and $Z$ states. 

We conclude that electro- and photoproduction facilities can complement the existing experiments that produce \XYZ.  In fact, such facilities will 
 give the  opportunity to study \XYZ in exclusive reactions that 
 provide valuable information about production mechanisms different from the reactions where the \XYZ have been seen so far. 
  This will further shed light on the  nature of several of these exotic candidates.
   
The code implementation to reproduce all results presented here can be accessed on the JPAC website~\cite{JPACweb}.

\acknowledgments
We thank M.~Battaglieri, S.~Dobbs, D.~Glazier, R.~Mitchell, and J.~Stevens  for useful discussions within the EIC Yellow Report Initiative. This work was supported by the U.S. Department of Energy under Grants No. DE-AC05-06OR23177 and No. DE-FG02-87ER40365. 
The work of C.F.-R. is supported by PAPIIT-DGAPA (UNAM, Mexico) under Grant No. IA101819 and by CONACYT (Mexico) under Grant No. A1-S-21389. 
 V.M. is supported by the Comunidad Aut\'onoma de Madrid through the Programa de Atracci\'on de Talento Investigador 2018-T1/TIC-10313 and by the Spanish national grant PID2019-106080GB-C21. 
 
\appendix
\section{\boldmath $X(3872)$ to $J/\psi\, \omega$ and $J/\psi\, \rho$ couplings}
\label{appendix:x3872couplings}

\begin{figure}[b]
\centering
\includegraphics[keepaspectratio]{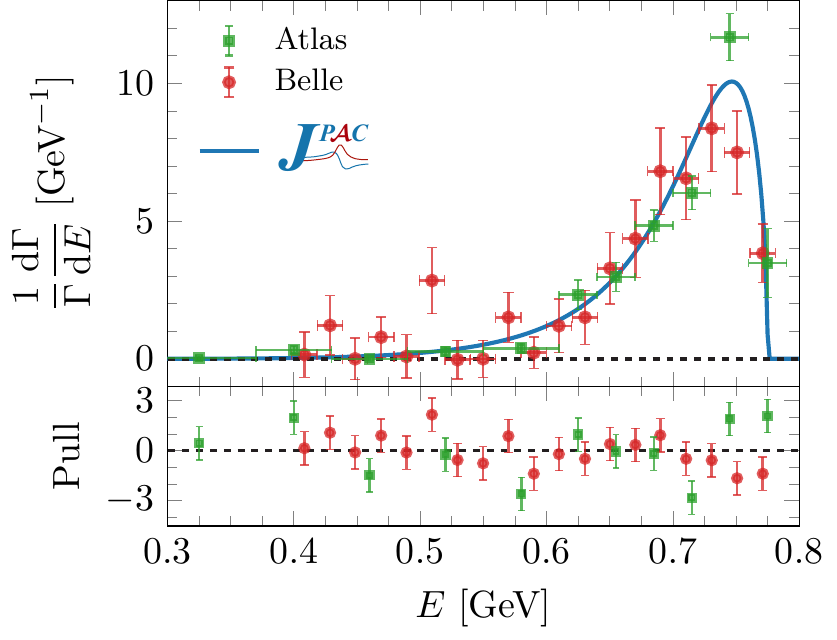}
\caption{Differential decay width for the process $\X \to J/\psi \pi \pi$. Our result in Eq.~\eqref{eq:XJpsiV} is given by the blue solid line, whereas the experimental data from Belle~\cite{Choi:2011fc} and ATLAS~\cite{Aaboud:2016vzw} are shown with red circles and green squares, respectively.\label{fig:chic1_dG_B}}
\end{figure}

We evaluate here the couplings of $\X\to\jpsi\, \mE$ with $\mE =\rho, \omega$. Experimentally, these are accessible through the decays $\X \to \jpsi\, \pi^+\pi^-$ and $\jpsi \,\pi^+\pi^-\pi^0$, respectively. We write the differential decay widths for these processes as:
\begin{align}\label{eq:XJpsiV}
&\frac{ \mathrm{d} \Gamma(X\to \jpsi \,n\pi)}{\mathrm{d} w^2 }  = \frac{1}{\pi} \frac{w}{( w^2 - m_\mE^2)^2 + m_\mE^2 \Gamma_\mE^2} \nn \\
&\qquad \times \Gamma[X \to \jpsi \,\mE(w^2)]\, \Gamma[\mE(w^2) \to n\pi]~, 
\end{align}
with $n=2,3$, $w$ is the invariant mass of the $n\pi$ system, \ie of the virtual vector \mE.  
The first width is given by:
\begin{align}
\Gamma[X \to \jpsi \,\mE(w^2)] &= \frac{\lambda^{1/2}\!\left(m_X^2,m_\psi^2,w^2\right)}{48\pi M_X^3}\nn\\ 
&\quad\times\!\!\sum_{\lambda_X\lambda_\psi\lambda_\mE}\!\! \left|\mel{\lambda_\psi\lambda_\mE}{T}{\lambda_X} \right|^2~,
\end{align}
where:
\begin{align}
\mel{\lambda_\psi\lambda_\mE}{T}{\lambda_X}  &= - i g_{\psi X \mE}\, \varepsilon_{\alpha\beta\gamma\mu} \,
 \varepsilon^\alpha (p_X,\lambda_X)\nn\\
&\qquad\times\varepsilon^{*\beta}(p_\psi,\lambda_\psi)\,
\varepsilon^{*\gamma}(p_\mE,\lambda_\mE)\,
p^\mu_\psi~.
\end{align}
For $\rho \to 2\pi$, we consider the standard amplitude (see \eg Ref.~\cite{Borasoy:1995ds}) dependent on the vector coupling constant $g_V = 0.086$ and the pion decay constant $f_\pi = 93\ \text{MeV}$, which leads to the $\rho$ width,
\begin{equation}
\Gamma[\rho(w^2) \to 2\pi] = \frac{1}{6\pi} \left( \frac{w^2 g_V }{f_\pi^2} \right)^2 \frac{(w^2 - 4m_\pi^2)^{3/2}}{8w^2}~. \label{eq:rhoto2pi}
\end{equation}
The shape of the differential decay width of the process $X \to J/\psi \,\pi\pi$ is completely fixed by Eqs.~\eqref{eq:XJpsiV} and~\eqref{eq:rhoto2pi}, and is independent of the value of the global coupling. In Fig.~\ref{fig:chic1_dG_B} we show the invariant $\pi\pi$ mass spectrum, which is completely dominated by the $\rho$, and agrees fairly well with the experimental data from Belle~\cite{Choi:2011fc} and ATLAS~\cite{Aaboud:2016vzw} (better agreement can be reached by giving more freedom to the $\rho$ lineshape, or including $\rho$-$\omega$ mixing~\cite{Hanhart:2011tn,*Faccini:2012zv,Abulencia:2005zc}). The coupling $g_{\psi X\rho}$ is extracted from the integrated width:
\begin{equation}
    \Gamma(X \to \jpsi \, \pi\pi) = \int_{4m_{\pi}^2}^{(m_X-m_\psi)^2} \!\!\mathrm {d} w^2
    \frac{ \mathrm{d}\Gamma(X \to \jpsi \, \pi\pi)}{\mathrm{d} w^2}~.
\end{equation}
As experimental input, we consider the branching ratio $\mathcal{B}[X \to J/\psi \pi \pi] = 4.1^{+1.9}_{-1.1}\, \%$~\cite{Li:2019kpj} and the total width $\Gamma_X = 1.19 \pm 0.19\ \text{MeV}$, the average of the recent LHCb measurements~\cite{Aaij:2020qga,Aaij:2020xjx}.

For the $\omega \to 3\pi$ decay, while more sophisticated approaches exist~\cite{Niecknig:2012sj,*Danilkin:2014cra,*Albaladejo:2020smb}, we take here the simple vertex:
\begin{equation}
\mel{3\pi}{T}{\lambda_\omega} = i g_{\omega 3\pi}\, \varepsilon_{\mu\nu\alpha\beta}\, \varepsilon^{\mu}(p_\omega,\lambda_\omega)\, p_+^{\nu}\, p_-^{\alpha}\, p_0^\beta\, 
\end{equation}
which results in the width:
\begin{align}
&\Gamma(\omega(w^2) \to 3\pi)  = \frac{1}{(2\pi)^3} \frac{1}{32 w^3 } \frac{g_{\omega 3\pi}^2}{72} \int_{4m_\pi^2}^{(w-m_\pi)^2}  \frac{\mathrm{d} {w'}^2}{w'} 
\nn \\
&\qquad \times \left({w'}^2 - 4m_\pi^2\right)^{3/2} \lambda^{3/2}(w^2,{w'}^2,m_\pi^2) ~, \label{eq:omegato3pi}
\end{align}
with $w'$ the invariant mass of a dipion subsystem. The coupling $g_{\omega 3\pi}$ is adjusted to reproduce the experimental $\omega \to 3\pi$ width, $\Gamma(\omega \to 3\pi) = \mB(\omega \to 3\pi) \,\Gamma_{\omega}$, where $\mB(\omega \to 3\pi) = 89.3 \pm 0.6\%$ and $\Gamma_{\omega} = 8.49 \pm 0.08\mev$~\cite{pdg}. The integrated width is given by:
\begin{equation}\label{eq:IntegratedWidthXJ3pi}
    \Gamma(X \to \jpsi \,3\pi) = \int^{(m_X-m_\psi)^2}_{9m_\pi^2} \!\!\mathrm {d} w^2
    \frac{ \mathrm{d} \Gamma(X \to \jpsi 3\pi)}{\mathrm{d} w^2}~,
\end{equation}
The coupling $g_{\psi X\omega}$ is obtained from Eq.~\eqref{eq:IntegratedWidthXJ3pi} and the branching ratio $\mB(X \to \jpsi \,\omega) = 4.4^{+2.3}_{-1.3}\%$ \cite{Li:2019kpj},
\begin{equation}
\mB(X \to \jpsi \,\omega) =  \frac{ \Gamma(X \to \jpsi \,3\pi) }{ \mB(\omega \to 3\pi)\, \Gamma_X}~.
\end{equation}
The resulting values for the couplings are reported in \cref{tab:X_couplings}.

\section{\boldmath Amplitudes for Primakoff effect}
\label{app:Primakoff_kinematics}

The calculations simplify in the nucleus rest frame, where
\begin{align}
    q &= \left(\nu,0,0,p_\gamma \right) \nn\\
    p &= \left(m_A,0,0,0\right)\nn\\
    q' &= \left(E_X, p_X \sin \theta_X,0,p_X \cos \theta_X\right) \nn \\
    k &= q-q'\nn\\
    \varepsilon(q,\lamgam = \pm 1) &= \left(0,\mp\frac{1}{\sqrt{2}},-\frac{i}{\sqrt{2}},0\right)\nn \\
    \varepsilon(q,\lamgam = 0) &= \left(\frac{p_\gamma}{Q},0,0,\frac{\nu}{Q}\right)\nn\\
        \varepsilon(q',\lamX = \pm 1) &= \left(0,\mp\frac{\cos \theta_X}{\sqrt{2}},-\frac{i}{\sqrt{2}},\pm\frac{\sin \theta_X}{\sqrt{2}}\right)\nn \\
    \varepsilon(q',\lamX = 0) &= \left(\frac{p_X}{m_X}, \frac{E_X}{m_X}\sin\theta_X,0,\frac{E_X}{m_X} \cos\theta_X\right)\,,
\end{align}
with $p_\gamma = \sqrt{\nu^2 + Q^2}$, $E_X = \sqrt{p_X^2 + m_X^2}$. The variables $p_X$, $\theta_X$, $\nu$ are related to the invariants $Q^2$, $x_A$ and $t$ by
\begin{align}
    \nu &= \frac{Q^2}{2m_A x_A}\nn\\
    p_X &= \frac{\sqrt{t^2 + 4 m_A \nu t + 4m_A^2 \left(\nu^2 - m_X^2\right)}}{2 m_A}\nn\\
    \cos\theta_X &= \frac{t - m_X^2 + Q^2 + 2 E_X \nu}{2 p_X p_\gamma}\,.
\end{align}
In this frame, the nuclear tensor in Eq.~\eqref{eq:Primakoff_msq} reduces to the timelike component only, $W_{00} = 64 Z^2 m_A^6 \,F^2_0(t)/(4m_A^2-t)^2$. The cross sections read
\begin{equation}
    \frac{d\sigma_{L,T}}{dt} = \frac{\alpha g_{X\gamma\gamma^*}^2 \mathcal{A}_{L,T} W_{00}}{8 m_A m_X^4 p_\gamma t^2 \left(2m_A\nu - Q^2\right)}\,,
\end{equation}
where the virtual photon flux is given in the Hand convention~\cite{Hand:1963bb}, and
\begin{align}
\mathcal{A}_T &= \frac{m_X^4}{2g_{X\gamma\gamma^*}^2}\sum_{\lamgam = \pm} \mT^{00}_{\lambda_\gamma} \nn\\
&=\cos^4\frac{\theta_X}{2} \Big(p_X p_\gamma (p_\gamma + p_X) + \nu E_X (p_\gamma - p_X) \nn\\
&\quad- 2 p_X p_\gamma^2 \cos \theta_X\Big)^2 + \sin^4\frac{\theta_X}{2} \Big(E_X \nu(p_X + p_\gamma) \nn\\
&\quad + p_X p_\gamma (p_X - p_\gamma) - 2 p_X p_\gamma^2 \cos\theta_X\Big)^2 + \frac{p_\gamma^2}{2m_X^2} \nn\\
&\quad\times  \sin^2\theta_X\left(\nu(E_X^2 + p_X^2)  - 2 E_X p_X p_\gamma \cos\theta_X\right)^2\,,\\
\mathcal{A}_L &= \frac{m_X^4}{g_{X\gamma\gamma^*}^2} \mathcal{T}^{00}_{0} = p_X^2 E_X^2 Q^2 \sin^2 \theta_X\,,
\end{align}
with $\mathcal{T}^{\mu\nu}_{\lambda_\gamma} = \sum_{\lambda_X} \mathcal{T}^\mu_{\lamgam\lamX} \mathcal{T}^{*\nu}_{\lamgam\lamX}$.
As customary, the electroproduction cross section is given by~\cite{Bedlinskiy:2014tvi}
\begin{equation}
    \frac{d\sigma(e A \to e X A)}{dt dQ^2 dx_A} = \mathcal{G}(x_A,Q^2,s)\left( \frac{d\sigma_T}{dt} + \epsilon \frac{d\sigma_L}{dt}\right)\,,
\end{equation}
with
\begin{align}
    \mathcal{G}(x_A,Q^2,s) &= \frac{\alpha}{2\pi} \frac{Q^2}{(s - m_A^2)^2 } \frac{1 - x_A}{x_A^3} \frac{1}{1 - \epsilon}\,,\nn\\
    \epsilon &= \frac{1-y-\frac{m_A^2 Q^2}{(s - m_A^2)^2}}{1-y+\frac{y^2}{2}+\frac{m_A^2 Q^2}{(s - m_A^2)^2}}\,,\nn\\
    y &= \frac{Q^2}{x_A(s-m_A^2)}\,,    
\end{align}
where by $s$ we mean the energy squared in the center-of-mass of the electron-nucleus collision. 

\section{\boldmath Extracting the \Y-over-$\psi(2S)$ signal ratios}
\label{app:hera}
We consider the data on diffractive production of $\psi(2S)$ at HERA, in a kinematics similar to the one covered by the EIC~\cite{Adloff:2002re}. We consider the invariant mass spectrum of $\ell^+\ell^-\pi^+\pi^-$.
The dataset includes also events where the proton dissociates into a jet of invariant mass $< 22\%\, W_{\gamma p}$ and is left mostly undetected in the beamline. We assume that the mass spectrum does neither depend on the other kinematical variables, nor on the specific class of events (whether elastic or dissociative). 

\begin{figure*}
\centering
\includegraphics[width=.8\textwidth]{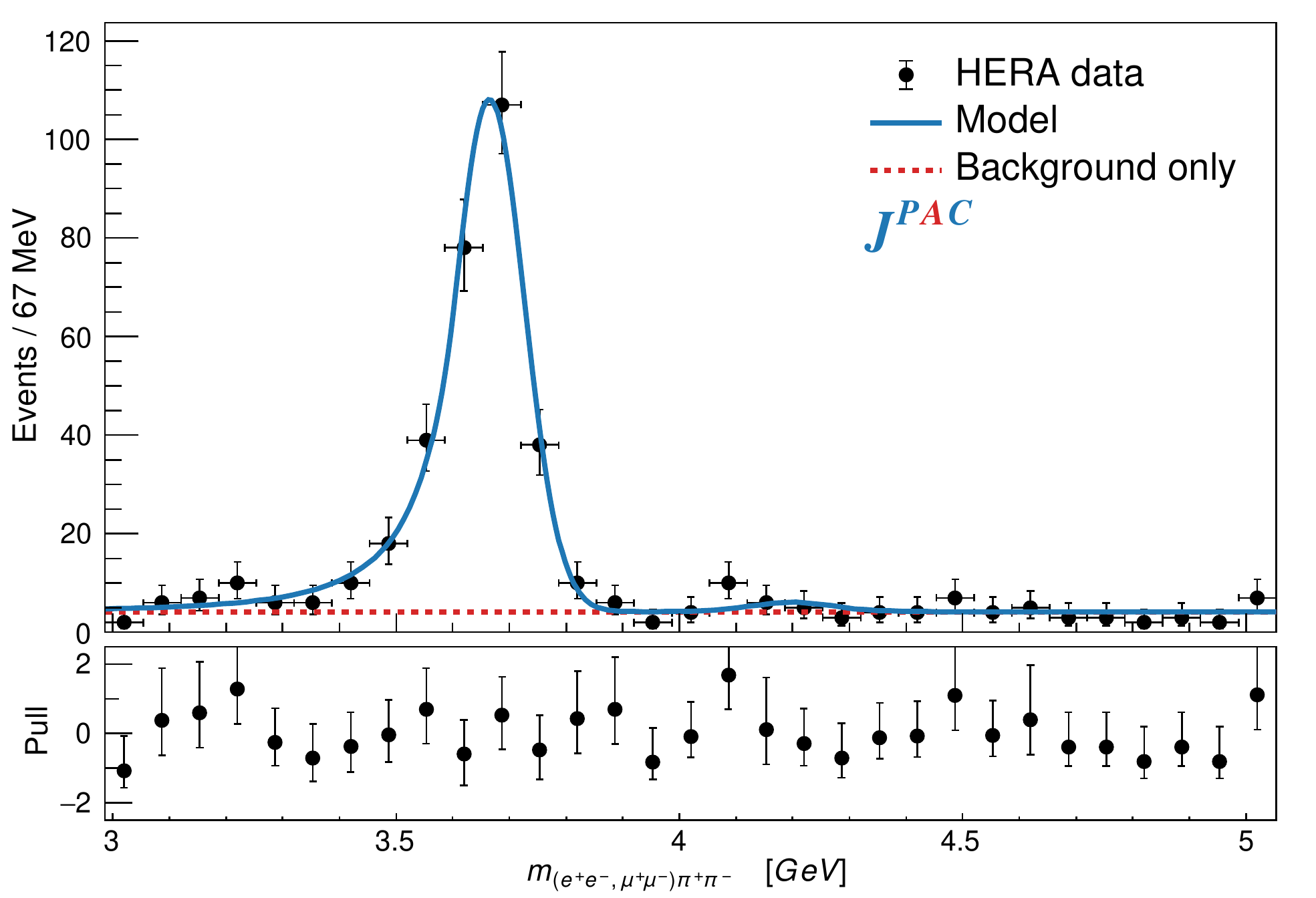}
\caption{Fit to the HERA data from~\cite{Adloff:2002re}. The blue curve shows the model discussed in the text. The dashed red line is the constant background. The $\psi(2S)$ peak is shifted to a value 21\mev lighter than the nominal mass; the same shift is applied to $m_Y$. A hint of \Y, compatible with zero would appear at $\sim 4.2\gev$. }\label{fig:hera}
\end{figure*}
Data show clearly a $\psi(2S)$ peak, together with some noise that might be interpreted as a hint of a \Y state. The $\psi(2S)$ is modeled as a Crystal Ball function,
\begin{align}
    &f_{\psi'}(m;m_{\psi'},\sigma,\alpha,n) \propto \nn\\
    &\left\{ \begin{matrix}\exp\left(-\frac{(m-m_{\psi'})^2}{2\sigma^2}\right) & \text{for }m > m_{\psi'}-\alpha \sigma\\
    e^{-\alpha^2/2} \left(1 - \frac{\alpha^2}{n} - \frac{(m-m_{\psi'})\alpha}{\sigma n}\right)^{-n} & \text{for }m < m_{\psi'}-\alpha \sigma
    \end{matrix}\right.,
\end{align}
while the \Y is modeled by the convolution of a Gaussian and a nonrelativistic Breit-Wigner,
\begin{align}
    f_{Y}(m;m_Y,\Gamma_Y,\sigma) &\propto \int_{-\infty}^\infty \exp\left(-\frac{(m'-m)^2}{2\sigma^2}\right) \nn\\
    &\qquad\times\frac{1}{(m'-m_Y)^2 + \Gamma_Y^2/4} dm'\,,
\end{align}
where the parameter $\sigma$ related to the experimental resolution is kept the same for the two curves. The two curves are added incoherently, together with a possible constant background.
We perform a maximum likelihood fit in the $m\in [3;5]\gev$ energy range, shown also in \cref{fig:hera}.
The fit returns a value of $m_{\psi'}$ that is 21\mev lighter than the nominal $\psi(2S)$ mass, so we fix $m_Y$ to the nominal value shifted by the same amount, $4220 - 21 \mev$. The width is fixed to the nominal value as well, $\Gamma_Y = 44\mev$. The fit shows no evidence for a \Y signal, the ratio between the two signals being compatible with zero. A Bayesian upper limit returns a ratio between the \Y and the $\psi(2S)$ signals of 
\begin{equation}
    R_\text{HERA} < 6.5\% \text{ at 95\% C.L.}
\end{equation}

\bibliographystyle{apsrev4-2.bst}
\bibliography{quattro}

\begin{thebibliography}{109}%
\makeatletter
\providecommand \@ifxundefined [1]{%
 \@ifx{#1\undefined}
}%
\providecommand \@ifnum [1]{%
 \ifnum #1\expandafter \@firstoftwo
 \else \expandafter \@secondoftwo
 \fi
}%
\providecommand \@ifx [1]{%
 \ifx #1\expandafter \@firstoftwo
 \else \expandafter \@secondoftwo
 \fi
}%
\providecommand \natexlab [1]{#1}%
\providecommand \enquote  [1]{``#1''}%
\providecommand \bibnamefont  [1]{#1}%
\providecommand \bibfnamefont [1]{#1}%
\providecommand \citenamefont [1]{#1}%
\providecommand \href@noop [0]{\@secondoftwo}%
\providecommand \href [0]{\begingroup \@sanitize@url \@href}%
\providecommand \@href[1]{\@@startlink{#1}\@@href}%
\providecommand \@@href[1]{\endgroup#1\@@endlink}%
\providecommand \@sanitize@url [0]{\catcode `\\12\catcode `\$12\catcode
  `\&12\catcode `\#12\catcode `\^12\catcode `\_12\catcode `\%12\relax}%
\providecommand \@@startlink[1]{}%
\providecommand \@@endlink[0]{}%
\providecommand \url  [0]{\begingroup\@sanitize@url \@url }%
\providecommand \@url [1]{\endgroup\@href {#1}{\urlprefix }}%
\providecommand \urlprefix  [0]{URL }%
\providecommand \Eprint [0]{\href }%
\providecommand \doibase [0]{https://doi.org/}%
\providecommand \selectlanguage [0]{\@gobble}%
\providecommand \bibinfo  [0]{\@secondoftwo}%
\providecommand \bibfield  [0]{\@secondoftwo}%
\providecommand \translation [1]{[#1]}%
\providecommand \BibitemOpen [0]{}%
\providecommand \bibitemStop [0]{}%
\providecommand \bibitemNoStop [0]{.\EOS\space}%
\providecommand \EOS [0]{\spacefactor3000\relax}%
\providecommand \BibitemShut  [1]{\csname bibitem#1\endcsname}%
\let\auto@bib@innerbib\@empty
\bibitem [{\citenamefont {Esposito}\ \emph {et~al.}(2017)\citenamefont
  {Esposito}, \citenamefont {Pilloni},\ and\ \citenamefont
  {Polosa}}]{Esposito:2016noz}%
  \BibitemOpen
  \bibfield  {author} {\bibinfo {author} {\bibfnamefont {A.}~\bibnamefont
  {Esposito}}, \bibinfo {author} {\bibfnamefont {A.}~\bibnamefont {Pilloni}},\
  and\ \bibinfo {author} {\bibfnamefont {A.~D.}\ \bibnamefont {Polosa}},\
  }\href {https://doi.org/10.1016/j.physrep.2016.11.002} {\bibfield  {journal}
  {\bibinfo  {journal} {Phys.Rept.}\ }\textbf {\bibinfo {volume} {668}},\
  \bibinfo {pages} {1} (\bibinfo {year} {2017})},\ \Eprint
  {https://arxiv.org/abs/1611.07920} {arXiv:1611.07920 [hep-ph]} \BibitemShut
  {NoStop}%
\bibitem [{\citenamefont {Guo}\ \emph {et~al.}(2018)\citenamefont {Guo},
  \citenamefont {Hanhart}, \citenamefont {Mei{\ss}ner}, \citenamefont {Wang},
  \citenamefont {Zhao},\ and\ \citenamefont {Zou}}]{Guo:2017jvc}%
  \BibitemOpen
  \bibfield  {author} {\bibinfo {author} {\bibfnamefont {F.-K.}\ \bibnamefont
  {Guo}}, \bibinfo {author} {\bibfnamefont {C.}~\bibnamefont {Hanhart}},
  \bibinfo {author} {\bibfnamefont {U.-G.}\ \bibnamefont {Mei{\ss}ner}},
  \bibinfo {author} {\bibfnamefont {Q.}~\bibnamefont {Wang}}, \bibinfo {author}
  {\bibfnamefont {Q.}~\bibnamefont {Zhao}},\ and\ \bibinfo {author}
  {\bibfnamefont {B.-S.}\ \bibnamefont {Zou}},\ }\href
  {https://doi.org/10.1103/RevModPhys.90.015004} {\bibfield  {journal}
  {\bibinfo  {journal} {Rev.Mod.Phys.}\ }\textbf {\bibinfo {volume} {90}},\
  \bibinfo {pages} {015004} (\bibinfo {year} {2018})},\ \Eprint
  {https://arxiv.org/abs/1705.00141} {arXiv:1705.00141 [hep-ph]} \BibitemShut
  {NoStop}%
\bibitem [{\citenamefont {Olsen}\ \emph {et~al.}(2018)\citenamefont {Olsen},
  \citenamefont {Skwarnicki},\ and\ \citenamefont {Zieminska}}]{Olsen:2017bmm}%
  \BibitemOpen
  \bibfield  {author} {\bibinfo {author} {\bibfnamefont {S.~L.}\ \bibnamefont
  {Olsen}}, \bibinfo {author} {\bibfnamefont {T.}~\bibnamefont {Skwarnicki}},\
  and\ \bibinfo {author} {\bibfnamefont {D.}~\bibnamefont {Zieminska}},\ }\href
  {https://doi.org/10.1103/RevModPhys.90.015003} {\bibfield  {journal}
  {\bibinfo  {journal} {Rev.Mod.Phys.}\ }\textbf {\bibinfo {volume} {90}},\
  \bibinfo {pages} {015003} (\bibinfo {year} {2018})},\ \Eprint
  {https://arxiv.org/abs/1708.04012} {arXiv:1708.04012 [hep-ph]} \BibitemShut
  {NoStop}%
\bibitem [{\citenamefont {Brambilla}\ \emph {et~al.}(2020)\citenamefont
  {Brambilla}, \citenamefont {Eidelman}, \citenamefont {Hanhart}, \citenamefont
  {Nefediev}, \citenamefont {Shen}, \citenamefont {Thomas}, \citenamefont
  {Vairo},\ and\ \citenamefont {Yuan}}]{Brambilla:2019esw}%
  \BibitemOpen
  \bibfield  {author} {\bibinfo {author} {\bibfnamefont {N.}~\bibnamefont
  {Brambilla}}, \bibinfo {author} {\bibfnamefont {S.}~\bibnamefont {Eidelman}},
  \bibinfo {author} {\bibfnamefont {C.}~\bibnamefont {Hanhart}}, \bibinfo
  {author} {\bibfnamefont {A.}~\bibnamefont {Nefediev}}, \bibinfo {author}
  {\bibfnamefont {C.-P.}\ \bibnamefont {Shen}}, \bibinfo {author}
  {\bibfnamefont {C.~E.}\ \bibnamefont {Thomas}}, \bibinfo {author}
  {\bibfnamefont {A.}~\bibnamefont {Vairo}},\ and\ \bibinfo {author}
  {\bibfnamefont {C.-Z.}\ \bibnamefont {Yuan}},\ }\href
  {https://doi.org/10.1016/j.physrep.2020.05.001} {\bibfield  {journal}
  {\bibinfo  {journal} {Phys.Rept.}\ }\textbf {\bibinfo {volume} {873}},\
  \bibinfo {pages} {1} (\bibinfo {year} {2020})},\ \Eprint
  {https://arxiv.org/abs/1907.07583} {arXiv:1907.07583 [hep-ex]} \BibitemShut
  {NoStop}%
\bibitem [{\citenamefont {Szczepaniak}(2015)}]{Szczepaniak:2015eza}%
  \BibitemOpen
  \bibfield  {author} {\bibinfo {author} {\bibfnamefont {A.~P.}\ \bibnamefont
  {Szczepaniak}},\ }\href {https://doi.org/10.1016/j.physletb.2015.06.029}
  {\bibfield  {journal} {\bibinfo  {journal} {Phys.Lett.}\ }\textbf {\bibinfo
  {volume} {B747}},\ \bibinfo {pages} {410} (\bibinfo {year} {2015})},\ \Eprint
  {https://arxiv.org/abs/1501.01691} {arXiv:1501.01691 [hep-ph]} \BibitemShut
  {NoStop}%
\bibitem [{\citenamefont {Albaladejo}\ \emph {et~al.}(2016)\citenamefont
  {Albaladejo}, \citenamefont {Guo}, \citenamefont {Hidalgo-Duque},\ and\
  \citenamefont {Nieves}}]{Albaladejo:2015lob}%
  \BibitemOpen
  \bibfield  {author} {\bibinfo {author} {\bibfnamefont {M.}~\bibnamefont
  {Albaladejo}}, \bibinfo {author} {\bibfnamefont {F.-K.}\ \bibnamefont {Guo}},
  \bibinfo {author} {\bibfnamefont {C.}~\bibnamefont {Hidalgo-Duque}},\ and\
  \bibinfo {author} {\bibfnamefont {J.}~\bibnamefont {Nieves}},\ }\href
  {https://doi.org/10.1016/j.physletb.2016.02.025} {\bibfield  {journal}
  {\bibinfo  {journal} {Phys.Lett.}\ }\textbf {\bibinfo {volume} {B755}},\
  \bibinfo {pages} {337} (\bibinfo {year} {2016})},\ \Eprint
  {https://arxiv.org/abs/1512.03638} {arXiv:1512.03638 [hep-ph]} \BibitemShut
  {NoStop}%
\bibitem [{\citenamefont {Guo}\ \emph {et~al.}(2016)\citenamefont {Guo},
  \citenamefont {Mei{\ss}ner}, \citenamefont {Nieves},\ and\ \citenamefont
  {Yang}}]{Guo:2016bkl}%
  \BibitemOpen
  \bibfield  {author} {\bibinfo {author} {\bibfnamefont {F.-K.}\ \bibnamefont
  {Guo}}, \bibinfo {author} {\bibfnamefont {U.-G.}\ \bibnamefont
  {Mei{\ss}ner}}, \bibinfo {author} {\bibfnamefont {J.}~\bibnamefont
  {Nieves}},\ and\ \bibinfo {author} {\bibfnamefont {Z.}~\bibnamefont {Yang}},\
  }\href {https://doi.org/10.1140/epja/i2016-16318-4} {\bibfield  {journal}
  {\bibinfo  {journal} {Eur.Phys.J.}\ }\textbf {\bibinfo {volume} {A52}},\
  \bibinfo {pages} {318} (\bibinfo {year} {2016})},\ \Eprint
  {https://arxiv.org/abs/1605.05113} {arXiv:1605.05113 [hep-ph]} \BibitemShut
  {NoStop}%
\bibitem [{\citenamefont {Pilloni}\ \emph {et~al.}(2017)\citenamefont
  {Pilloni}, \citenamefont {Fern\'andez-Ram\'irez}, \citenamefont {Jackura},
  \citenamefont {Mathieu}, \citenamefont {Mikhasenko}, \citenamefont {Nys},\
  and\ \citenamefont {Szczepaniak}}]{Pilloni:2016obd}%
  \BibitemOpen
  \bibfield  {author} {\bibinfo {author} {\bibfnamefont {A.}~\bibnamefont
  {Pilloni}}, \bibinfo {author} {\bibfnamefont {C.}~\bibnamefont
  {Fern\'andez-Ram\'irez}}, \bibinfo {author} {\bibfnamefont {A.}~\bibnamefont
  {Jackura}}, \bibinfo {author} {\bibfnamefont {V.}~\bibnamefont {Mathieu}},
  \bibinfo {author} {\bibfnamefont {M.}~\bibnamefont {Mikhasenko}}, \bibinfo
  {author} {\bibfnamefont {J.}~\bibnamefont {Nys}},\ and\ \bibinfo {author}
  {\bibfnamefont {A.~P.}\ \bibnamefont {Szczepaniak}} (\bibinfo {collaboration}
  {JPAC}),\ }\href {https://doi.org/10.1016/j.physletb.2017.06.030} {\bibfield
  {journal} {\bibinfo  {journal} {Phys.Lett.}\ }\textbf {\bibinfo {volume}
  {B772}},\ \bibinfo {pages} {200} (\bibinfo {year} {2017})},\ \Eprint
  {https://arxiv.org/abs/1612.06490} {arXiv:1612.06490 [hep-ph]} \BibitemShut
  {NoStop}%
\bibitem [{\citenamefont {Nakamura}\ and\ \citenamefont
  {Tsushima}(2019)}]{Nakamura:2019btl}%
  \BibitemOpen
  \bibfield  {author} {\bibinfo {author} {\bibfnamefont {S.}~\bibnamefont
  {Nakamura}}\ and\ \bibinfo {author} {\bibfnamefont {K.}~\bibnamefont
  {Tsushima}},\ }\href {https://doi.org/10.1103/PhysRevD.100.051502} {\bibfield
   {journal} {\bibinfo  {journal} {Phys.Rev.}\ }\textbf {\bibinfo {volume}
  {D100}},\ \bibinfo {pages} {051502} (\bibinfo {year} {2019})},\ \Eprint
  {https://arxiv.org/abs/1901.07385} {arXiv:1901.07385 [hep-ph]} \BibitemShut
  {NoStop}%
\bibitem [{\citenamefont {Guo}\ \emph {et~al.}(2020)\citenamefont {Guo},
  \citenamefont {Liu},\ and\ \citenamefont {Sakai}}]{Guo:2019twa}%
  \BibitemOpen
  \bibfield  {author} {\bibinfo {author} {\bibfnamefont {F.-K.}\ \bibnamefont
  {Guo}}, \bibinfo {author} {\bibfnamefont {X.-H.}\ \bibnamefont {Liu}},\ and\
  \bibinfo {author} {\bibfnamefont {S.}~\bibnamefont {Sakai}},\ }\href
  {https://doi.org/10.1016/j.ppnp.2020.103757} {\bibfield  {journal} {\bibinfo
  {journal} {Prog.Part.Nucl.Phys.}\ }\textbf {\bibinfo {volume} {112}},\
  \bibinfo {pages} {103757} (\bibinfo {year} {2020})},\ \Eprint
  {https://arxiv.org/abs/1912.07030} {arXiv:1912.07030 [hep-ph]} \BibitemShut
  {NoStop}%
\bibitem [{\citenamefont {Adolph}\ \emph {et~al.}(2015)\citenamefont {Adolph}
  \emph {et~al.}}]{Adolph:2014hba}%
  \BibitemOpen
  \bibfield  {author} {\bibinfo {author} {\bibfnamefont {C.}~\bibnamefont
  {Adolph}} \emph {et~al.} (\bibinfo {collaboration} {COMPASS}),\ }\href
  {https://doi.org/10.1016/j.physletb.2015.01.042} {\bibfield  {journal}
  {\bibinfo  {journal} {Phys.Lett.}\ }\textbf {\bibinfo {volume} {B742}},\
  \bibinfo {pages} {330} (\bibinfo {year} {2015})},\ \Eprint
  {https://arxiv.org/abs/1407.6186} {arXiv:1407.6186 [hep-ex]} \BibitemShut
  {NoStop}%
\bibitem [{\citenamefont {Aghasyan}\ \emph {et~al.}(2018)\citenamefont
  {Aghasyan} \emph {et~al.}}]{Aghasyan:2017utv}%
  \BibitemOpen
  \bibfield  {author} {\bibinfo {author} {\bibfnamefont {M.}~\bibnamefont
  {Aghasyan}} \emph {et~al.} (\bibinfo {collaboration} {COMPASS}),\ }\href
  {https://doi.org/10.1016/j.physletb.2018.07.008} {\bibfield  {journal}
  {\bibinfo  {journal} {Phys.Lett.}\ }\textbf {\bibinfo {volume} {B783}},\
  \bibinfo {pages} {334} (\bibinfo {year} {2018})},\ \Eprint
  {https://arxiv.org/abs/1707.01796} {arXiv:1707.01796} \BibitemShut {NoStop}%
\bibitem [{\citenamefont {Meziani}\ \emph {et~al.}(2016)\citenamefont {Meziani}
  \emph {et~al.}}]{Meziani:2016lhg}%
  \BibitemOpen
  \bibfield  {author} {\bibinfo {author} {\bibfnamefont {Z.~E.}\ \bibnamefont
  {Meziani}} \emph {et~al.},\ }\Eprint {https://arxiv.org/abs/1609.00676}
  {arXiv:1609.00676 [hep-ex]}  (\bibinfo {year} {2016})\BibitemShut {NoStop}%
\bibitem [{\citenamefont {Ali}\ \emph {et~al.}(2019)\citenamefont {Ali} \emph
  {et~al.}}]{Ali:2019lzf}%
  \BibitemOpen
  \bibfield  {author} {\bibinfo {author} {\bibfnamefont {A.}~\bibnamefont
  {Ali}} \emph {et~al.} (\bibinfo {collaboration} {GlueX}),\ }\href
  {https://doi.org/10.1103/PhysRevLett.123.072001} {\bibfield  {journal}
  {\bibinfo  {journal} {Phys.Rev.Lett.}\ }\textbf {\bibinfo {volume} {123}},\
  \bibinfo {pages} {072001} (\bibinfo {year} {2019})},\ \Eprint
  {https://arxiv.org/abs/1905.10811} {arXiv:1905.10811 [nucl-ex]} \BibitemShut
  {NoStop}%
\bibitem [{Bro(2020)}]{Brodsky:2020vco}%
  \BibitemOpen
  \href {https://doi.org/10.1142/S0218301320300064} {\emph {\bibinfo {title}
  {{Strong QCD from Hadron Structure Experiments}}}}\ (\bibinfo {year} {2020})\
  \Eprint {https://arxiv.org/abs/2006.06802} {arXiv:2006.06802 [hep-ph]}
  \BibitemShut {NoStop}%
\bibitem [{\citenamefont {Accardi}\ \emph {et~al.}(2016)\citenamefont {Accardi}
  \emph {et~al.}}]{Accardi:2012qut}%
  \BibitemOpen
  \bibfield  {author} {\bibinfo {author} {\bibfnamefont {A.}~\bibnamefont
  {Accardi}} \emph {et~al.},\ }\href
  {https://doi.org/10.1140/epja/i2016-16268-9} {\bibfield  {journal} {\bibinfo
  {journal} {Eur.Phys.J.}\ }\textbf {\bibinfo {volume} {A52}},\ \bibinfo
  {pages} {268} (\bibinfo {year} {2016})},\ \Eprint
  {https://arxiv.org/abs/1212.1701} {arXiv:1212.1701 [nucl-ex]} \BibitemShut
  {NoStop}%
\bibitem [{eic()}]{eic}%
  \BibitemOpen
  \href@noop {} {\bibinfo {title} {The electron-ion collider}},\ \bibinfo
  {note} {{\href{https://www.bnl.gov/eic/}{\tt https://www.bnl.gov/eic/}
  }}\BibitemShut {NoStop}%
\bibitem [{\citenamefont {Chen}(2018)}]{Chen:2018wyz}%
  \BibitemOpen
  \bibfield  {author} {\bibinfo {author} {\bibfnamefont {X.}~\bibnamefont
  {Chen}},\ }\href {https://doi.org/10.22323/1.316.0170} {\bibfield  {journal}
  {\bibinfo  {journal} {PoS}\ }\textbf {\bibinfo {volume} {DIS2018}},\ \bibinfo
  {pages} {170} (\bibinfo {year} {2018})},\ \Eprint
  {https://arxiv.org/abs/1809.00448} {arXiv:1809.00448 [nucl-ex]} \BibitemShut
  {NoStop}%
\bibitem [{\citenamefont {{JPAC Collaboration}}(2020)}]{jpacinpreparation}%
  \BibitemOpen
  \bibfield  {author} {\bibinfo {author} {\bibnamefont {{JPAC Collaboration}}}}
  (\bibinfo {year} {2020}),\ \bibinfo {note} {in preparation}\BibitemShut
  {NoStop}%
\bibitem [{\citenamefont {Martin}\ and\ \citenamefont
  {Spearman}(1970)}]{Martin:1970}%
  \BibitemOpen
  \bibfield  {author} {\bibinfo {author} {\bibfnamefont {A.}~\bibnamefont
  {Martin}}\ and\ \bibinfo {author} {\bibfnamefont {T.}~\bibnamefont
  {Spearman}},\ }\href {https://books.google.com/books?id=sxAzAAAAMAAJ} {\emph
  {\bibinfo {title} {Elementary particle theory}}}\ (\bibinfo  {publisher}
  {North-Holland Pub. Co.},\ \bibinfo {year} {1970})\BibitemShut {NoStop}%
\bibitem [{\citenamefont {Collins}(2009)}]{Collins:1977jy}%
  \BibitemOpen
  \bibfield  {author} {\bibinfo {author} {\bibfnamefont {P.~D.~B.}\
  \bibnamefont {Collins}},\ }\href
  {http://www-spires.fnal.gov/spires/find/books/www?cl=QC793.3.R4C695} {\emph
  {\bibinfo {title} {{An Introduction to Regge Theory and High-Energy
  Physics}}}},\ Cambridge Monographs on Mathematical Physics\ (\bibinfo
  {publisher} {Cambridge Univ. Press},\ \bibinfo {address} {Cambridge, UK},\
  \bibinfo {year} {2009})\BibitemShut {NoStop}%
\bibitem [{\citenamefont {Nys}\ \emph {et~al.}(2018)\citenamefont {Nys},
  \citenamefont {Hiller~Blin}, \citenamefont {Mathieu}, \citenamefont
  {Fern\'andez-Ram\'irez}, \citenamefont {Jackura}, \citenamefont {Pilloni},
  \citenamefont {Ryckebusch}, \citenamefont {Szczepaniak},\ and\ \citenamefont
  {Fox}}]{Nys:2018vck}%
  \BibitemOpen
  \bibfield  {author} {\bibinfo {author} {\bibfnamefont {J.}~\bibnamefont
  {Nys}}, \bibinfo {author} {\bibfnamefont {A.}~\bibnamefont {Hiller~Blin}},
  \bibinfo {author} {\bibfnamefont {V.}~\bibnamefont {Mathieu}}, \bibinfo
  {author} {\bibfnamefont {C.}~\bibnamefont {Fern\'andez-Ram\'irez}}, \bibinfo
  {author} {\bibfnamefont {A.}~\bibnamefont {Jackura}}, \bibinfo {author}
  {\bibfnamefont {A.}~\bibnamefont {Pilloni}}, \bibinfo {author} {\bibfnamefont
  {J.}~\bibnamefont {Ryckebusch}}, \bibinfo {author} {\bibfnamefont
  {A.}~\bibnamefont {Szczepaniak}},\ and\ \bibinfo {author} {\bibfnamefont
  {G.}~\bibnamefont {Fox}} (\bibinfo {collaboration} {JPAC}),\ }\href
  {https://doi.org/10.1103/PhysRevD.98.034020} {\bibfield  {journal} {\bibinfo
  {journal} {Phys.Rev.}\ }\textbf {\bibinfo {volume} {D98}},\ \bibinfo {pages}
  {034020} (\bibinfo {year} {2018})},\ \Eprint
  {https://arxiv.org/abs/1806.01891} {arXiv:1806.01891 [hep-ph]} \BibitemShut
  {NoStop}%
\bibitem [{\citenamefont {Zyla}\ \emph {et~al.}(2020)\citenamefont {Zyla} \emph
  {et~al.}}]{pdg}%
  \BibitemOpen
  \bibfield  {author} {\bibinfo {author} {\bibfnamefont {P.}~\bibnamefont
  {Zyla}} \emph {et~al.} (\bibinfo {collaboration} {Particle Data Group}),\
  }\href {https://doi.org/10.1093/ptep/ptaa104} {\bibfield  {journal} {\bibinfo
   {journal} {PTEP}\ }\textbf {\bibinfo {volume} {2020}},\ \bibinfo {pages}
  {083C01} (\bibinfo {year} {2020})}\BibitemShut {NoStop}%
\bibitem [{\citenamefont {Bondar}\ \emph {et~al.}(2012)\citenamefont {Bondar}
  \emph {et~al.}}]{Belle:2011aa}%
  \BibitemOpen
  \bibfield  {author} {\bibinfo {author} {\bibfnamefont {A.}~\bibnamefont
  {Bondar}} \emph {et~al.} (\bibinfo {collaboration} {\belle}),\ }\href
  {https://doi.org/10.1103/PhysRevLett.108.122001} {\bibfield  {journal}
  {\bibinfo  {journal} {Phys.Rev.Lett.}\ }\textbf {\bibinfo {volume} {108}},\
  \bibinfo {pages} {122001} (\bibinfo {year} {2012})},\ \Eprint
  {https://arxiv.org/abs/1110.2251} {arXiv:1110.2251 [hep-ex]} \BibitemShut
  {NoStop}%
\bibitem [{\citenamefont {Ablikim}\ \emph
  {et~al.}(2017{\natexlab{a}})\citenamefont {Ablikim} \emph
  {et~al.}}]{Collaboration:2017njt}%
  \BibitemOpen
  \bibfield  {author} {\bibinfo {author} {\bibfnamefont {M.}~\bibnamefont
  {Ablikim}} \emph {et~al.} (\bibinfo {collaboration} {BESIII}),\ }\href
  {https://doi.org/10.1103/PhysRevLett.119.072001} {\bibfield  {journal}
  {\bibinfo  {journal} {Phys.Rev.Lett.}\ }\textbf {\bibinfo {volume} {119}},\
  \bibinfo {pages} {072001} (\bibinfo {year} {2017}{\natexlab{a}})},\ \Eprint
  {https://arxiv.org/abs/1706.04100} {arXiv:1706.04100 [hep-ex]} \BibitemShut
  {NoStop}%
\bibitem [{\citenamefont {Wang}\ \emph
  {et~al.}(2015{\natexlab{a}})\citenamefont {Wang}, \citenamefont {Chen},\ and\
  \citenamefont {Guskov}}]{Wang:2015lwa}%
  \BibitemOpen
  \bibfield  {author} {\bibinfo {author} {\bibfnamefont {X.-Y.}\ \bibnamefont
  {Wang}}, \bibinfo {author} {\bibfnamefont {X.-R.}\ \bibnamefont {Chen}},\
  and\ \bibinfo {author} {\bibfnamefont {A.}~\bibnamefont {Guskov}},\ }\href
  {https://doi.org/10.1103/PhysRevD.92.094017} {\bibfield  {journal} {\bibinfo
  {journal} {Phys.Rev.}\ }\textbf {\bibinfo {volume} {D92}},\ \bibinfo {pages}
  {094017} (\bibinfo {year} {2015}{\natexlab{a}})},\ \Eprint
  {https://arxiv.org/abs/1503.02125} {arXiv:1503.02125 [hep-ph]} \BibitemShut
  {NoStop}%
\bibitem [{\citenamefont {Galat\`a}(2011)}]{Galata:2011bi}%
  \BibitemOpen
  \bibfield  {author} {\bibinfo {author} {\bibfnamefont {G.}~\bibnamefont
  {Galat\`a}},\ }\href {https://doi.org/10.1103/PhysRevC.83.065203} {\bibfield
  {journal} {\bibinfo  {journal} {Phys.Rev.}\ }\textbf {\bibinfo {volume}
  {C83}},\ \bibinfo {pages} {065203} (\bibinfo {year} {2011})},\ \Eprint
  {https://arxiv.org/abs/1102.2070} {arXiv:1102.2070 [hep-ph]} \BibitemShut
  {NoStop}%
\bibitem [{\citenamefont {Lin}\ \emph {et~al.}(2013)\citenamefont {Lin},
  \citenamefont {Liu},\ and\ \citenamefont {Xu}}]{Lin:2013mka}%
  \BibitemOpen
  \bibfield  {author} {\bibinfo {author} {\bibfnamefont {Q.-Y.}\ \bibnamefont
  {Lin}}, \bibinfo {author} {\bibfnamefont {X.}~\bibnamefont {Liu}},\ and\
  \bibinfo {author} {\bibfnamefont {H.-S.}\ \bibnamefont {Xu}},\ }\href
  {https://doi.org/10.1103/PhysRevD.88.114009} {\bibfield  {journal} {\bibinfo
  {journal} {Phys.Rev.}\ }\textbf {\bibinfo {volume} {D88}},\ \bibinfo {pages}
  {114009} (\bibinfo {year} {2013})},\ \Eprint
  {https://arxiv.org/abs/1308.6345} {arXiv:1308.6345 [hep-ph]} \BibitemShut
  {NoStop}%
\bibitem [{\citenamefont {Wang}\ \emph {et~al.}(2020)\citenamefont {Wang},
  \citenamefont {Kou}, \citenamefont {Lin}, \citenamefont {Xie}, \citenamefont
  {Chen},\ and\ \citenamefont {Guskov}}]{Wang:2020stx}%
  \BibitemOpen
  \bibfield  {author} {\bibinfo {author} {\bibfnamefont {X.-Y.}\ \bibnamefont
  {Wang}}, \bibinfo {author} {\bibfnamefont {W.}~\bibnamefont {Kou}}, \bibinfo
  {author} {\bibfnamefont {Q.-Y.}\ \bibnamefont {Lin}}, \bibinfo {author}
  {\bibfnamefont {Y.-P.}\ \bibnamefont {Xie}}, \bibinfo {author} {\bibfnamefont
  {X.}~\bibnamefont {Chen}},\ and\ \bibinfo {author} {\bibfnamefont
  {A.}~\bibnamefont {Guskov}},\ }\Eprint {https://arxiv.org/abs/2009.05789}
  {arXiv:2009.05789 [hep-ph]}  (\bibinfo {year} {2020})\BibitemShut {NoStop}%
\bibitem [{\citenamefont {Chiang}\ \emph {et~al.}(2002)\citenamefont {Chiang},
  \citenamefont {Yang}, \citenamefont {Tiator},\ and\ \citenamefont
  {Drechsel}}]{Chiang:2001as}%
  \BibitemOpen
  \bibfield  {author} {\bibinfo {author} {\bibfnamefont {W.-T.}\ \bibnamefont
  {Chiang}}, \bibinfo {author} {\bibfnamefont {S.-N.}\ \bibnamefont {Yang}},
  \bibinfo {author} {\bibfnamefont {L.}~\bibnamefont {Tiator}},\ and\ \bibinfo
  {author} {\bibfnamefont {D.}~\bibnamefont {Drechsel}},\ }\href
  {https://doi.org/10.1016/S0375-9474(01)01325-2} {\bibfield  {journal}
  {\bibinfo  {journal} {Nucl.Phys.}\ }\textbf {\bibinfo {volume} {A700}},\
  \bibinfo {pages} {429} (\bibinfo {year} {2002})},\ \Eprint
  {https://arxiv.org/abs/nucl-th/0110034} {arXiv:nucl-th/0110034 [nucl-th]}
  \BibitemShut {NoStop}%
\bibitem [{\citenamefont {Hiller~Blin}(2017)}]{Blin:2017hez}%
  \BibitemOpen
  \bibfield  {author} {\bibinfo {author} {\bibfnamefont {A.}~\bibnamefont
  {Hiller~Blin}},\ }\href {https://doi.org/10.1103/PhysRevD.96.093008}
  {\bibfield  {journal} {\bibinfo  {journal} {Phys.Rev.}\ }\textbf {\bibinfo
  {volume} {D96}},\ \bibinfo {pages} {093008} (\bibinfo {year} {2017})},\
  \Eprint {https://arxiv.org/abs/1707.02255} {arXiv:1707.02255 [hep-ph]}
  \BibitemShut {NoStop}%
\bibitem [{\citenamefont {Matsinos}(2019)}]{Matsinos:2019kqi}%
  \BibitemOpen
  \bibfield  {author} {\bibinfo {author} {\bibfnamefont {E.}~\bibnamefont
  {Matsinos}},\ }\Eprint {https://arxiv.org/abs/1901.01204} {arXiv:1901.01204
  [nucl-th]}  (\bibinfo {year} {2019})\BibitemShut {NoStop}%
\bibitem [{\citenamefont {Gasparyan}\ \emph {et~al.}(2003)\citenamefont
  {Gasparyan}, \citenamefont {Haidenbauer}, \citenamefont {Hanhart},\ and\
  \citenamefont {Speth}}]{Gasparyan:2003fp}%
  \BibitemOpen
  \bibfield  {author} {\bibinfo {author} {\bibfnamefont {A.}~\bibnamefont
  {Gasparyan}}, \bibinfo {author} {\bibfnamefont {J.}~\bibnamefont
  {Haidenbauer}}, \bibinfo {author} {\bibfnamefont {C.}~\bibnamefont
  {Hanhart}},\ and\ \bibinfo {author} {\bibfnamefont {J.}~\bibnamefont
  {Speth}},\ }\href {https://doi.org/10.1103/PhysRevC.68.045207} {\bibfield
  {journal} {\bibinfo  {journal} {Phys.Rev.}\ }\textbf {\bibinfo {volume}
  {C68}},\ \bibinfo {pages} {045207} (\bibinfo {year} {2003})},\ \Eprint
  {https://arxiv.org/abs/nucl-th/0307072} {arXiv:nucl-th/0307072} \BibitemShut
  {NoStop}%
\bibitem [{\citenamefont {Irving}\ and\ \citenamefont
  {Worden}(1977)}]{Irving:1977ea}%
  \BibitemOpen
  \bibfield  {author} {\bibinfo {author} {\bibfnamefont {A.~C.}\ \bibnamefont
  {Irving}}\ and\ \bibinfo {author} {\bibfnamefont {R.~P.}\ \bibnamefont
  {Worden}},\ }\href {https://doi.org/10.1016/0370-1573(77)90010-2} {\bibfield
  {journal} {\bibinfo  {journal} {Phys.Rept.}\ }\textbf {\bibinfo {volume}
  {34}},\ \bibinfo {pages} {117} (\bibinfo {year} {1977})}\BibitemShut
  {NoStop}%
\bibitem [{\citenamefont {Aaij}\ \emph
  {et~al.}(2020{\natexlab{a}})\citenamefont {Aaij} \emph
  {et~al.}}]{Aaij:2020qga}%
  \BibitemOpen
  \bibfield  {author} {\bibinfo {author} {\bibfnamefont {R.}~\bibnamefont
  {Aaij}} \emph {et~al.} (\bibinfo {collaboration} {LHCb}),\ }\Eprint
  {https://arxiv.org/abs/2005.13419} {arXiv:2005.13419 [hep-ex]}  (\bibinfo
  {year} {2020}{\natexlab{a}})\BibitemShut {NoStop}%
\bibitem [{\citenamefont {Aaij}\ \emph
  {et~al.}(2020{\natexlab{b}})\citenamefont {Aaij} \emph
  {et~al.}}]{Aaij:2020xjx}%
  \BibitemOpen
  \bibfield  {author} {\bibinfo {author} {\bibfnamefont {R.}~\bibnamefont
  {Aaij}} \emph {et~al.} (\bibinfo {collaboration} {LHCb}),\ }\href
  {https://doi.org/10.1007/JHEP08(2020)123} {\bibfield  {journal} {\bibinfo
  {journal} {JHEP}\ }\textbf {\bibinfo {volume} {08}},\ \bibinfo {pages}
  {123}},\ \Eprint {https://arxiv.org/abs/2005.13422} {arXiv:2005.13422
  [hep-ex]} \BibitemShut {NoStop}%
\bibitem [{\citenamefont {Aaij}\ \emph {et~al.}(2013)\citenamefont {Aaij} \emph
  {et~al.}}]{Aaij:2013zoa}%
  \BibitemOpen
  \bibfield  {author} {\bibinfo {author} {\bibfnamefont {R.}~\bibnamefont
  {Aaij}} \emph {et~al.} (\bibinfo {collaboration} {LHCb}),\ }\href
  {https://doi.org/10.1103/PhysRevLett.110.222001} {\bibfield  {journal}
  {\bibinfo  {journal} {Phys.Rev.Lett.}\ }\textbf {\bibinfo {volume} {110}},\
  \bibinfo {pages} {222001} (\bibinfo {year} {2013})},\ \Eprint
  {https://arxiv.org/abs/1302.6269} {arXiv:1302.6269 [hep-ex]} \BibitemShut
  {NoStop}%
\bibitem [{\citenamefont {Aaij}\ \emph {et~al.}(2015)\citenamefont {Aaij} \emph
  {et~al.}}]{Aaij:2015eva}%
  \BibitemOpen
  \bibfield  {author} {\bibinfo {author} {\bibfnamefont {R.}~\bibnamefont
  {Aaij}} \emph {et~al.} (\bibinfo {collaboration} {LHCb}),\ }\href
  {https://doi.org/10.1103/PhysRevD.92.011102} {\bibfield  {journal} {\bibinfo
  {journal} {Phys.Rev.}\ }\textbf {\bibinfo {volume} {D92}},\ \bibinfo {pages}
  {011102} (\bibinfo {year} {2015})},\ \Eprint
  {https://arxiv.org/abs/1504.06339} {arXiv:1504.06339 [hep-ex]} \BibitemShut
  {NoStop}%
\bibitem [{\citenamefont {Lees}\ \emph {et~al.}(2020)\citenamefont {Lees} \emph
  {et~al.}}]{Lees:2019xea}%
  \BibitemOpen
  \bibfield  {author} {\bibinfo {author} {\bibfnamefont {J.}~\bibnamefont
  {Lees}} \emph {et~al.} (\bibinfo {collaboration} {\babar}),\ }\href
  {https://doi.org/10.1103/PhysRevLett.124.152001} {\bibfield  {journal}
  {\bibinfo  {journal} {Phys.Rev.Lett.}\ }\textbf {\bibinfo {volume} {124}},\
  \bibinfo {pages} {152001} (\bibinfo {year} {2020})},\ \Eprint
  {https://arxiv.org/abs/1911.11740} {arXiv:1911.11740 [hep-ex]} \BibitemShut
  {NoStop}%
\bibitem [{\citenamefont {Li}\ and\ \citenamefont {Yuan}(2019)}]{Li:2019kpj}%
  \BibitemOpen
  \bibfield  {author} {\bibinfo {author} {\bibfnamefont {C.}~\bibnamefont
  {Li}}\ and\ \bibinfo {author} {\bibfnamefont {C.-Z.}\ \bibnamefont {Yuan}},\
  }\href {https://doi.org/10.1103/PhysRevD.100.094003} {\bibfield  {journal}
  {\bibinfo  {journal} {Phys.Rev.}\ }\textbf {\bibinfo {volume} {D100}},\
  \bibinfo {pages} {094003} (\bibinfo {year} {2019})},\ \Eprint
  {https://arxiv.org/abs/1907.09149} {arXiv:1907.09149 [hep-ex]} \BibitemShut
  {NoStop}%
\bibitem [{\citenamefont {Hanhart}\ \emph {et~al.}(2012)\citenamefont
  {Hanhart}, \citenamefont {Kalashnikova}, \citenamefont {Kudryavtsev},\ and\
  \citenamefont {Nefediev}}]{Hanhart:2011tn}%
  \BibitemOpen
  \bibfield  {author} {\bibinfo {author} {\bibfnamefont {C.}~\bibnamefont
  {Hanhart}}, \bibinfo {author} {\bibfnamefont {Y.~S.}\ \bibnamefont
  {Kalashnikova}}, \bibinfo {author} {\bibfnamefont {A.~E.}\ \bibnamefont
  {Kudryavtsev}},\ and\ \bibinfo {author} {\bibfnamefont {A.~V.}\ \bibnamefont
  {Nefediev}},\ }\href {https://doi.org/10.1103/PhysRevD.85.011501} {\bibfield
  {journal} {\bibinfo  {journal} {Phys.Rev.}\ }\textbf {\bibinfo {volume}
  {D85}},\ \bibinfo {pages} {011501} (\bibinfo {year} {2012})},\ \Eprint
  {https://arxiv.org/abs/1111.6241} {arXiv:1111.6241 [hep-ph]} \BibitemShut
  {NoStop}%
\bibitem [{\citenamefont {Brazzi}\ \emph {et~al.}(2011)\citenamefont {Brazzi},
  \citenamefont {Grinstein}, \citenamefont {Piccinini}, \citenamefont
  {Polosa},\ and\ \citenamefont {Sabelli}}]{Brazzi:2011fq}%
  \BibitemOpen
  \bibfield  {author} {\bibinfo {author} {\bibfnamefont {F.}~\bibnamefont
  {Brazzi}}, \bibinfo {author} {\bibfnamefont {B.}~\bibnamefont {Grinstein}},
  \bibinfo {author} {\bibfnamefont {F.}~\bibnamefont {Piccinini}}, \bibinfo
  {author} {\bibfnamefont {A.~D.}\ \bibnamefont {Polosa}},\ and\ \bibinfo
  {author} {\bibfnamefont {C.}~\bibnamefont {Sabelli}},\ }\href
  {https://doi.org/10.1103/PhysRevD.84.014003} {\bibfield  {journal} {\bibinfo
  {journal} {Phys.Rev.}\ }\textbf {\bibinfo {volume} {D84}},\ \bibinfo {pages}
  {014003} (\bibinfo {year} {2011})},\ \Eprint
  {https://arxiv.org/abs/1103.3155} {arXiv:1103.3155 [hep-ph]} \BibitemShut
  {NoStop}%
\bibitem [{\citenamefont {Faccini}\ \emph {et~al.}(2012)\citenamefont
  {Faccini}, \citenamefont {Piccinini}, \citenamefont {Pilloni},\ and\
  \citenamefont {Polosa}}]{Faccini:2012zv}%
  \BibitemOpen
  \bibfield  {author} {\bibinfo {author} {\bibfnamefont {R.}~\bibnamefont
  {Faccini}}, \bibinfo {author} {\bibfnamefont {F.}~\bibnamefont {Piccinini}},
  \bibinfo {author} {\bibfnamefont {A.}~\bibnamefont {Pilloni}},\ and\ \bibinfo
  {author} {\bibfnamefont {A.}~\bibnamefont {Polosa}},\ }\href
  {https://doi.org/10.1103/PhysRevD.86.054012} {\bibfield  {journal} {\bibinfo
  {journal} {Phys.Rev.}\ }\textbf {\bibinfo {volume} {D86}},\ \bibinfo {pages}
  {054012} (\bibinfo {year} {2012})},\ \Eprint
  {https://arxiv.org/abs/1204.1223} {arXiv:1204.1223 [hep-ph]} \BibitemShut
  {NoStop}%
\bibitem [{\citenamefont {Landau}(1948)}]{Landau:1948kw}%
  \BibitemOpen
  \bibfield  {author} {\bibinfo {author} {\bibfnamefont {L.}~\bibnamefont
  {Landau}},\ }\href {https://doi.org/10.1016/B978-0-08-010586-4.50070-5}
  {\bibfield  {journal} {\bibinfo  {journal} {Dokl.Akad.Nauk SSSR}\ }\textbf
  {\bibinfo {volume} {60}},\ \bibinfo {pages} {207} (\bibinfo {year}
  {1948})}\BibitemShut {NoStop}%
\bibitem [{\citenamefont {Yang}(1950)}]{Yang:1950rg}%
  \BibitemOpen
  \bibfield  {author} {\bibinfo {author} {\bibfnamefont {C.-N.}\ \bibnamefont
  {Yang}},\ }\href {https://doi.org/10.1103/PhysRev.77.242} {\bibfield
  {journal} {\bibinfo  {journal} {Phys.Rev.}\ }\textbf {\bibinfo {volume}
  {77}},\ \bibinfo {pages} {242} (\bibinfo {year} {1950})}\BibitemShut
  {NoStop}%
\bibitem [{\citenamefont {Teramoto}\ \emph {et~al.}(2020)\citenamefont
  {Teramoto} \emph {et~al.}}]{Teramoto:2020ezr}%
  \BibitemOpen
  \bibfield  {author} {\bibinfo {author} {\bibfnamefont {Y.}~\bibnamefont
  {Teramoto}} \emph {et~al.} (\bibinfo {collaboration} {Belle}),\ }\Eprint
  {https://arxiv.org/abs/2007.05696} {arXiv:2007.05696 [hep-ex]}  (\bibinfo
  {year} {2020})\BibitemShut {NoStop}%
\bibitem [{\citenamefont {Donnelly}\ \emph {et~al.}(2017)\citenamefont
  {Donnelly}, \citenamefont {Formaggio}, \citenamefont {Holstein},
  \citenamefont {Milner},\ and\ \citenamefont {Surrow}}]{Donnelly:2017aaa}%
  \BibitemOpen
  \bibfield  {author} {\bibinfo {author} {\bibfnamefont {T.~W.}\ \bibnamefont
  {Donnelly}}, \bibinfo {author} {\bibfnamefont {J.~A.}\ \bibnamefont
  {Formaggio}}, \bibinfo {author} {\bibfnamefont {B.~R.}\ \bibnamefont
  {Holstein}}, \bibinfo {author} {\bibfnamefont {R.~G.}\ \bibnamefont
  {Milner}},\ and\ \bibinfo {author} {\bibfnamefont {B.}~\bibnamefont
  {Surrow}},\ }\href@noop {} {\emph {\bibinfo {title} {{Foundations of Nuclear
  and Particle Physics}}}}\ (\bibinfo  {publisher} {Cambridge University
  Press},\ \bibinfo {year} {2017})\BibitemShut {NoStop}%
\bibitem [{\citenamefont {Aloni}\ \emph {et~al.}(2019)\citenamefont {Aloni},
  \citenamefont {Fanelli}, \citenamefont {Soreq},\ and\ \citenamefont
  {Williams}}]{Aloni:2019ruo}%
  \BibitemOpen
  \bibfield  {author} {\bibinfo {author} {\bibfnamefont {D.}~\bibnamefont
  {Aloni}}, \bibinfo {author} {\bibfnamefont {C.}~\bibnamefont {Fanelli}},
  \bibinfo {author} {\bibfnamefont {Y.}~\bibnamefont {Soreq}},\ and\ \bibinfo
  {author} {\bibfnamefont {M.}~\bibnamefont {Williams}},\ }\href
  {https://doi.org/10.1103/PhysRevLett.123.071801} {\bibfield  {journal}
  {\bibinfo  {journal} {Phys.Rev.Lett.}\ }\textbf {\bibinfo {volume} {123}},\
  \bibinfo {pages} {071801} (\bibinfo {year} {2019})},\ \Eprint
  {https://arxiv.org/abs/1903.03586} {arXiv:1903.03586 [hep-ph]} \BibitemShut
  {NoStop}%
\bibitem [{\citenamefont {De~Jager}\ \emph {et~al.}(1974)\citenamefont
  {De~Jager}, \citenamefont {De~Vries},\ and\ \citenamefont
  {De~Vries}}]{DeJager:1974liz}%
  \BibitemOpen
  \bibfield  {author} {\bibinfo {author} {\bibfnamefont {C.}~\bibnamefont
  {De~Jager}}, \bibinfo {author} {\bibfnamefont {H.}~\bibnamefont {De~Vries}},\
  and\ \bibinfo {author} {\bibfnamefont {C.}~\bibnamefont {De~Vries}},\ }\href
  {https://doi.org/10.1016/S0092-640X(74)80002-1} {\bibfield  {journal}
  {\bibinfo  {journal} {Atom.Data Nucl.Data Tabl.}\ }\textbf {\bibinfo {volume}
  {14}},\ \bibinfo {pages} {479} (\bibinfo {year} {1974})},\ \bibinfo {note}
  {[Erratum: Atom.Data Nucl.Data Tabl. 16, 580-580 (1975)]}\BibitemShut
  {NoStop}%
\bibitem [{\citenamefont {Aaij}\ \emph
  {et~al.}(2020{\natexlab{c}})\citenamefont {Aaij} \emph
  {et~al.}}]{Aaij:2020fnh}%
  \BibitemOpen
  \bibfield  {author} {\bibinfo {author} {\bibfnamefont {R.}~\bibnamefont
  {Aaij}} \emph {et~al.} (\bibinfo {collaboration} {LHCb}),\ }\Eprint
  {https://arxiv.org/abs/2006.16957} {arXiv:2006.16957 [hep-ex]}  (\bibinfo
  {year} {2020}{\natexlab{c}})\BibitemShut {NoStop}%
\bibitem [{\citenamefont {Karliner}\ \emph {et~al.}(2017)\citenamefont
  {Karliner}, \citenamefont {Nussinov},\ and\ \citenamefont
  {Rosner}}]{Karliner:2016zzc}%
  \BibitemOpen
  \bibfield  {author} {\bibinfo {author} {\bibfnamefont {M.}~\bibnamefont
  {Karliner}}, \bibinfo {author} {\bibfnamefont {S.}~\bibnamefont {Nussinov}},\
  and\ \bibinfo {author} {\bibfnamefont {J.~L.}\ \bibnamefont {Rosner}},\
  }\href {https://doi.org/10.1103/PhysRevD.95.034011} {\bibfield  {journal}
  {\bibinfo  {journal} {Phys.Rev.}\ }\textbf {\bibinfo {volume} {D95}},\
  \bibinfo {pages} {034011} (\bibinfo {year} {2017})},\ \Eprint
  {https://arxiv.org/abs/1611.00348} {arXiv:1611.00348 [hep-ph]} \BibitemShut
  {NoStop}%
\bibitem [{\citenamefont {Debastiani}\ and\ \citenamefont
  {Navarra}(2019)}]{Debastiani:2017msn}%
  \BibitemOpen
  \bibfield  {author} {\bibinfo {author} {\bibfnamefont {V.}~\bibnamefont
  {Debastiani}}\ and\ \bibinfo {author} {\bibfnamefont {F.}~\bibnamefont
  {Navarra}},\ }\href {https://doi.org/10.1088/1674-1137/43/1/013105}
  {\bibfield  {journal} {\bibinfo  {journal} {Chin.Phys.}\ }\textbf {\bibinfo
  {volume} {C43}},\ \bibinfo {pages} {013105} (\bibinfo {year} {2019})},\
  \Eprint {https://arxiv.org/abs/1706.07553} {arXiv:1706.07553 [hep-ph]}
  \BibitemShut {NoStop}%
\bibitem [{\citenamefont {Bedolla}\ \emph {et~al.}(2019)\citenamefont
  {Bedolla}, \citenamefont {Ferretti}, \citenamefont {Roberts},\ and\
  \citenamefont {Santopinto}}]{Bedolla:2019zwg}%
  \BibitemOpen
  \bibfield  {author} {\bibinfo {author} {\bibfnamefont {M.~A.}\ \bibnamefont
  {Bedolla}}, \bibinfo {author} {\bibfnamefont {J.}~\bibnamefont {Ferretti}},
  \bibinfo {author} {\bibfnamefont {C.~D.}\ \bibnamefont {Roberts}},\ and\
  \bibinfo {author} {\bibfnamefont {E.}~\bibnamefont {Santopinto}},\ }\Eprint
  {https://arxiv.org/abs/1911.00960} {arXiv:1911.00960 [hep-ph]}  (\bibinfo
  {year} {2019})\BibitemShut {NoStop}%
\bibitem [{\citenamefont {Becchi}\ \emph {et~al.}(2020)\citenamefont {Becchi},
  \citenamefont {Giachino}, \citenamefont {Maiani},\ and\ \citenamefont
  {Santopinto}}]{Becchi:2020uvq}%
  \BibitemOpen
  \bibfield  {author} {\bibinfo {author} {\bibfnamefont {C.}~\bibnamefont
  {Becchi}}, \bibinfo {author} {\bibfnamefont {A.}~\bibnamefont {Giachino}},
  \bibinfo {author} {\bibfnamefont {L.}~\bibnamefont {Maiani}},\ and\ \bibinfo
  {author} {\bibfnamefont {E.}~\bibnamefont {Santopinto}},\ }\Eprint
  {https://arxiv.org/abs/2006.14388} {arXiv:2006.14388 [hep-ph]}  (\bibinfo
  {year} {2020})\BibitemShut {NoStop}%
\bibitem [{\citenamefont {Hiller~Blin}\ \emph {et~al.}(2016)\citenamefont
  {Hiller~Blin}, \citenamefont {Fern\'andez-Ram\'irez}, \citenamefont
  {Jackura}, \citenamefont {Mathieu}, \citenamefont {Mokeev}, \citenamefont
  {Pilloni},\ and\ \citenamefont {Szczepaniak}}]{Blin:2016dlf}%
  \BibitemOpen
  \bibfield  {author} {\bibinfo {author} {\bibfnamefont {A.~N.}\ \bibnamefont
  {Hiller~Blin}}, \bibinfo {author} {\bibfnamefont {C.}~\bibnamefont
  {Fern\'andez-Ram\'irez}}, \bibinfo {author} {\bibfnamefont {A.}~\bibnamefont
  {Jackura}}, \bibinfo {author} {\bibfnamefont {V.}~\bibnamefont {Mathieu}},
  \bibinfo {author} {\bibfnamefont {V.~I.}\ \bibnamefont {Mokeev}}, \bibinfo
  {author} {\bibfnamefont {A.}~\bibnamefont {Pilloni}},\ and\ \bibinfo {author}
  {\bibfnamefont {A.~P.}\ \bibnamefont {Szczepaniak}},\ }\href
  {https://doi.org/10.1103/PhysRevD.94.034002} {\bibfield  {journal} {\bibinfo
  {journal} {Phys.Rev.}\ }\textbf {\bibinfo {volume} {D94}},\ \bibinfo {pages}
  {034002} (\bibinfo {year} {2016})},\ \Eprint
  {https://arxiv.org/abs/1606.08912} {arXiv:1606.08912 [hep-th]} \BibitemShut
  {NoStop}%
\bibitem [{\citenamefont {Winney}\ \emph {et~al.}(2019)\citenamefont {Winney},
  \citenamefont {Fanelli}, \citenamefont {Pilloni}, \citenamefont
  {Hiller~Blin}, \citenamefont {Fern\'andez-Ram\'irez}, \citenamefont
  {Albaladejo}, \citenamefont {Mathieu}, \citenamefont {Mokeev},\ and\
  \citenamefont {Szczepaniak}}]{Winney:2019edt}%
  \BibitemOpen
  \bibfield  {author} {\bibinfo {author} {\bibfnamefont {D.}~\bibnamefont
  {Winney}}, \bibinfo {author} {\bibfnamefont {C.}~\bibnamefont {Fanelli}},
  \bibinfo {author} {\bibfnamefont {A.}~\bibnamefont {Pilloni}}, \bibinfo
  {author} {\bibfnamefont {A.~N.}\ \bibnamefont {Hiller~Blin}}, \bibinfo
  {author} {\bibfnamefont {C.}~\bibnamefont {Fern\'andez-Ram\'irez}}, \bibinfo
  {author} {\bibfnamefont {M.}~\bibnamefont {Albaladejo}}, \bibinfo {author}
  {\bibfnamefont {V.}~\bibnamefont {Mathieu}}, \bibinfo {author} {\bibfnamefont
  {V.~I.}\ \bibnamefont {Mokeev}},\ and\ \bibinfo {author} {\bibfnamefont
  {A.~P.}\ \bibnamefont {Szczepaniak}} (\bibinfo {collaboration} {JPAC}),\
  }\href {https://doi.org/10.1103/PhysRevD.100.034019} {\bibfield  {journal}
  {\bibinfo  {journal} {Phys.Rev.}\ }\textbf {\bibinfo {volume} {D100}},\
  \bibinfo {pages} {034019} (\bibinfo {year} {2019})},\ \Eprint
  {https://arxiv.org/abs/1907.09393} {arXiv:1907.09393 [hep-ph]} \BibitemShut
  {NoStop}%
\bibitem [{\citenamefont {Ablikim}\ \emph
  {et~al.}(2017{\natexlab{b}})\citenamefont {Ablikim} \emph
  {et~al.}}]{Ablikim:2016qzw}%
  \BibitemOpen
  \bibfield  {author} {\bibinfo {author} {\bibfnamefont {M.}~\bibnamefont
  {Ablikim}} \emph {et~al.} (\bibinfo {collaboration} {BESIII}),\ }\href
  {https://doi.org/10.1103/PhysRevLett.118.092001} {\bibfield  {journal}
  {\bibinfo  {journal} {Phys.Rev.Lett.}\ }\textbf {\bibinfo {volume} {118}},\
  \bibinfo {pages} {092001} (\bibinfo {year} {2017}{\natexlab{b}})},\ \Eprint
  {https://arxiv.org/abs/1611.01317} {arXiv:1611.01317 [hep-ex]} \BibitemShut
  {NoStop}%
\bibitem [{\citenamefont {Ablikim}\ \emph
  {et~al.}(2017{\natexlab{c}})\citenamefont {Ablikim} \emph
  {et~al.}}]{BESIII:2016adj}%
  \BibitemOpen
  \bibfield  {author} {\bibinfo {author} {\bibfnamefont {M.}~\bibnamefont
  {Ablikim}} \emph {et~al.} (\bibinfo {collaboration} {BESIII}),\ }\href
  {https://doi.org/10.1103/PhysRevLett.118.092002} {\bibfield  {journal}
  {\bibinfo  {journal} {Phys.Rev.Lett.}\ }\textbf {\bibinfo {volume} {118}},\
  \bibinfo {pages} {092002} (\bibinfo {year} {2017}{\natexlab{c}})},\ \Eprint
  {https://arxiv.org/abs/1610.07044} {arXiv:1610.07044 [hep-ex]} \BibitemShut
  {NoStop}%
\bibitem [{\citenamefont {Ablikim}\ \emph
  {et~al.}(2017{\natexlab{d}})\citenamefont {Ablikim} \emph
  {et~al.}}]{Ablikim:2017oaf}%
  \BibitemOpen
  \bibfield  {author} {\bibinfo {author} {\bibfnamefont {M.}~\bibnamefont
  {Ablikim}} \emph {et~al.} (\bibinfo {collaboration} {BESIII}),\ }\href
  {https://doi.org/10.1103/PhysRevD.96.032004} {\bibfield  {journal} {\bibinfo
  {journal} {Phys.Rev.}\ }\textbf {\bibinfo {volume} {D96}},\ \bibinfo {pages}
  {032004} (\bibinfo {year} {2017}{\natexlab{d}})},\ \Eprint
  {https://arxiv.org/abs/1703.08787} {arXiv:1703.08787 [hep-ex]} \BibitemShut
  {NoStop}%
\bibitem [{\citenamefont {Ablikim}\ \emph
  {et~al.}(2019{\natexlab{a}})\citenamefont {Ablikim} \emph
  {et~al.}}]{Ablikim:2018vxx}%
  \BibitemOpen
  \bibfield  {author} {\bibinfo {author} {\bibfnamefont {M.}~\bibnamefont
  {Ablikim}} \emph {et~al.} (\bibinfo {collaboration} {BESIII}),\ }\href
  {https://doi.org/10.1103/PhysRevLett.122.102002} {\bibfield  {journal}
  {\bibinfo  {journal} {Phys.Rev.Lett.}\ }\textbf {\bibinfo {volume} {122}},\
  \bibinfo {pages} {102002} (\bibinfo {year} {2019}{\natexlab{a}})},\ \Eprint
  {https://arxiv.org/abs/1808.02847} {arXiv:1808.02847 [hep-ex]} \BibitemShut
  {NoStop}%
\bibitem [{\citenamefont {Ablikim}\ \emph
  {et~al.}(2019{\natexlab{b}})\citenamefont {Ablikim} \emph
  {et~al.}}]{Ablikim:2019apl}%
  \BibitemOpen
  \bibfield  {author} {\bibinfo {author} {\bibfnamefont {M.}~\bibnamefont
  {Ablikim}} \emph {et~al.} (\bibinfo {collaboration} {BESIII}),\ }\href
  {https://doi.org/10.1103/PhysRevD.99.091103} {\bibfield  {journal} {\bibinfo
  {journal} {Phys.Rev.}\ }\textbf {\bibinfo {volume} {D99}},\ \bibinfo {pages}
  {091103} (\bibinfo {year} {2019}{\natexlab{b}})},\ \Eprint
  {https://arxiv.org/abs/1903.02359} {arXiv:1903.02359 [hep-ex]} \BibitemShut
  {NoStop}%
\bibitem [{\citenamefont {Ablikim}\ \emph
  {et~al.}(2020{\natexlab{a}})\citenamefont {Ablikim} \emph
  {et~al.}}]{Ablikim:2020cyd}%
  \BibitemOpen
  \bibfield  {author} {\bibinfo {author} {\bibfnamefont {M.}~\bibnamefont
  {Ablikim}} \emph {et~al.} (\bibinfo {collaboration} {BESIII}),\ }\href
  {https://doi.org/10.1103/PhysRevD.102.031101} {\bibfield  {journal} {\bibinfo
   {journal} {Phys.Rev.}\ }\textbf {\bibinfo {volume} {D102}},\ \bibinfo
  {pages} {031101} (\bibinfo {year} {2020}{\natexlab{a}})},\ \Eprint
  {https://arxiv.org/abs/2003.03705} {arXiv:2003.03705 [hep-ex]} \BibitemShut
  {NoStop}%
\bibitem [{\citenamefont {Ablikim}\ \emph {et~al.}(2008)\citenamefont {Ablikim}
  \emph {et~al.}}]{Ablikim:2007gd}%
  \BibitemOpen
  \bibfield  {author} {\bibinfo {author} {\bibfnamefont {M.}~\bibnamefont
  {Ablikim}} \emph {et~al.} (\bibinfo {collaboration} {BES}),\ }\href
  {https://doi.org/10.1016/j.physletb.2007.11.100} {\bibfield  {journal}
  {\bibinfo  {journal} {Phys.Lett.}\ }\textbf {\bibinfo {volume} {B660}},\
  \bibinfo {pages} {315} (\bibinfo {year} {2008})},\ \Eprint
  {https://arxiv.org/abs/0705.4500} {arXiv:0705.4500 [hep-ex]} \BibitemShut
  {NoStop}%
\bibitem [{\citenamefont {Mo}\ \emph {et~al.}(2006)\citenamefont {Mo},
  \citenamefont {Li}, \citenamefont {Yuan}, \citenamefont {He}, \citenamefont
  {Hu} \emph {et~al.}}]{Mo:2006ss}%
  \BibitemOpen
  \bibfield  {author} {\bibinfo {author} {\bibfnamefont {X.~H.}\ \bibnamefont
  {Mo}}, \bibinfo {author} {\bibfnamefont {G.}~\bibnamefont {Li}}, \bibinfo
  {author} {\bibfnamefont {C.~Z.}\ \bibnamefont {Yuan}}, \bibinfo {author}
  {\bibfnamefont {K.~L.}\ \bibnamefont {He}}, \bibinfo {author} {\bibfnamefont
  {H.~M.}\ \bibnamefont {Hu}}, \emph {et~al.},\ }\href
  {https://doi.org/10.1016/j.physletb.2006.07.060} {\bibfield  {journal}
  {\bibinfo  {journal} {Phys.Lett.}\ }\textbf {\bibinfo {volume} {B640}},\
  \bibinfo {pages} {182} (\bibinfo {year} {2006})},\ \Eprint
  {https://arxiv.org/abs/hep-ex/0603024} {arXiv:hep-ex/0603024 [hep-ex]}
  \BibitemShut {NoStop}%
\bibitem [{\citenamefont {Cao}\ \emph {et~al.}(2020{\natexlab{a}})\citenamefont
  {Cao}, \citenamefont {Qi}, \citenamefont {Tang}, \citenamefont {Xue},\ and\
  \citenamefont {Zheng}}]{Cao:2020vab}%
  \BibitemOpen
  \bibfield  {author} {\bibinfo {author} {\bibfnamefont {Q.-F.}\ \bibnamefont
  {Cao}}, \bibinfo {author} {\bibfnamefont {H.-R.}\ \bibnamefont {Qi}},
  \bibinfo {author} {\bibfnamefont {G.-Y.}\ \bibnamefont {Tang}}, \bibinfo
  {author} {\bibfnamefont {Y.-F.}\ \bibnamefont {Xue}},\ and\ \bibinfo {author}
  {\bibfnamefont {H.-Q.}\ \bibnamefont {Zheng}},\ }\Eprint
  {https://arxiv.org/abs/2002.05641} {arXiv:2002.05641 [hep-ph]}  (\bibinfo
  {year} {2020}{\natexlab{a}})\BibitemShut {NoStop}%
\bibitem [{\citenamefont {Donnachie}\ \emph {et~al.}(2005)\citenamefont
  {Donnachie}, \citenamefont {Dosch}, \citenamefont {Nachtmann},\ and\
  \citenamefont {Landshoff}}]{Donnachie:2002en}%
  \BibitemOpen
  \bibfield  {author} {\bibinfo {author} {\bibfnamefont {S.}~\bibnamefont
  {Donnachie}}, \bibinfo {author} {\bibfnamefont {H.~G.}\ \bibnamefont
  {Dosch}}, \bibinfo {author} {\bibfnamefont {O.}~\bibnamefont {Nachtmann}},\
  and\ \bibinfo {author} {\bibfnamefont {P.}~\bibnamefont {Landshoff}},\ }\href
  {https://books.google.it/books?id=WpGHPwAACAAJ} {\emph {\bibinfo {title}
  {{Pomeron physics and QCD}}}},\ Cambridge Monographs on Particle Physics,
  Nuclear Physics and Cosmology\ (\bibinfo  {publisher} {Cambridge University
  Press},\ \bibinfo {year} {2005})\BibitemShut {NoStop}%
\bibitem [{\citenamefont {Laget}(2020)}]{Laget:2019tou}%
  \BibitemOpen
  \bibfield  {author} {\bibinfo {author} {\bibfnamefont {J.}~\bibnamefont
  {Laget}},\ }\href {https://doi.org/10.1016/j.ppnp.2019.103737} {\bibfield
  {journal} {\bibinfo  {journal} {Prog.Part.Nucl.Phys.}\ }\textbf {\bibinfo
  {volume} {111}},\ \bibinfo {pages} {103737} (\bibinfo {year} {2020})},\
  \Eprint {https://arxiv.org/abs/1911.04825} {arXiv:1911.04825 [hep-ph]}
  \BibitemShut {NoStop}%
\bibitem [{\citenamefont {Mathieu}\ \emph {et~al.}(2018)\citenamefont
  {Mathieu}, \citenamefont {Nys}, \citenamefont {Fern\'andez-Ram\'irez},
  \citenamefont {Jackura}, \citenamefont {Pilloni}, \citenamefont {Sherrill},
  \citenamefont {Szczepaniak},\ and\ \citenamefont {Fox}}]{Mathieu:2018xyc}%
  \BibitemOpen
  \bibfield  {author} {\bibinfo {author} {\bibfnamefont {V.}~\bibnamefont
  {Mathieu}}, \bibinfo {author} {\bibfnamefont {J.}~\bibnamefont {Nys}},
  \bibinfo {author} {\bibfnamefont {C.}~\bibnamefont {Fern\'andez-Ram\'irez}},
  \bibinfo {author} {\bibfnamefont {A.}~\bibnamefont {Jackura}}, \bibinfo
  {author} {\bibfnamefont {A.}~\bibnamefont {Pilloni}}, \bibinfo {author}
  {\bibfnamefont {N.}~\bibnamefont {Sherrill}}, \bibinfo {author}
  {\bibfnamefont {A.~P.}\ \bibnamefont {Szczepaniak}},\ and\ \bibinfo {author}
  {\bibfnamefont {G.}~\bibnamefont {Fox}} (\bibinfo {collaboration} {JPAC}),\
  }\href {https://doi.org/10.1103/PhysRevD.97.094003} {\bibfield  {journal}
  {\bibinfo  {journal} {Phys.Rev.}\ }\textbf {\bibinfo {volume} {D97}},\
  \bibinfo {pages} {094003} (\bibinfo {year} {2018})},\ \Eprint
  {https://arxiv.org/abs/1802.09403} {arXiv:1802.09403 [hep-ph]} \BibitemShut
  {NoStop}%
\bibitem [{\citenamefont {Brodsky}\ \emph {et~al.}(2001)\citenamefont
  {Brodsky}, \citenamefont {Chudakov}, \citenamefont {Hoyer},\ and\
  \citenamefont {Laget}}]{Brodsky:2000zc}%
  \BibitemOpen
  \bibfield  {author} {\bibinfo {author} {\bibfnamefont {S.~J.}\ \bibnamefont
  {Brodsky}}, \bibinfo {author} {\bibfnamefont {E.}~\bibnamefont {Chudakov}},
  \bibinfo {author} {\bibfnamefont {P.}~\bibnamefont {Hoyer}},\ and\ \bibinfo
  {author} {\bibfnamefont {J.~M.}\ \bibnamefont {Laget}},\ }\href
  {https://doi.org/10.1016/S0370-2693(00)01373-3} {\bibfield  {journal}
  {\bibinfo  {journal} {Phys.Lett.}\ }\textbf {\bibinfo {volume} {B498}},\
  \bibinfo {pages} {23} (\bibinfo {year} {2001})},\ \Eprint
  {https://arxiv.org/abs/hep-ph/0010343} {arXiv:hep-ph/0010343 [hep-ph]}
  \BibitemShut {NoStop}%
\bibitem [{\citenamefont {Du}\ \emph {et~al.}(2020)\citenamefont {Du},
  \citenamefont {Baru}, \citenamefont {Guo}, \citenamefont {Hanhart},
  \citenamefont {Mei\ss{}ner}, \citenamefont {Nefediev},\ and\ \citenamefont
  {Strakovsky}}]{Du:2020bqj}%
  \BibitemOpen
  \bibfield  {author} {\bibinfo {author} {\bibfnamefont {M.-L.}\ \bibnamefont
  {Du}}, \bibinfo {author} {\bibfnamefont {V.}~\bibnamefont {Baru}}, \bibinfo
  {author} {\bibfnamefont {F.-K.}\ \bibnamefont {Guo}}, \bibinfo {author}
  {\bibfnamefont {C.}~\bibnamefont {Hanhart}}, \bibinfo {author} {\bibfnamefont
  {U.-G.}\ \bibnamefont {Mei\ss{}ner}}, \bibinfo {author} {\bibfnamefont
  {A.}~\bibnamefont {Nefediev}},\ and\ \bibinfo {author} {\bibfnamefont
  {I.}~\bibnamefont {Strakovsky}},\ }\Eprint {https://arxiv.org/abs/2009.08345}
  {arXiv:2009.08345 [hep-ph]}  (\bibinfo {year} {2020})\BibitemShut {NoStop}%
\bibitem [{\citenamefont {Chekanov}\ \emph {et~al.}(2002)\citenamefont
  {Chekanov} \emph {et~al.}}]{Chekanov:2002xi}%
  \BibitemOpen
  \bibfield  {author} {\bibinfo {author} {\bibfnamefont {S.}~\bibnamefont
  {Chekanov}} \emph {et~al.} (\bibinfo {collaboration} {ZEUS}),\ }\href
  {https://doi.org/10.1007/s10052-002-0953-7} {\bibfield  {journal} {\bibinfo
  {journal} {Eur.Phys.J.}\ }\textbf {\bibinfo {volume} {C24}},\ \bibinfo
  {pages} {345} (\bibinfo {year} {2002})},\ \Eprint
  {https://arxiv.org/abs/hep-ex/0201043} {arXiv:hep-ex/0201043 [hep-ex]}
  \BibitemShut {NoStop}%
\bibitem [{\citenamefont {Aktas}\ \emph {et~al.}(2006)\citenamefont {Aktas}
  \emph {et~al.}}]{Aktas:2005xu}%
  \BibitemOpen
  \bibfield  {author} {\bibinfo {author} {\bibfnamefont {A.}~\bibnamefont
  {Aktas}} \emph {et~al.} (\bibinfo {collaboration} {H1}),\ }\href
  {https://doi.org/10.1140/epjc/s2006-02519-5} {\bibfield  {journal} {\bibinfo
  {journal} {Eur.Phys.J.}\ }\textbf {\bibinfo {volume} {C46}},\ \bibinfo
  {pages} {585} (\bibinfo {year} {2006})},\ \Eprint
  {https://arxiv.org/abs/hep-ex/0510016} {arXiv:hep-ex/0510016 [hep-ex]}
  \BibitemShut {NoStop}%
\bibitem [{\citenamefont {Camerini}\ \emph {et~al.}(1975)\citenamefont
  {Camerini}, \citenamefont {Learned}, \citenamefont {Prepost}, \citenamefont
  {Spencer}, \citenamefont {Wiser}, \citenamefont {Ash}, \citenamefont
  {Anderson}, \citenamefont {Ritson}, \citenamefont {Sherden},\ and\
  \citenamefont {Sinclair}}]{Camerini:1975cy}%
  \BibitemOpen
  \bibfield  {author} {\bibinfo {author} {\bibfnamefont {U.}~\bibnamefont
  {Camerini}}, \bibinfo {author} {\bibfnamefont {J.~G.}\ \bibnamefont
  {Learned}}, \bibinfo {author} {\bibfnamefont {R.}~\bibnamefont {Prepost}},
  \bibinfo {author} {\bibfnamefont {C.~M.}\ \bibnamefont {Spencer}}, \bibinfo
  {author} {\bibfnamefont {D.~E.}\ \bibnamefont {Wiser}}, \bibinfo {author}
  {\bibfnamefont {W.}~\bibnamefont {Ash}}, \bibinfo {author} {\bibfnamefont
  {R.~L.}\ \bibnamefont {Anderson}}, \bibinfo {author} {\bibfnamefont
  {D.}~\bibnamefont {Ritson}}, \bibinfo {author} {\bibfnamefont
  {D.}~\bibnamefont {Sherden}},\ and\ \bibinfo {author} {\bibfnamefont {C.~K.}\
  \bibnamefont {Sinclair}},\ }\href
  {https://doi.org/10.1103/PhysRevLett.35.483} {\bibfield  {journal} {\bibinfo
  {journal} {Phys.Rev.Lett.}\ }\textbf {\bibinfo {volume} {35}},\ \bibinfo
  {pages} {483} (\bibinfo {year} {1975})}\BibitemShut {NoStop}%
\bibitem [{\citenamefont {Close}\ and\ \citenamefont
  {Schuler}(1999)}]{Close:1999bi}%
  \BibitemOpen
  \bibfield  {author} {\bibinfo {author} {\bibfnamefont {F.~E.}\ \bibnamefont
  {Close}}\ and\ \bibinfo {author} {\bibfnamefont {G.~A.}\ \bibnamefont
  {Schuler}},\ }\href {https://doi.org/10.1016/S0370-2693(99)00875-8}
  {\bibfield  {journal} {\bibinfo  {journal} {Phys.Lett.}\ }\textbf {\bibinfo
  {volume} {B464}},\ \bibinfo {pages} {279} (\bibinfo {year} {1999})},\ \Eprint
  {https://arxiv.org/abs/hep-ph/9905305} {arXiv:hep-ph/9905305 [hep-ph]}
  \BibitemShut {NoStop}%
\bibitem [{\citenamefont {Lesniak}\ and\ \citenamefont
  {Szczepaniak}(2003)}]{Lesniak:2003gf}%
  \BibitemOpen
  \bibfield  {author} {\bibinfo {author} {\bibfnamefont {L.}~\bibnamefont
  {Lesniak}}\ and\ \bibinfo {author} {\bibfnamefont {A.~P.}\ \bibnamefont
  {Szczepaniak}},\ }\href@noop {} {\bibfield  {journal} {\bibinfo  {journal}
  {Acta Phys.Polon.}\ }\textbf {\bibinfo {volume} {B34}},\ \bibinfo {pages}
  {3389} (\bibinfo {year} {2003})},\ \Eprint
  {https://arxiv.org/abs/hep-ph/0304007} {arXiv:hep-ph/0304007 [hep-ph]}
  \BibitemShut {NoStop}%
\bibitem [{\citenamefont {Besson}\ \emph {et~al.}(2008)\citenamefont {Besson}
  \emph {et~al.}}]{Besson:2008pr}%
  \BibitemOpen
  \bibfield  {author} {\bibinfo {author} {\bibfnamefont {D.}~\bibnamefont
  {Besson}} \emph {et~al.} (\bibinfo {collaboration} {CLEO}),\ }\href
  {https://doi.org/10.1103/PhysRevD.78.032012} {\bibfield  {journal} {\bibinfo
  {journal} {Phys.Rev.}\ }\textbf {\bibinfo {volume} {D78}},\ \bibinfo {pages}
  {032012} (\bibinfo {year} {2008})},\ \Eprint
  {https://arxiv.org/abs/0806.0315} {arXiv:0806.0315 [hep-ex]} \BibitemShut
  {NoStop}%
\bibitem [{\citenamefont {Libby}\ \emph {et~al.}(2009)\citenamefont {Libby}
  \emph {et~al.}}]{Libby:2009qb}%
  \BibitemOpen
  \bibfield  {author} {\bibinfo {author} {\bibfnamefont {J.}~\bibnamefont
  {Libby}} \emph {et~al.} (\bibinfo {collaboration} {CLEO}),\ }\href
  {https://doi.org/10.1103/PhysRevD.80.072002} {\bibfield  {journal} {\bibinfo
  {journal} {Phys.Rev.}\ }\textbf {\bibinfo {volume} {D80}},\ \bibinfo {pages}
  {072002} (\bibinfo {year} {2009})},\ \Eprint
  {https://arxiv.org/abs/0909.0193} {arXiv:0909.0193 [hep-ex]} \BibitemShut
  {NoStop}%
\bibitem [{\citenamefont {Adloff}\ \emph {et~al.}(1998)\citenamefont {Adloff}
  \emph {et~al.}}]{Adloff:1997yv}%
  \BibitemOpen
  \bibfield  {author} {\bibinfo {author} {\bibfnamefont {C.}~\bibnamefont
  {Adloff}} \emph {et~al.} (\bibinfo {collaboration} {H1}),\ }\href
  {https://doi.org/10.1016/S0370-2693(97)01529-3} {\bibfield  {journal}
  {\bibinfo  {journal} {Phys.Lett.}\ }\textbf {\bibinfo {volume} {B421}},\
  \bibinfo {pages} {385} (\bibinfo {year} {1998})},\ \Eprint
  {https://arxiv.org/abs/hep-ex/9711012} {arXiv:hep-ex/9711012} \BibitemShut
  {NoStop}%
\bibitem [{\citenamefont {Grzelak}(2019)}]{Grzelak:2018srl}%
  \BibitemOpen
  \bibfield  {author} {\bibinfo {author} {\bibfnamefont {G.}~\bibnamefont
  {Grzelak}} (\bibinfo {collaboration} {ZEUS}),\ }\href
  {https://doi.org/10.22323/1.340.0921} {\bibfield  {journal} {\bibinfo
  {journal} {PoS}\ }\textbf {\bibinfo {volume} {ICHEP2018}},\ \bibinfo {pages}
  {921} (\bibinfo {year} {2019})}\BibitemShut {NoStop}%
\bibitem [{\citenamefont {Voloshin}\ and\ \citenamefont
  {Zakharov}(1980)}]{Voloshin:1980zf}%
  \BibitemOpen
  \bibfield  {author} {\bibinfo {author} {\bibfnamefont {M.~B.}\ \bibnamefont
  {Voloshin}}\ and\ \bibinfo {author} {\bibfnamefont {V.~I.}\ \bibnamefont
  {Zakharov}},\ }\href {https://doi.org/10.1103/PhysRevLett.45.688} {\bibfield
  {journal} {\bibinfo  {journal} {Phys.Rev.Lett.}\ }\textbf {\bibinfo {volume}
  {45}},\ \bibinfo {pages} {688} (\bibinfo {year} {1980})}\BibitemShut
  {NoStop}%
\bibitem [{\citenamefont {Novikov}\ and\ \citenamefont
  {Shifman}(1981)}]{Novikov:1980fa}%
  \BibitemOpen
  \bibfield  {author} {\bibinfo {author} {\bibfnamefont {V.~A.}\ \bibnamefont
  {Novikov}}\ and\ \bibinfo {author} {\bibfnamefont {M.~A.}\ \bibnamefont
  {Shifman}},\ }\href {https://doi.org/10.1007/BF01429829} {\bibfield
  {journal} {\bibinfo  {journal} {Z.Phys.}\ }\textbf {\bibinfo {volume} {C8}},\
  \bibinfo {pages} {43} (\bibinfo {year} {1981})}\BibitemShut {NoStop}%
\bibitem [{\citenamefont {He}\ \emph {et~al.}(2006)\citenamefont {He} \emph
  {et~al.}}]{He:2006kg}%
  \BibitemOpen
  \bibfield  {author} {\bibinfo {author} {\bibfnamefont {Q.}~\bibnamefont {He}}
  \emph {et~al.} (\bibinfo {collaboration} {\cleo}),\ }\href
  {https://doi.org/10.1103/PhysRevD.74.091104} {\bibfield  {journal} {\bibinfo
  {journal} {Phys.Rev.}\ }\textbf {\bibinfo {volume} {D74}},\ \bibinfo {pages}
  {091104} (\bibinfo {year} {2006})},\ \Eprint
  {https://arxiv.org/abs/hep-ex/0611021} {arXiv:hep-ex/0611021 [hep-ex]}
  \BibitemShut {NoStop}%
\bibitem [{\citenamefont {Lees}\ \emph {et~al.}(2012)\citenamefont {Lees} \emph
  {et~al.}}]{Lees:2012cn}%
  \BibitemOpen
  \bibfield  {author} {\bibinfo {author} {\bibfnamefont {J.~P.}\ \bibnamefont
  {Lees}} \emph {et~al.} (\bibinfo {collaboration} {\babar}),\ }\href
  {https://doi.org/10.1103/PhysRevD.86.051102} {\bibfield  {journal} {\bibinfo
  {journal} {Phys.Rev.}\ }\textbf {\bibinfo {volume} {D86}},\ \bibinfo {pages}
  {051102} (\bibinfo {year} {2012})},\ \Eprint
  {https://arxiv.org/abs/1204.2158} {arXiv:1204.2158 [hep-ex]} \BibitemShut
  {NoStop}%
\bibitem [{\citenamefont {Kou}\ and\ \citenamefont {Pene}(2005)}]{Kou:2005gt}%
  \BibitemOpen
  \bibfield  {author} {\bibinfo {author} {\bibfnamefont {E.}~\bibnamefont
  {Kou}}\ and\ \bibinfo {author} {\bibfnamefont {O.}~\bibnamefont {Pene}},\
  }\href {https://doi.org/10.1016/j.physletb.2005.09.013} {\bibfield  {journal}
  {\bibinfo  {journal} {Phys.Lett.}\ }\textbf {\bibinfo {volume} {B631}},\
  \bibinfo {pages} {164} (\bibinfo {year} {2005})},\ \Eprint
  {https://arxiv.org/abs/hep-ph/0507119} {arXiv:hep-ph/0507119 [hep-ph]}
  \BibitemShut {NoStop}%
\bibitem [{\citenamefont {Liu}\ \emph {et~al.}(2012)\citenamefont {Liu} \emph
  {et~al.}}]{Liu:2012ze}%
  \BibitemOpen
  \bibfield  {author} {\bibinfo {author} {\bibfnamefont {L.}~\bibnamefont
  {Liu}} \emph {et~al.} (\bibinfo {collaboration} {Hadron Spectrum}),\ }\href
  {https://doi.org/10.1007/JHEP07(2012)126} {\bibfield  {journal} {\bibinfo
  {journal} {JHEP}\ }\textbf {\bibinfo {volume} {1207}},\ \bibinfo {pages}
  {126}},\ \Eprint {https://arxiv.org/abs/1204.5425} {arXiv:1204.5425 [hep-ph]}
  \BibitemShut {NoStop}%
\bibitem [{\citenamefont {Guo}\ \emph {et~al.}(2014)\citenamefont {Guo},
  \citenamefont {Y\'epez-Mart\'inez},\ and\ \citenamefont
  {Szczepaniak}}]{Guo:2014zva}%
  \BibitemOpen
  \bibfield  {author} {\bibinfo {author} {\bibfnamefont {P.}~\bibnamefont
  {Guo}}, \bibinfo {author} {\bibfnamefont {T.}~\bibnamefont
  {Y\'epez-Mart\'inez}},\ and\ \bibinfo {author} {\bibfnamefont {A.~P.}\
  \bibnamefont {Szczepaniak}},\ }\href
  {https://doi.org/10.1103/PhysRevD.89.116005} {\bibfield  {journal} {\bibinfo
  {journal} {Phys.Rev.}\ }\textbf {\bibinfo {volume} {D89}},\ \bibinfo {pages}
  {116005} (\bibinfo {year} {2014})},\ \Eprint
  {https://arxiv.org/abs/1402.5863} {arXiv:1402.5863 [hep-ph]} \BibitemShut
  {NoStop}%
\bibitem [{\citenamefont {Oncala}\ and\ \citenamefont
  {Soto}(2017)}]{Oncala:2017hop}%
  \BibitemOpen
  \bibfield  {author} {\bibinfo {author} {\bibfnamefont {R.}~\bibnamefont
  {Oncala}}\ and\ \bibinfo {author} {\bibfnamefont {J.}~\bibnamefont {Soto}},\
  }\href {https://doi.org/10.1103/PhysRevD.96.014004} {\bibfield  {journal}
  {\bibinfo  {journal} {Phys.Rev.}\ }\textbf {\bibinfo {volume} {D96}},\
  \bibinfo {pages} {014004} (\bibinfo {year} {2017})},\ \Eprint
  {https://arxiv.org/abs/1702.03900} {arXiv:1702.03900 [hep-ph]} \BibitemShut
  {NoStop}%
\bibitem [{\citenamefont {Karliner}\ and\ \citenamefont
  {Rosner}(2016)}]{Karliner:2015voa}%
  \BibitemOpen
  \bibfield  {author} {\bibinfo {author} {\bibfnamefont {M.}~\bibnamefont
  {Karliner}}\ and\ \bibinfo {author} {\bibfnamefont {J.~L.}\ \bibnamefont
  {Rosner}},\ }\href {https://doi.org/10.1016/j.physletb.2015.11.068}
  {\bibfield  {journal} {\bibinfo  {journal} {Phys.Lett.}\ }\textbf {\bibinfo
  {volume} {B752}},\ \bibinfo {pages} {329} (\bibinfo {year} {2016})},\ \Eprint
  {https://arxiv.org/abs/1508.01496} {arXiv:1508.01496 [hep-ph]} \BibitemShut
  {NoStop}%
\bibitem [{\citenamefont {Kubarovsky}\ and\ \citenamefont
  {Voloshin}(2015)}]{Kubarovsky:2015aaa}%
  \BibitemOpen
  \bibfield  {author} {\bibinfo {author} {\bibfnamefont {V.}~\bibnamefont
  {Kubarovsky}}\ and\ \bibinfo {author} {\bibfnamefont {M.~B.}\ \bibnamefont
  {Voloshin}},\ }\href {https://doi.org/10.1103/PhysRevD.92.031502} {\bibfield
  {journal} {\bibinfo  {journal} {Phys.Rev.}\ }\textbf {\bibinfo {volume}
  {D92}},\ \bibinfo {pages} {031502} (\bibinfo {year} {2015})},\ \Eprint
  {https://arxiv.org/abs/1508.00888} {arXiv:1508.00888 [hep-ph]} \BibitemShut
  {NoStop}%
\bibitem [{\citenamefont {Wang}\ \emph
  {et~al.}(2015{\natexlab{b}})\citenamefont {Wang}, \citenamefont {Liu},\ and\
  \citenamefont {Zhao}}]{Wang:2015jsa}%
  \BibitemOpen
  \bibfield  {author} {\bibinfo {author} {\bibfnamefont {Q.}~\bibnamefont
  {Wang}}, \bibinfo {author} {\bibfnamefont {X.-H.}\ \bibnamefont {Liu}},\ and\
  \bibinfo {author} {\bibfnamefont {Q.}~\bibnamefont {Zhao}},\ }\href
  {https://doi.org/10.1103/PhysRevD.92.034022} {\bibfield  {journal} {\bibinfo
  {journal} {Phys.Rev.}\ }\textbf {\bibinfo {volume} {D92}},\ \bibinfo {pages}
  {034022} (\bibinfo {year} {2015}{\natexlab{b}})},\ \Eprint
  {https://arxiv.org/abs/1508.00339} {arXiv:1508.00339 [hep-ph]} \BibitemShut
  {NoStop}%
\bibitem [{\citenamefont {Cao}\ and\ \citenamefont {Dai}(2019)}]{Cao:2019kst}%
  \BibitemOpen
  \bibfield  {author} {\bibinfo {author} {\bibfnamefont {X.}~\bibnamefont
  {Cao}}\ and\ \bibinfo {author} {\bibfnamefont {J.-p.}\ \bibnamefont {Dai}},\
  }\href {https://doi.org/10.1103/PhysRevD.100.054033} {\bibfield  {journal}
  {\bibinfo  {journal} {Phys.Rev.}\ }\textbf {\bibinfo {volume} {D100}},\
  \bibinfo {pages} {054033} (\bibinfo {year} {2019})},\ \Eprint
  {https://arxiv.org/abs/1904.06015} {arXiv:1904.06015 [hep-ph]} \BibitemShut
  {NoStop}%
\bibitem [{\citenamefont {Braaten}\ \emph {et~al.}(2019)\citenamefont
  {Braaten}, \citenamefont {He},\ and\ \citenamefont
  {Ingles}}]{Braaten:2019sxh}%
  \BibitemOpen
  \bibfield  {author} {\bibinfo {author} {\bibfnamefont {E.}~\bibnamefont
  {Braaten}}, \bibinfo {author} {\bibfnamefont {L.-P.}\ \bibnamefont {He}},\
  and\ \bibinfo {author} {\bibfnamefont {K.}~\bibnamefont {Ingles}},\ }\href
  {https://doi.org/10.1103/PhysRevD.100.094006} {\bibfield  {journal} {\bibinfo
   {journal} {Phys.Rev.}\ }\textbf {\bibinfo {volume} {D100}},\ \bibinfo
  {pages} {094006} (\bibinfo {year} {2019})},\ \Eprint
  {https://arxiv.org/abs/1903.04355} {arXiv:1903.04355 [hep-ph]} \BibitemShut
  {NoStop}%
\bibitem [{\citenamefont {Ablikim}\ \emph
  {et~al.}(2020{\natexlab{b}})\citenamefont {Ablikim} \emph
  {et~al.}}]{Ablikim:2020xpq}%
  \BibitemOpen
  \bibfield  {author} {\bibinfo {author} {\bibfnamefont {M.}~\bibnamefont
  {Ablikim}} \emph {et~al.} (\bibinfo {collaboration} {BESIII}),\ }\href
  {https://doi.org/10.1103/PhysRevLett.124.242001} {\bibfield  {journal}
  {\bibinfo  {journal} {Phys.Rev.Lett.}\ }\textbf {\bibinfo {volume} {124}},\
  \bibinfo {pages} {242001} (\bibinfo {year} {2020}{\natexlab{b}})},\ \Eprint
  {https://arxiv.org/abs/2001.01156} {arXiv:2001.01156 [hep-ex]} \BibitemShut
  {NoStop}%
\bibitem [{\citenamefont {Garmash}\ \emph {et~al.}(2015)\citenamefont {Garmash}
  \emph {et~al.}}]{Garmash:2014dhx}%
  \BibitemOpen
  \bibfield  {author} {\bibinfo {author} {\bibfnamefont {A.}~\bibnamefont
  {Garmash}} \emph {et~al.} (\bibinfo {collaboration} {Belle}),\ }\href
  {https://doi.org/10.1103/PhysRevD.91.072003} {\bibfield  {journal} {\bibinfo
  {journal} {Phys.Rev.}\ }\textbf {\bibinfo {volume} {D91}},\ \bibinfo {pages}
  {072003} (\bibinfo {year} {2015})},\ \Eprint
  {https://arxiv.org/abs/1403.0992} {arXiv:1403.0992 [hep-ex]} \BibitemShut
  {NoStop}%
\bibitem [{\citenamefont {Aaij}\ \emph {et~al.}(2019)\citenamefont {Aaij} \emph
  {et~al.}}]{Aaij:2019vzc}%
  \BibitemOpen
  \bibfield  {author} {\bibinfo {author} {\bibfnamefont {R.}~\bibnamefont
  {Aaij}} \emph {et~al.} (\bibinfo {collaboration} {LHCb}),\ }\href
  {https://doi.org/10.1103/PhysRevLett.122.222001} {\bibfield  {journal}
  {\bibinfo  {journal} {Phys.Rev.Lett}\ }\textbf {\bibinfo {volume} {122}},\
  \bibinfo {pages} {222001} (\bibinfo {year} {2019})},\ \Eprint
  {https://arxiv.org/abs/1904.03947} {arXiv:1904.03947 [hep-ex]} \BibitemShut
  {NoStop}%
\bibitem [{\citenamefont {Cao}\ \emph {et~al.}(2020{\natexlab{b}})\citenamefont
  {Cao}, \citenamefont {Guo}, \citenamefont {Liang}, \citenamefont {Wu},
  \citenamefont {Xie}, \citenamefont {Xie}, \citenamefont {Yang},\ and\
  \citenamefont {Zou}}]{Cao:2019gqo}%
  \BibitemOpen
  \bibfield  {author} {\bibinfo {author} {\bibfnamefont {X.}~\bibnamefont
  {Cao}}, \bibinfo {author} {\bibfnamefont {F.-K.}\ \bibnamefont {Guo}},
  \bibinfo {author} {\bibfnamefont {Y.-T.}\ \bibnamefont {Liang}}, \bibinfo
  {author} {\bibfnamefont {J.-J.}\ \bibnamefont {Wu}}, \bibinfo {author}
  {\bibfnamefont {J.-J.}\ \bibnamefont {Xie}}, \bibinfo {author} {\bibfnamefont
  {Y.-P.}\ \bibnamefont {Xie}}, \bibinfo {author} {\bibfnamefont
  {Z.}~\bibnamefont {Yang}},\ and\ \bibinfo {author} {\bibfnamefont {B.-S.}\
  \bibnamefont {Zou}},\ }\href {https://doi.org/10.1103/PhysRevD.101.074010}
  {\bibfield  {journal} {\bibinfo  {journal} {Phys.Rev.}\ }\textbf {\bibinfo
  {volume} {D101}},\ \bibinfo {pages} {074010} (\bibinfo {year}
  {2020}{\natexlab{b}})},\ \Eprint {https://arxiv.org/abs/1912.12054}
  {arXiv:1912.12054 [hep-ph]} \BibitemShut {NoStop}%
\bibitem [{\citenamefont {Paryev}(2020)}]{Paryev:2020jkp}%
  \BibitemOpen
  \bibfield  {author} {\bibinfo {author} {\bibfnamefont {E.}~\bibnamefont
  {Paryev}},\ }\Eprint {https://arxiv.org/abs/2007.01172} {arXiv:2007.01172
  [nucl-th]}  (\bibinfo {year} {2020})\BibitemShut {NoStop}%
\bibitem [{\citenamefont {{GlueX Collaboration}}()}]{gluextcr}%
  \BibitemOpen
  \bibfield  {author} {\bibinfo {author} {\bibnamefont {{GlueX
  Collaboration}}},\ }\bibinfo {note}
  {{\href{https://halldweb.jlab.org/DocDB/0025/002511/006/tcr.pdf}{https://halldweb.jlab.org/DocDB/
  0025/002511/006/tcr.pdf}}}\BibitemShut {NoStop}%
\bibitem [{\citenamefont {{JPAC Collaboration}}()}]{JPACweb}%
  \BibitemOpen
  \bibfield  {author} {\bibinfo {author} {\bibnamefont {{JPAC
  Collaboration}}},\ }\bibinfo {note}
  {{\href{http://cgl.soic.indiana.edu/jpac/}{http://cgl.soic.indiana.edu/jpac/}}}\BibitemShut
  {NoStop}%
\bibitem [{\citenamefont {Choi}\ \emph {et~al.}(2011)\citenamefont {Choi},
  \citenamefont {Olsen}, \citenamefont {Trabelsi}, \citenamefont {Adachi},
  \citenamefont {Aihara} \emph {et~al.}}]{Choi:2011fc}%
  \BibitemOpen
  \bibfield  {author} {\bibinfo {author} {\bibfnamefont {S.-K.}\ \bibnamefont
  {Choi}}, \bibinfo {author} {\bibfnamefont {S.~L.}\ \bibnamefont {Olsen}},
  \bibinfo {author} {\bibfnamefont {K.}~\bibnamefont {Trabelsi}}, \bibinfo
  {author} {\bibfnamefont {I.}~\bibnamefont {Adachi}}, \bibinfo {author}
  {\bibfnamefont {H.}~\bibnamefont {Aihara}}, \emph {et~al.},\ }\href
  {https://doi.org/10.1103/PhysRevD.84.052004} {\bibfield  {journal} {\bibinfo
  {journal} {Phys.Rev.}\ }\textbf {\bibinfo {volume} {D84}},\ \bibinfo {pages}
  {052004} (\bibinfo {year} {2011})},\ \Eprint
  {https://arxiv.org/abs/1107.0163} {arXiv:1107.0163 [hep-ex]} \BibitemShut
  {NoStop}%
\bibitem [{\citenamefont {Aaboud}\ \emph {et~al.}(2017)\citenamefont {Aaboud}
  \emph {et~al.}}]{Aaboud:2016vzw}%
  \BibitemOpen
  \bibfield  {author} {\bibinfo {author} {\bibfnamefont {M.}~\bibnamefont
  {Aaboud}} \emph {et~al.} (\bibinfo {collaboration} {ATLAS}),\ }\href
  {https://doi.org/10.1007/JHEP01(2017)117} {\bibfield  {journal} {\bibinfo
  {journal} {JHEP}\ }\textbf {\bibinfo {volume} {01}},\ \bibinfo {pages}
  {117}},\ \Eprint {https://arxiv.org/abs/1610.09303} {arXiv:1610.09303
  [hep-ex]} \BibitemShut {NoStop}%
\bibitem [{\citenamefont {Borasoy}\ and\ \citenamefont
  {Mei{\ss}ner}(1996)}]{Borasoy:1995ds}%
  \BibitemOpen
  \bibfield  {author} {\bibinfo {author} {\bibfnamefont {B.}~\bibnamefont
  {Borasoy}}\ and\ \bibinfo {author} {\bibfnamefont {U.-G.}\ \bibnamefont
  {Mei{\ss}ner}},\ }\href {https://doi.org/10.1142/S0217751X96002376}
  {\bibfield  {journal} {\bibinfo  {journal} {Int.J.Mod.Phys.}\ }\textbf
  {\bibinfo {volume} {A11}},\ \bibinfo {pages} {5183} (\bibinfo {year}
  {1996})},\ \Eprint {https://arxiv.org/abs/hep-ph/9511320}
  {arXiv:hep-ph/9511320} \BibitemShut {NoStop}%
\bibitem [{\citenamefont {Abulencia}\ \emph {et~al.}(2006)\citenamefont
  {Abulencia} \emph {et~al.}}]{Abulencia:2005zc}%
  \BibitemOpen
  \bibfield  {author} {\bibinfo {author} {\bibfnamefont {A.}~\bibnamefont
  {Abulencia}} \emph {et~al.} (\bibinfo {collaboration} {CDF}),\ }\href
  {https://doi.org/10.1103/PhysRevLett.96.102002} {\bibfield  {journal}
  {\bibinfo  {journal} {Phys.Rev.Lett.}\ }\textbf {\bibinfo {volume} {96}},\
  \bibinfo {pages} {102002} (\bibinfo {year} {2006})},\ \Eprint
  {https://arxiv.org/abs/hep-ex/0512074} {arXiv:hep-ex/0512074 [hep-ex]}
  \BibitemShut {NoStop}%
\bibitem [{\citenamefont {Niecknig}\ \emph {et~al.}(2012)\citenamefont
  {Niecknig}, \citenamefont {Kubis},\ and\ \citenamefont
  {Schneider}}]{Niecknig:2012sj}%
  \BibitemOpen
  \bibfield  {author} {\bibinfo {author} {\bibfnamefont {F.}~\bibnamefont
  {Niecknig}}, \bibinfo {author} {\bibfnamefont {B.}~\bibnamefont {Kubis}},\
  and\ \bibinfo {author} {\bibfnamefont {S.~P.}\ \bibnamefont {Schneider}},\
  }\href {https://doi.org/10.1140/epjc/s10052-012-2014-1} {\bibfield  {journal}
  {\bibinfo  {journal} {Eur.Phys.J.}\ }\textbf {\bibinfo {volume} {C72}},\
  \bibinfo {pages} {2014} (\bibinfo {year} {2012})},\ \Eprint
  {https://arxiv.org/abs/1203.2501} {arXiv:1203.2501 [hep-ph]} \BibitemShut
  {NoStop}%
\bibitem [{\citenamefont {Danilkin}\ \emph {et~al.}(2015)\citenamefont
  {Danilkin}, \citenamefont {Fern\'andez-Ram\'irez}, \citenamefont {Guo},
  \citenamefont {Mathieu}, \citenamefont {Schott}, \citenamefont {Shi},\ and\
  \citenamefont {Szczepaniak}}]{Danilkin:2014cra}%
  \BibitemOpen
  \bibfield  {author} {\bibinfo {author} {\bibfnamefont {I.~V.}\ \bibnamefont
  {Danilkin}}, \bibinfo {author} {\bibfnamefont {C.}~\bibnamefont
  {Fern\'andez-Ram\'irez}}, \bibinfo {author} {\bibfnamefont {P.}~\bibnamefont
  {Guo}}, \bibinfo {author} {\bibfnamefont {V.}~\bibnamefont {Mathieu}},
  \bibinfo {author} {\bibfnamefont {D.}~\bibnamefont {Schott}}, \bibinfo
  {author} {\bibfnamefont {M.}~\bibnamefont {Shi}},\ and\ \bibinfo {author}
  {\bibfnamefont {A.~P.}\ \bibnamefont {Szczepaniak}},\ }\href
  {https://doi.org/10.1103/PhysRevD.91.094029} {\bibfield  {journal} {\bibinfo
  {journal} {Phys.Rev.}\ }\textbf {\bibinfo {volume} {D91}},\ \bibinfo {pages}
  {094029} (\bibinfo {year} {2015})},\ \Eprint
  {https://arxiv.org/abs/1409.7708} {arXiv:1409.7708 [hep-ph]} \BibitemShut
  {NoStop}%
\bibitem [{\citenamefont {Albaladejo}\ \emph {et~al.}(2020)\citenamefont
  {Albaladejo}, \citenamefont {Danilkin}, \citenamefont {Gonz\`alez-Sol\'is},
  \citenamefont {Winney}, \citenamefont {Fern\'andez-Ram\'irez}, \citenamefont
  {Hiller~Blin}, \citenamefont {Mathieu}, \citenamefont {Mikhasenko},
  \citenamefont {Pilloni},\ and\ \citenamefont
  {Szczepaniak}}]{Albaladejo:2020smb}%
  \BibitemOpen
  \bibfield  {author} {\bibinfo {author} {\bibfnamefont {M.}~\bibnamefont
  {Albaladejo}}, \bibinfo {author} {\bibfnamefont {I.}~\bibnamefont
  {Danilkin}}, \bibinfo {author} {\bibfnamefont {S.}~\bibnamefont
  {Gonz\`alez-Sol\'is}}, \bibinfo {author} {\bibfnamefont {D.}~\bibnamefont
  {Winney}}, \bibinfo {author} {\bibfnamefont {C.}~\bibnamefont
  {Fern\'andez-Ram\'irez}}, \bibinfo {author} {\bibfnamefont {A.}~\bibnamefont
  {Hiller~Blin}}, \bibinfo {author} {\bibfnamefont {V.}~\bibnamefont
  {Mathieu}}, \bibinfo {author} {\bibfnamefont {M.}~\bibnamefont {Mikhasenko}},
  \bibinfo {author} {\bibfnamefont {A.}~\bibnamefont {Pilloni}},\ and\ \bibinfo
  {author} {\bibfnamefont {A.}~\bibnamefont {Szczepaniak}},\ }\Eprint
  {https://arxiv.org/abs/2006.01058} {arXiv:2006.01058 [hep-ph]}  (\bibinfo
  {year} {2020})\BibitemShut {NoStop}%
\bibitem [{\citenamefont {Hand}(1963)}]{Hand:1963bb}%
  \BibitemOpen
  \bibfield  {author} {\bibinfo {author} {\bibfnamefont {L.}~\bibnamefont
  {Hand}},\ }\href {https://doi.org/10.1103/PhysRev.129.1834} {\bibfield
  {journal} {\bibinfo  {journal} {Phys.Rev.}\ }\textbf {\bibinfo {volume}
  {129}},\ \bibinfo {pages} {1834} (\bibinfo {year} {1963})}\BibitemShut
  {NoStop}%
\bibitem [{\citenamefont {Bedlinskiy}\ \emph {et~al.}(2014)\citenamefont
  {Bedlinskiy} \emph {et~al.}}]{Bedlinskiy:2014tvi}%
  \BibitemOpen
  \bibfield  {author} {\bibinfo {author} {\bibfnamefont {I.}~\bibnamefont
  {Bedlinskiy}} \emph {et~al.} (\bibinfo {collaboration} {CLAS}),\ }\href
  {https://doi.org/10.1103/PhysRevC.90.039901} {\bibfield  {journal} {\bibinfo
  {journal} {Phys.Rev.}\ }\textbf {\bibinfo {volume} {C90}},\ \bibinfo {pages}
  {025205} (\bibinfo {year} {2014})},\ \bibinfo {note} {[Addendum: Phys.Rev.C
  90, 039901 (2014)]},\ \Eprint {https://arxiv.org/abs/1405.0988}
  {arXiv:1405.0988 [nucl-ex]} \BibitemShut {NoStop}%
\bibitem [{\citenamefont {Adloff}\ \emph {et~al.}(2002)\citenamefont {Adloff}
  \emph {et~al.}}]{Adloff:2002re}%
  \BibitemOpen
  \bibfield  {author} {\bibinfo {author} {\bibfnamefont {C.}~\bibnamefont
  {Adloff}} \emph {et~al.} (\bibinfo {collaboration} {H1}),\ }\href
  {https://doi.org/10.1016/S0370-2693(02)02275-X} {\bibfield  {journal}
  {\bibinfo  {journal} {Phys.Lett.}\ }\textbf {\bibinfo {volume} {B541}},\
  \bibinfo {pages} {251} (\bibinfo {year} {2002})},\ \Eprint
  {https://arxiv.org/abs/hep-ex/0205107} {arXiv:hep-ex/0205107} \BibitemShut
  {NoStop}%
\end{thebibliography}%
\end{document}